\documentclass[11pt]{article}

\usepackage[margin=1in]{geometry}

\usepackage[T1]{fontenc}

\usepackage{amsfonts,amssymb,amsmath,amsthm}
\usepackage{mathtools}
\usepackage{nccmath}
\usepackage{thmtools}
\usepackage{booktabs}

\declaretheorem[style=definition]{definition}

\declaretheorem{theorem}
\declaretheorem{lemma}
\declaretheorem{corollary}

\usepackage{dsfont}
\usepackage[numbers,sort]{natbib}

\usepackage{graphicx}
\usepackage[dvipsnames]{xcolor}
\usepackage[colorlinks=true,pdfpagemode=UseNone,urlcolor=RoyalBlue,linkcolor=RoyalBlue,citecolor=OliveGreen,pdfstartview=FitH]{hyperref}
\usepackage{mathtools}
\usepackage{enumitem}

\usepackage{cleveref}

\usepackage{tcolorbox}
\newtcolorbox{algorithm}[1]
{
	adjusted title = {#1},
	fonttitle = \bfseries,
	before skip = 12pt,
	after skip = 18pt,
	before upper = {\setlength{\abovedisplayskip}{8pt}\setlength{\belowdisplayskip}{8pt}},
}

\newcommand{\R}{\mathbb{R}}
\newcommand{\Z}{\mathbb{Z}}
\newcommand{\Exp}{\mathrm{Exp}}
\newcommand{\Unif}{\mathrm{Unif}}
\newcommand{\E}{\mathbf{E}}
\renewcommand{\Pr}{\mathbf{Pr}}
\renewcommand{\d}{\mathrm{d}}

\newcommand{\Poisson}{\operatorname{Pois}}

\newcommand{\Bernoulli}{\operatorname{Bern}}

\newcommand{\Disjoint}{\mathbin{\Box}}
\newcommand{\1}{\mathbf{1}}
\newcommand{\Ind}{\mathds{1}}
\newcommand{\defeq}{\coloneqq}

\newcommand{\Generator}{\mathcal{G}}
\newcommand{\MatchKernel}{\mathsf{M}}

\newcommand{\WinKernel}{\mathsf{W}}
\newcommand{\AgeCone}{\mathcal{C}_\lambda}
\newcommand{\GapKernel}{\mathsf{G}}
\newcommand{\IdKernel}{\mathsf{Id}}
\newcommand{\InfGenerator}{\mathcal{L}}
\newcommand{\FirstAge}{\widehat A}
\newcommand{\OfflineObj}{\mathcal I}
\newcommand{\OnlineObj}{\mathcal J}
\newcommand{\NormalizedBid}{\bar b}
\newcommand{\AdwordsRatio}{\Gamma^\star_\mathrm{AW}}

\title{Online Correlated Exponential Clocks
    \thanks{The code of the numerical verification in this paper is provided at \href{https://github.com/Jiahao566/Online-Correlated-Exponential-Clocks.git}{Online-Correlated-Exponential-Clocks.git}.}
}

\author{
    Zhiyi Huang\thanks{The University of Hong Kong. Email: zhiyi@cs.hku.hk.}
    \and
    Kalen Patton\thanks{École polytechnique fédérale de Lausanne (EPFL). Email: kalen.patton@epfl.ch.}
    \and
    Peilin Yang\thanks{The University of Hong Kong. Email: peilinyang@connect.hku.hk.}
    \and
    Jiahao Zhao\thanks{The University of Hong Kong. Email: zjiahao@connect.hku.hk.}
}

\date{November 7, 2026}

\begin{document}

\begin{titlepage}

\maketitle
\thispagestyle{empty}
\begin{abstract}
In online resource allocation, we seek to allocate an incoming sequence of requests $j \in \OnlineObj$ among a set of offline resources $i \in \OfflineObj$, where decisions must be made immediately without knowledge of the future. State-of-the-art results typically follow from relax-and-round methods. For many problems, recent advances have come from improved methods of online rounding with negative correlations. However, such rounding schemes have largely been implemented on a case-by-case basis, making it difficult to transfer improvements and new ideas between problem settings. This leads us to ask the question: \textit{Can we develop unified methods of online rounding with negative correlations which can obtain state-of-the-art guarantees across different resource allocation settings?}
    
We answer this question affirmatively by introducing the framework of \textit{Online Correlated Exponential Clocks (OCEC)}. In this framework, fractional allocations $x_{ij}$ are assigned exponential clocks $Y_{ij} \sim \Exp(x_{ij})$, and each request $j$ is allocated to the resource $i$ with smallest $Y_{ij}$. The key idea is that the sequence $(Y_{ij})_j$ for each fixed $i$ is negatively correlated via an \textit{OCEC generator}, resulting in better-than-baseline guarantees.

Our contributions are two-fold: First, we introduce OCEC rounding as a problem-independent interface which any online algorithm can use. 
This isolates the challenge of online rounding into the self-contained problem of designing an effective OCEC generator. 
Second, we demonstrate the flexibility of this method by giving a single OCEC generator which yields improved competitive bounds in three distinct problems: 0.556 for Display Ads, 0.530 for AdWords, and 4.69 for squared-load balancing. Previous best results were obtained by Blanc and Charikar (FOCS 2021), Chen, Huang, and Sun (FOCS 2024), and Borst and Kashaev (SODA 2027), respectively.

\end{abstract}

\end{titlepage}

\section{Introduction}
\label{sec:introduction}

Online resource allocation has broad applications, such as assigning impressions to advertisers and jobs to machines.
In these applications, the resources (e.g., advertisers or machines) are known in advance, while the requests (e.g., impressions or jobs) arrive over time.
The arrival of a request reveals its values or processing sizes on the resources; the algorithm then assigns it immediately and irrevocably, without knowing future requests.
The objective depends on the set of requests assigned to each resource.
In matching~\cite{KarpVV90}, Display Ads~\cite{FeldmanKMMP09}, and AdWords~\cite{MehtaSVV07}, we seek to maximize the value of allocation, while in load balancing~\cite{AwerbuchAGKKV95}, we seek to minimize cost.

A common approach for such problems is to compute a fractional allocation for each arriving request and round it randomly.
These fractional decisions are guided by online primal-dual analysis~\cite{BuchbinderJN07,DevanurJK13}, while the rounding scheme seeks to convert the fractional allocations into integral ones while minimizing additional loss incurred. The simplest choice, independent rounding, assigns each request according to its fractional allocation, independently of earlier decisions.
This clearly preserves fractional allocations as marginal probabilities, but as it is forgetful of its rounding decisions, it may repeatedly overlook the same advertiser or overload the same machine.

By correlating rounding decisions, we can improve the objective while still preserving marginal probability.
For example, suppose an advertiser pays one dollar if it receives at least one of two impressions, each assigned to it with probability $1/2$.
Independent rounding earns $75$ cents in expectation, whereas correlated rounding earns one dollar by giving the advertiser the second impression exactly when it misses the first.
However, achieving such gains online is difficult.
An advertiser that missed earlier impressions should be favored next, but two such advertisers may compete for the same impression.
Moreover, the algorithm cannot plan the correlation in advance, since it does not see future requests.

Despite these difficulties, dependent rounding has driven much recent progress in online advertising and scheduling.
In online advertising, Display Ads and AdWords admit $(1-1/e)$-competitive algorithms under large-market assumptions~\cite{FeldmanKMMP09,MehtaSVV07}, whereas the optimal randomized competitive ratio for AdWords with general bids is strictly below $1-1/e$~\cite{PanigrahiZhang26}.
Without large-market assumptions, the greedy ratio of $1/2$ remained the best known for both problems for over a decade.
\cite{FahrbachHTZ22} introduced online correlated selection (OCS) to obtain the first ratio above $1/2$ for Display Ads, and \cite{HuangZZ20} adapted it to obtain the first such ratio for AdWords.
Subsequent work improved the OCS guarantees and developed multiway and stochastic variants~\cite{ShinAn21,GaoEtAl21,BlancCharikar21,ChenHuangSun2024,FarhadiHajiaghayi26}.\footnote{See \Cref{sec:further-related} for concurrent work \cite{FarhadiHajiaghayi26} on Display Ads.}

In scheduling, independent rounding is $5$-competitive for online squared-load balancing~\cite{Caragiannis08}, and \cite{BorstKashaev25} improved the ratio to $4.9843$ by implementing the correlated rounding of \cite{ImS20} online.
That rounding scheme was developed for the offline minimization of total weighted completion time on unrelated machines.
There, \cite{BansalSS16} first used negatively correlated rounding to break the long-standing $1.5$-approximation barrier of independent rounding, and \cite{ImS20} and subsequent work~\cite{ImL23,BavejaQS24,Harris24,Li25,HarrisLRV26} improved the approximation ratio further.

These online dependent rounding schemes are problem specific, and advances do not readily transfer.
For example, multiway OCS improved Display Ads~\cite{BlancCharikar21}, but the first multiway OCS for AdWords came three years later~\cite{ChenHuangSun2024}.

\subsection{Our Results}

We introduce \emph{online correlated exponential clocks} (OCEC) as a unified framework for online dependent rounding through exponential races.
Each resource $i$ runs an independent \emph{OCEC generator}, an online algorithm that takes the fractional allocation $x_{ij}$ from each request $j$ as input and returns a clock $Y_{ij}$.
For each request, the fractional allocations sum to one and are chosen independently of the rounding randomness.
We impose two requirements on these clocks:
\begin{enumerate}[topsep=4pt,itemsep=2pt]
  \item Each clock $Y_{ij}$ follows the exponential distribution of rate $x_{ij}$.
  \item The clocks $Y_{ij}$ of different requests $j$ are negatively associated.
\end{enumerate}
Each request $j$ goes to the resource with the smallest clock; we call this \emph{OCEC rounding}.
These requirements yield the following rounding guarantees, respectively:
\begin{enumerate}[topsep=4pt,itemsep=2pt]
  \item Each resource wins with probability exactly its fractional allocation $x_{ij}$.
  \item For a submodular reward or a supermodular cost of a resource's assigned requests, it suffices to analyze that resource against fresh competing clocks~\cite{QiuSingla2022}.
\end{enumerate}
A generator that draws fresh clocks yields independent rounding.
By correlating a resource's clocks across requests, a generator can reduce the chance of repeatedly overlooking it.

With this framework, we isolate a common pattern that has appeared both in earlier OCS~\cite{BlancCharikar21} and in dependent rounding schemes for scheduling~\cite{ImS20,Harris24,HarrisLRV26}. These works each build around a particular clock-generation mechanism,\footnote{
  In OCEC terms, the iterative fair contention resolution of \cite{ImS20} can be recast as an exponential race once its Poisson tickets are read as arrival times; its online implementation by \cite{BorstKashaev25} then fits our definition of OCEC rounding.
  \cite{Harris24} and \cite{HarrisLRV26} explicitly use exponential races based on multivariate geometric and Dirichlet variables; the online Dirichlet construction likewise fits our definition.
  For request masses that are multiples of $1/m$, the multiway OCS of \cite{BlancCharikar21} can also be expressed in our framework.
} but this common underlying tool is somewhat obscured by their different problem-specific implementations. We formalize this tool for the first time in the abstract, providing an interface that any online algorithm can implement. This also separates OCEC design into a standalone problem. Different applications can then reuse an OCEC generator while tailoring the fractional allocation and analysis to their objectives.

Our second contribution is an OCEC generator that simultaneously improves the competitive ratios of three online problems.
This generator runs a Poisson process of rate $\lambda$ on an internal timeline and assigns each request $j$ the next interval $I_j$ of length $x_{ij}$.
The \emph{first age} of $I_j$ is the time between its first Poisson arrival and the preceding Poisson arrival, or zero if $I_j$ contains no arrival.
The generator returns a clock coupled with the first age of $I_j$ through their distribution functions, so that larger first ages give smaller clocks and thus higher winning probabilities.
Intuitively, a large first age for one request requires a long preceding gap without arrivals.
Requests whose intervals lie within that gap have first age zero, so the coupling favors the request at the end of the gap while disfavoring those inside it.
All three improvements, for Display Ads, AdWords, and squared-load balancing, use this single generator with the same arrival rate $\lambda=4.5$.\footnote{This $\lambda$ is approximately optimal for each of the three applications under our analyses.}
The applications differ in their fractional allocation rules and in when they use correlated rather than independent clocks.

\begin{table}[htbp]
  \centering
  \medskip
  \renewcommand{\arraystretch}{1.15}
  \begin{tabular}{@{}l@{\hspace{2em}}l@{~~}l@{\hspace{2em}}l@{~~}l@{\hspace{2em}}l@{~~}l@{}}
    \toprule
    Problem                & \multicolumn{2}{@{}l@{\hspace{2em}}}{Baseline}                             & \multicolumn{2}{@{}l@{\hspace{2em}}}{Previous ratio} & \multicolumn{2}{@{}l@{}}{\textbf{This paper}}                                                                          \\[-1ex]
                           & \multicolumn{2}{@{}l@{\hspace{2em}}}{\footnotesize (independent rounding)} & \multicolumn{2}{l}{}                                 & \multicolumn{2}{l}{}                                                                                                   \\
    \midrule
    Display Ads            & $0.5$                                                                      &                                                      & $0.5368$                                      & \cite{BlancCharikar21}  & \textbf{0.556} & (\Cref{thm:display-ads})    \\
    AdWords                & $0.5$                                                                      &                                                      & $0.504$                                       & \cite{ChenHuangSun2024} & \textbf{0.530} & (\Cref{thm:adwords})        \\
    Squared-load balancing & $5$                                                                        & \cite{Caragiannis08}                                 & $4.9843$                                      & \cite{BorstKashaev25}   & \textbf{4.69}  & (\Cref{thm:load-balancing}) \\
    \bottomrule
  \end{tabular}

\end{table}

\subsection{Overview of Techniques}
Here we discuss our main techniques and their relationship to the work of \cite{BlancCharikar21} and \cite{BorstKashaev25}.

\paragraph{Negative association}
The first crucial step in our analysis is the requirement that the clocks in our OCEC generators be negatively associated. This property lets us reduce the complex dynamics of multiple competing generators to an analysis of each generator in isolation. By the submodular dominance property of negatively associated random variables~\cite{QiuSingla2022}, replacing a resource's competitors with fresh, independent clocks can only decrease its expected submodular reward (or for load balancing, increase its expected supermodular cost). Hence, we can analyze a ``fresh-competitor process'' for each resource separately, obtaining a lower bound on reward for the advertising problems and an upper bound on cost for squared-load balancing.

\paragraph{State tracking with age-measures}
For our advertising problems, we track each generator's state in the fresh-competitor process using an \emph{age-measure}. Recall that the age is the time elapsed since the generator's last Poisson arrival. For example, in matching, the measure records both the probability that a resource is still unmatched and the distribution of its age in those outcomes. These measures evolve predictably as requests arrive. This extra information matters because a longer wait since the last arrival makes the resource more likely to win when the next Poisson arrival occurs. For Display Ads and AdWords, we extend this state tracking to account for impression values and remaining budgets, respectively.

\paragraph{Primal-dual analysis via a potential function}
We use a potential function on these age-measures to turn this state information into an online allocation rule and a competitive guarantee. Intuitively, the potential assigns a value to a resource's remaining opportunities for reward. Each resource makes an offer for an arriving request based on this potential and the request's value, and we allocate the request fractionally to the resources with the highest current offers. The potential also supplies the accounting for our primal-dual analysis: spent potential contributes to the resource's dual variable, while the offers determine the request's dual variable. We choose the potential so that the reward credited to each request pays for both contributions. Together with the allocation rule, this yields a bound comparing our expected reward to the offline optimum.

\paragraph{Relation with \cite{BlancCharikar21}.}
Our OCEC generator is inspired by the continuous limit of the multiway OCS of \cite{BlancCharikar21}, in which their geometric arrival process becomes a Poisson process.
Viewed through OCEC, their construction first converts arrival ages into desired winning probabilities, then couples these probabilities to exponential clocks.
Ensuring that the clocks achieve the desired winning probabilities imposes an additional feasibility condition.
Our generator maps the first arrival age directly to an exponential clock, avoiding this intermediate step and its associated restriction.

For Display Ads, we also retain more information across \emph{gaps}, in which low-value requests interrupt the rounding guarantee at a higher level.
Such low-value requests cannot earn a reward at that level, but they can affect future winning chances by disrupting the constructed negative correlations.
The gap analysis of \cite{BlancCharikar21} simply discards correlation between requests before and after a gap. On the other hand, tracking the age-measure through the gap lets us account for some of the surviving correlation, limiting the resulting loss in our potential bound.

\paragraph{Covariance savings for load balancing.}
For squared-load balancing, negative correlation reduces the probability of assigning two jobs to the same machine.
Let $X_{ij}$ indicate whether job $j$ is assigned to machine $i$, and let $I_{ij}$ be job $j$'s interval on machine $i$'s timeline.
For two jobs $j<k$ that machine $i$ rounds with the OCEC generator ($\lambda=4.5$), we prove the covariance bound
$$
  \operatorname{Cov}(X_{ij},X_{ik})
  \le -\int_{I_{ij}}\int_{I_{ik}}\delta(v-u)\,dv\,du,
  \qquad
  \delta(d)=0.86\exp\!\left(-d-\frac{25}{3}d^2\right).
$$
Thus every pair of such jobs contributes a saving, which decays with their separation on the timeline.

\paragraph{Relation with \cite{BorstKashaev25}.}
Both \cite{BorstKashaev25} and this paper follow the online primal-dual framework, with some implementation differences.\footnote{Both algorithms apply online dependent rounding only to small fractional allocations.
  \cite{BorstKashaev25} explicitly require the job size to be comparable to a dual variable tracking the machine's expected load.
  We instead invoke OCEC rounding when independent rounding fails the local primal-dual inequality, which implies analogous size bounds.
}
Our improved competitive ratio comes mainly from our covariance-saving bound.
They partition jobs on each machine into groups of total fractional allocation at most one.
For two jobs with infinitesimal fractional allocations, their guaranteed covariance-saving rate is $1-2/(e+1)\approx0.462$ within a group and zero across groups.
Our bound has $\delta(0)=0.86$ and decays with distance, while remaining positive at every finite separation.

\subsection{Further Related Work}
\label{sec:further-related}

\paragraph{Online matching.}
\cite{KarpVV90} introduced online bipartite matching and gave the optimal $(1-1/e)$-competitive Ranking algorithm for unweighted graphs under adversarial arrivals.
\cite{AggarwalGKM11} extended this guarantee to vertex-weighted matching by randomly perturbing vertex weights.
The Display Ads~\cite{FeldmanKMMP09} and AdWords~\cite{MehtaSVV07} problems considered in this paper also generalize online matching.
For further background and references, see the surveys~\cite{Mehta13,HuangTW24}.

\paragraph{Online rounding for matching.}
Besides the OCS literature, a related line of work studies online rounding of fractional bipartite matchings with vertexwise guarantees: each offline vertex $i$ is matched with probability at least a $\rho$ fraction of its total fractional allocation.
This line overlaps with OCS as the algorithm of \cite{GaoEtAl21} provides such guarantees.
However, improving the rounding ratio does not automatically yield an improved competitive ratio for online matching.
\cite{BuchbinderNW23} identified structured fractional matchings that admit lossless online rounding.
\cite{NaorSW25} consider the stronger requirement of retaining a $\rho$ fraction of every edge's marginal, which implies the vertexwise guarantee.
They achieve $\rho=0.652$, which \cite{HarrisLRV26} improve to $0.68$.

\paragraph{Online load balancing.}
\cite{AwerbuchAGKKV95} studied online load balancing under $\ell_p$ norms, giving $O(p)$-competitive algorithms for finite $p$ and showing that greedy is $(3+2\sqrt{2})$-competitive for the sum of squared loads.
\cite{Caragiannis08} proved that this ratio is optimal for deterministic algorithms and gave a $5$-competitive randomized algorithm based on independent rounding.
\cite{BorstKashaev25} later showed that $5$ is optimal among independent-rounding algorithms.
\cite{GuptaKP13} developed convex-duality analyses for load balancing with power costs and linear assignment costs, connecting these problems to online speed scaling.
More recent work extends online load balancing to symmetric norms~\cite{KesselheimMS23,KesselheimMS24} and develops integral online algorithms for broader classes of convex objectives~\cite{KesselheimMPS25}.

\paragraph{Offline dependent rounding.}
\cite{Srinivasan01,GandhiKPS06} study offline dependent rounding that preserves marginals with negative correlation under cardinality and bipartite degree constraints, respectively.
\cite{AgeevS04} introduced pipage rounding, which \cite{CalinescuCPV11} extended to matroid constraints for monotone submodular maximization.
Swap rounding~\cite{ChekuriVZ10} provides analogous marginal and correlation guarantees under matroid constraints, with concentration bounds for linear functions and lower-tail bounds for monotone submodular functions.
These methods have access to the full fractional solution, whereas our framework must round each request upon arrival.

\paragraph{Concurrent Work on Display Ads.}
In parallel with this work, \cite{FarhadiHajiaghayi26} obtained a $0.556195$-competitive algorithm for Display Ads, compared with our guarantee of $0.556297$ (calculated in \Cref{app:gamma-star}).
Both primal-dual analyses use a similar idea of tracking an age measure at each weight threshold.
Their work follows the multiway OCS framework of~\cite{BlancCharikar21}, in which they map discrete ages to target winning probabilities for a tournament among sampled candidates. On the other hand, our generator uses continuous ages, which are mapped directly to exponential clocks.

\section{Preliminaries on Negative Association}

\label{sec:negative-association}

This section collects the standard facts about negative association that we use.
For any $X = (X_k)_{k \in K}$ indexed by a countable set $K$, we write $X_A = (X_k)_{k \in A}$ for any $A \subseteq K$.

\begin{definition}[Negative Association \cite{JoagDevProschan83,QiuSingla2022}]
    Consider countably many random variables $X=(X_k)_{k\in K}$ with values in $\R\cup\{\infty\}$.
    They are \emph{negatively associated} if for all finite disjoint $A,B\subseteq K$ and all bounded Borel functions $f$ and $g$ that are both nondecreasing or both nonincreasing in each coordinate:
    \begin{equation}
        \E\bigl[f(X_A)\,g(X_B)\bigr]
        \le\E\bigl[f(X_A)\bigr]\,\E\bigl[g(X_B)\bigr].
        \label{eqn:negative-association}
    \end{equation}
    A random subset $S \subseteq K$ is negatively associated if the indicators $\Ind\{k\in S\}$ are.
\end{definition}

For example, mutually independent random variables are negatively associated, and so are random indicators that sum to at most one~\cite{Wajc17}.

An event is \emph{upward closed} and \emph{determined by $X_A$} if it has the form $\{X_A\in U\}$ for an upward-closed Borel set $U$, i.e., $x \in U$ and $y \ge x$ coordinate-wise imply $y \in U$.

\begin{lemma}[Characterization by Upward-Closed Events {\cite[Observation~1(b)]{DubhashiR94}}]
    \label{lem:na-events}
    Random variables $X=(X_k)_{k\in K}$ are negatively associated if and only if:
    \[
        \Pr[E_A \cap E_B] \le \Pr[E_A]\,\Pr[E_B]
    \]
    for all finite disjoint $A, B\subseteq K$ and all upward-closed events $E_A, E_B$ determined by $X_A, X_B$.
\end{lemma}

We will verify the condition of \Cref{lem:na-events} using the following inequality for product distributions.

\begin{lemma}[van den Berg--Kesten--Reimer (BKR) Inequality \cite{Reimer00}]
    \label{lem:bkr}
    Consider a product distribution on $\Omega = \prod_{k \in K} \Omega_k$, where $K$ and each $\Omega_k$ are finite.
    For $\omega\in\Omega$ and $A\subseteq K$, we say that $A$ \emph{certifies} an event $E$ at $\omega$ if every $\omega'\in\Omega$ with $\omega'_A=\omega_A$ lies in $E$.
    For events $E,F\subseteq\Omega$, their \emph{disjoint occurrence} is:
    \[
        E\Disjoint F
        =\bigl\{\omega\in\Omega:\text{there exist disjoint }A,B\subseteq K\text{ that certify }E\text{ and }F\text{ at }\omega\text{, respectively}\bigr\}.
    \]
    We have:
    \[
        \Pr[E\Disjoint F]\le\Pr[E]\,\Pr[F].
    \]
\end{lemma}

We will use the following closure properties to construct negatively associated clocks.

\begin{lemma}[Closure Properties]
    \label{lem:na-closure}
    Negative association is preserved by the following operations.
    \begin{enumerate}
        \item \emph{Independent unions~\cite{JoagDevProschan83}:~} If each of $X$ and $X'$ is negatively associated and they are independent of each other, then their union is negatively associated.
        \item \emph{Concordant monotone maps~\cite{JoagDevProschan83,BlancCharikar21}:~}
              Suppose $X = (X_k)_{k\in K}$ is negatively associated and $(A_\ell)_{\ell\in L}$ are finite disjoint subsets of $K$.
              Let $(h_\ell)_{\ell\in L}$ be Borel randomized maps that are all stochastically nondecreasing or all stochastically nonincreasing in each coordinate, with fresh randomness independent across $\ell\in L$ and of $X$.
              Then, $(h_\ell(X_{A_\ell}))_{\ell\in L}$ is negatively associated.
        \item \emph{Limits~\cite{LastSzekliYogeshwaran20}:~} If $X^{(n)}$ is negatively associated for every $n$ and $X^{(n)}$ converges to $X$ in distribution, then $X$ is negatively associated.
    \end{enumerate}
\end{lemma}

Finally, the following lemma compares negatively associated random sets with independent random sets having the same marginals.

\begin{lemma}[Submodular Dominance \cite{QiuSingla2022}]
    \label{lem:submodular-dominance}
    Let $S$ be a negatively associated random subset of a finite set $K$.
    Let $S^{\mathrm{Ind}}$ include each $k \in K$ independently with probability $\Pr[k \in S]$.
    For every submodular function $f : 2^K \to \R$, we have
    $
        \E \bigl[ f(S) \bigr] \ge \E \bigl[ f(S^{\mathrm{Ind}}) \bigr].
    $
\end{lemma}

\section{Online Correlated Exponential Clocks}
\label{sec:ocec}
In this section, we formally define OCEC generators, rounding, and the ``First-Age Poisson generator'' that we will use across all our applications.
\subsection{Definitions and Properties}

Let $\OfflineObj$ be a finite set of offline resources (e.g., advertisers or machines) and $\OnlineObj$ a finite set of online requests (e.g., impressions or jobs), with $\OnlineObj$ indexed in arrival order.
Throughout, $\Exp(0)$ is a point mass at $\infty$ and $\Exp(\infty)$ is a point mass at $0$ by convention.

\begin{definition}[OCEC Generator]
    An \emph{OCEC generator} $\Generator$ is an online algorithm.
    It receives as input a sequence of fractional allocations $x \in [0, 1]^\OnlineObj$, one for each request, fixed independently of the algorithm's internal randomness.
    Upon receiving each allocation $x_j$, it returns a clock $Y_j\in\R_+\cup\{\infty\}$.
    For every such input sequence, we require the following properties:
    \begin{enumerate}
        \item \emph{Exponential marginals:~} For every $j \in \OnlineObj$, $Y_j\sim\Exp(x_j)$.
        \item \emph{Negative association:~} Clocks $(Y_j)_{j\in \OnlineObj}$ are negatively associated.
    \end{enumerate}
\end{definition}

\begin{definition}[OCEC Rounding]\label{def:ocec-rounding}
    Given an OCEC generator $\Generator$ and fractional allocations $x \in[0,1]^{\OfflineObj \times \OnlineObj}$ satisfying $\sum_{i \in \OfflineObj} x_{ij} \le 1$ for every $j \in \OnlineObj$, fixed independently of the generators' randomness, the corresponding \emph{OCEC rounding} is as follows.
    \begin{enumerate}
        \item For each resource $i \in \OfflineObj$, run an independent copy $\Generator_i$ of $\Generator$.
        \item For each request $j \in \OnlineObj$ and its fractional allocation $(x_{ij})_{i \in \OfflineObj}$:
              \begin{enumerate}
                  \item Send $x_{ij}$ to the generator $\Generator_i$ for each $i$, and let $Y_{ij}$ be the returned clock.
                  \item Draw a fresh discard clock $\bar Y_j\sim\Exp(1-\sum_i x_{ij})$.
                  \item Allocate $j$ to the resource with the minimum exponential clock $Y_{ij}$, if it is below $\bar{Y}_j$.
              \end{enumerate}
    \end{enumerate}
    Let $S_i$ be the set of requests allocated to resource $i$.
\end{definition}

Since the clocks $Y_{ij}\sim\Exp(x_{ij})$ and the discard clocks $\bar Y_j\sim\Exp(1-\sum_i x_{ij})$ are independent across resources, and their total rate is 1, each $i$ has the smallest clock with probability $x_{ij}$.
That is, OCEC rounding preserves the marginals.

\begin{lemma}
    \label{lem:marginal}
    For any $i \in \OfflineObj$ and any $j \in \OnlineObj$, $\Pr \bigl[ j \in S_i \bigr] = x_{ij}$.
\end{lemma}

The next lemma localizes the analysis of each resource $i$ to its own generator.
Replacing the clocks of $i$'s competitors with fresh ones cannot increase the expectation of any submodular function of $S_i$, nor decrease that of any supermodular one.

\begin{lemma}
    \label{lem:ocec-dominance}
    Fix any $i\in\OfflineObj$, and let $S_i^{\mathrm{Ind}}=\{j\in\OnlineObj:Y_{ij}<Z_j\}$ for fresh competing clocks $Z_j\sim\Exp(1-x_{ij})$.
    Then, for every submodular function $f:2^\OnlineObj\to\R$,
    \[
        \E\bigl[f(S_i)\bigr]\ge\E\bigl[f(S_i^{\mathrm{Ind}})\bigr],
    \]
    and the inequality reverses for every supermodular $f$.
\end{lemma}

\begin{proof}
    It suffices to prove the inequality conditional on the clocks $Y_i=(Y_{ij})_{j\in\OnlineObj}$ of $\Generator_i$.
    We verify the hypotheses of \Cref{lem:submodular-dominance} for $S_i$ and $S_i^{\mathrm{Ind}}$ given $Y_i$.

    First, $S_i$ is negatively associated.
    The competing clocks $(Y_{i'j})_{i'\ne i,\,j\in\OnlineObj}$ are negatively associated by the definition of OCEC generators and the closure under independent unions (\Cref{lem:na-closure}), and remain so given $Y_i$ because they are independent of it.
    For each $j$, the indicator of $j\in S_i$, i.e., of $\min_{i'\ne i}Y_{i'j}>Y_{ij}$ and $\bar{Y}_j>Y_{ij}$, is nondecreasing in the block $(Y_{i'j})_{i'\ne i}$ and $\bar{Y}_j$, and these blocks are disjoint across $j$.
    The closure under concordant monotone maps completes this step.

    Second, $S_i^{\mathrm{Ind}}$ includes each $j$ independently, since $Z_j$ is fresh, and with the same probability as $S_i$.
    Indeed, the clocks $Y_{i'j}$, $i'\ne i$, and the discard clock $\bar{Y}_j$ are independent exponentials with total rate $1-x_{ij}$, so their minimum, like $Z_j$, follows $\Exp(1-x_{ij})$ and is independent of $Y_i$.

    \Cref{lem:submodular-dominance} then gives the conditional inequality for submodular $f$; applying it to $-f$ gives the supermodular case.
\end{proof}

\subsection{First-Age Poisson Generator}

We now describe an OCEC generator driven by a stationary Poisson process of rate $\lambda$ on $\R$, with arrivals $(T_k)_{k\in\Z}$ indexed in increasing order so that $T_0\le 0<T_1$.
The \emph{age} at time $t$ is $A(t) = t-\max\{T_k:T_k \leq t\}$, the time elapsed since the most recent arrival no later than $t$.
The age of an arrival $T_k$ is $T_k-T_{k-1}$, i.e., the age at the moment right before $T_k$.
For a bounded interval $I$, let $\FirstAge(I)$ denote the \emph{first age} of $I$, i.e., the age of the first arrival in $I$, or $0$ if $I$ contains no arrivals.
Let $F_x$ denote the cumulative distribution function of $\FirstAge(I)$ for an interval $I$ of length $x$; by stationarity, this distribution depends only on $x$.
A direct calculation, deferred to \Cref{app:ocec}, gives the following closed form.

\begin{lemma}
    \label{lem:first-age-cdf}
    For every $x>0$ and $a\ge0$:
    \begin{equation}
        F_x(a)
        =\begin{cases}
            1+e^{-\lambda x}-(1+\lambda a)e^{-\lambda a}, & 0\le a<x; \\[.5ex]
            1-\lambda x e^{-\lambda a},                   & x \leq a.
        \end{cases}
        \label{eqn:first-age-cdf}
    \end{equation}
\end{lemma}

The \emph{First-Age Poisson generator} lays the incoming allocations consecutively on the timeline $[0,\infty)$: each $x_j$ receives the next interval $I_j$ of length $x_j$.
It transforms each $\FirstAge(I_j)$ into an exponential clock by a probability integral transform with $F_{x_j}$, randomized when $I_j$ contains no arrivals.
The generator samples $-T_0$ and $T_1$ independently from $\Exp(\lambda)$ and generates subsequent arrivals online by successively adding independent $\Exp(\lambda)$ gaps.

\begin{algorithm}{First-Age Poisson Generator}
    \label{alg:first-age-poisson-generator}
    \textbf{Parameter:~} Arrival rate $\lambda>0$.

    \medskip

    \textbf{Online clock generation upon receiving allocation $x_j$:}
    \begin{enumerate}
        \item If $x_j=0$, return $Y_j = \infty$.
        \item Take the next interval
              $I_j=[\sum_{j'<j}x_{j'},\sum_{j'\le j}x_{j'})$,
              and compute $\FirstAge(I_j)$.
        \item Draw a fresh $V_j\sim\mathrm{Unif}[0,1]$ and set
              \begin{equation}
                  U_j=\begin{cases}
                      e^{-\lambda x_j}V_j,     & \FirstAge(I_j)=0; \\[.5ex]
                      F_{x_j}(\FirstAge(I_j)), & \FirstAge(I_j)>0.
                  \end{cases}
                  \label{eqn:first-age-uniform}
              \end{equation}
        \item Return $Y_j=-\mfrac{\log U_j}{x_j}$.
    \end{enumerate}
\end{algorithm}

Note that $U_j \sim \Unif[0,1]$ by construction, hence $Y_j \sim \Exp(x_j)$.
Each clock is also nonincreasing in the first age of its interval, so the clocks are negatively associated if the first ages are.
Intuitively, the first ages are negatively associated because a large first age in one interval requires a long empty stretch before its first arrival, which leaves less room for large first ages in the other intervals.

\begin{lemma}
    \label{lem:first-age-negative-association}
    For any pairwise disjoint intervals $I_1,\ldots,I_n$ of finite positive length, the first ages $\FirstAge(I_1),\ldots, \FirstAge(I_n)$ are negatively associated.
\end{lemma}

\begin{proof}
    We discretize time into cells $[\ell\delta,(\ell+1)\delta)$, $\ell\in\Z$, whose occupancies by Poisson arrivals are independent.
    The \emph{priority} $P_\ell$ of an occupied cell $\ell$ is one plus the number of consecutive empty cells before it; empty cells have priority $0$.
    For small $\delta$, the first age of an interval is approximately $\delta$ times the priority of its first occupied cell.
    Since the BKR inequality (\Cref{lem:bkr}) applies to finitely many independent variables, we first cap the priorities and consider $\min\{ P_\ell, \bar{p}\}$ for a positive integer $\bar{p}$, so that each depends on finitely many cells.

    We show that the capped priorities are negatively associated.
    Let $E$ and $F$ be upward-closed events determined by finite disjoint sets of capped priorities.
    A capped priority of at least $p\ge1$ is certified by its occupied cell together with the $p-1$ empty cells before it, and these certificates are disjoint across occupied cells, since no occupied cell lies among the empty cells preceding another.
    Since $E$ and $F$ are upward closed, these certificates show that whenever $E$ and $F$ both occur, they occur on disjoint sets of cells.
    The BKR inequality (\Cref{lem:bkr}) then gives
    \[
        \Pr[E\cap F]=\Pr[E\Disjoint F]\le\Pr[E]\,\Pr[F],
    \]
    which characterizes negative association (\Cref{lem:na-events}).

    As the cap grows, the capped priorities converge to the priorities, which are therefore negatively associated by the closure under limits (\Cref{lem:na-closure}).
    For each $j$, let $K_j$ be the set of cells contained in $I_j$, nonempty for small $\delta$, and let $\ell_j=\min K_j$.
    The discretized first age, $\delta$ times the priority of the first occupied cell in $K_j$, or $0$ if there is none, can be expressed as
    \[
        \FirstAge_\delta(I_j)=\delta\sum_{\ell\in K_j}P_\ell\,\Ind\{P_\ell\ge\ell-\ell_j+1\},
    \]
    since $P_\ell\ge\ell-\ell_j+1$ exactly when $\ell$ is the first occupied cell in $K_j$.
    This expression is nondecreasing in the priorities of the cells in $K_j$, and the sets $K_j$ are disjoint across $j$, so the discretized first ages are negatively associated by the closure under concordant monotone maps.
    Finally, $\FirstAge_\delta(I_j)\to\FirstAge(I_j)$ almost surely as $\delta\to0$, and the closure under limits completes the proof.
\end{proof}

In fact, the proof shows that, for every $\delta$, any functions of $I_1,\ldots,I_n$, each nondecreasing in the priorities of its interval's cells, are negatively associated, and so are their almost-sure limits as $\delta\to0$, such as the largest age or the total age of the arrivals in each interval.
Mapping any such function of $I_j$ to an $\Exp(x_j)$ clock by a nonincreasing map with fresh randomness thus also yields an OCEC generator.

\begin{lemma}
    \label{lem:first-age-ocec}
    The First-Age Poisson generator is an OCEC generator.
\end{lemma}

\begin{proof}
    Coordinates with $x_j=0$ have clocks $Y_j=\infty$, which follow $\Exp(0)$ by convention.
    We omit such coordinates since adjoining constants preserves negative association by the closure under independent unions (\Cref{lem:na-closure}).

    Recall that the input sequence and the corresponding intervals are fixed independently of the generator's randomness.
    For each $j$, $\FirstAge(I_j)$ has distribution function $F_{x_j}$, which is continuous except for an atom of mass $e^{-\lambda x_j}$ at zero (\Cref{lem:first-age-cdf}).
    Hence, the randomized probability integral transform in \eqref{eqn:first-age-uniform} makes $U_j$ uniform on $[0,1]$, so $Y_j=-\log U_j/x_j\sim\Exp(x_j)$.

    The intervals $I_j$ are pairwise disjoint, so their first ages are negatively associated by \Cref{lem:first-age-negative-association}.
    For every fixed $V_j$, $U_j$ is nondecreasing in $\FirstAge(I_j)$, since $F_{x_j}$ is nondecreasing and $e^{-\lambda x_j}V_j\le e^{-\lambda x_j}=F_{x_j}(0)$.
    Hence, each $Y_j$ is a nonincreasing map of $\FirstAge(I_j)$ with fresh randomness $V_j$, and the clocks are negatively associated by the closure under concordant monotone maps.
\end{proof}

\paragraph{Choice of $\lambda$.}
The rest of the paper fixes $\lambda= 4.5$, a rounded value that approximately optimizes the competitive ratios of all applications under our analysis.

\section{Warm-Up: Unweighted Matching}
\label{sec:matching}

This section applies OCEC to unweighted matching, the simplest setting for our analysis, which \Cref{sec:display-ads,sec:adwords} extend to Display Ads and AdWords.
Our goal here is to present the analysis in its simplest form rather than to optimize the ratio. Ranking already achieves the optimal ratio $1-1/e$ for this problem~\cite{KarpVV90}.

Consider a bipartite graph $G=(\OfflineObj, \OnlineObj,E)$.
Offline vertices $i\in \OfflineObj$ are given upfront, and online vertices $j\in \OnlineObj$ arrive one at a time.
When $j$ arrives, the algorithm observes its edges and immediately and irrevocably matches it to at most one neighbor.
The objective is the number of matched offline vertices.
We assume without loss of generality that every online vertex has at least one neighbor, since an online vertex without neighbors would be irrelevant.
The optimal matching size is at most the value of the standard matching LP, and hence, by weak duality, at most the objective of any feasible solution to its dual:
\begin{align*}
    \max_{y\ge0} \quad            & \textstyle \sum_{(i,j)\in E}y_{ij}
                                  &
    \min_{\alpha,\beta\ge0} \quad & \textstyle \sum_{i \in \OfflineObj} \alpha_i + \sum_{j \in \OnlineObj} \beta_j   \\
    \text{s.t.} \quad             & \textstyle \sum_{j:(i,j)\in E} y_{ij}\le1 \quad \forall i \in \OfflineObj      &
    \text{s.t.} \quad             & \alpha_i+\beta_j\ge1 \quad \forall(i,j)\in E                                     \\[1ex]
                                  & \textstyle \sum_{i:(i,j)\in E} y_{ij}\le1 \quad \forall j \in \OnlineObj
\end{align*}

We use $y$ for the LP variables and reserve $x$ for the algorithm's allocation masses.
An offline vertex $i$'s total fractional allocation $\sum_{j:(i,j)\in E}x_{ij}$ may exceed $1$, so $x$ need not be feasible for the matching LP.
The same distinction applies to Display Ads and AdWords.

\begin{theorem}
    \label{thm:matching}
    With the parameter $\lambda = 4.5$, Online Primal-Dual Allocation with First-Age Poisson OCEC in \Cref{sec:matching-primal-dual} is $0.556$-competitive for unweighted matching.
\end{theorem}

We first analyze OCEC rounding for a fixed sequence of fractional allocations, and then design the fractional allocation rule via an online primal-dual analysis to prove \Cref{thm:matching}.

\subsection{Matching Kernel}
\label{sec:matching-kernel}

\paragraph{Fresh competitor process.}
Fix fractional allocations, rounded by OCEC with the First-Age Poisson generator.
Since an offline vertex $i$ is matched exactly when $S_i \ne \varnothing$, and $S_i \mapsto \Ind\{ S_i \ne \varnothing \}$ is submodular, \Cref{lem:ocec-dominance} shows that $i$ is unmatched with probability at most that of losing every comparison against fresh competing clocks of rates $1-x_{ij}$.
We track this probability one request at a time.
We use this comparison process in all applications
considered in this paper.

\paragraph{Ages as states.}
Given the current age of $i$'s generator, its future arrivals and competing clocks are independent of the past, so the age is the state of $i$ between requests.
We describe this state by a measure $\nu$ on ages, where $\nu(H)$ denotes the probability that $i$ has lost every comparison so far and has current age in a measurable set $H$.
Initially, $\nu$ is the stationary distribution $\pi=\Exp(\lambda)$.
Its total mass $\nu([0,\infty))$ may be less than $1$, so $\nu$ need not be a probability measure.

\paragraph{Requests as transitions.}
Consider a request $x$, placed on $[0,x)$ of $i$'s timeline so that $A(0)$ and $A(x)$ are the ages before and after it.
Let $Y$ be its clock.
Given $Y$, $i$ loses to the fresh competing clock with probability $1-e^{-(1-x)Y}$.
The \emph{matching kernel} $\MatchKernel_x$ maps the measure $\nu$ of ages before the request to the measure after it:
\[
    (\nu\MatchKernel_x)(H)\defeq\int\E\bigl[\Ind\{A(x)\in H\}\bigl(1-e^{-(1-x)Y}\bigr)\bigm|A(0)=a\bigr]\,\nu(\d a).
\]
A request $x=0$ changes nothing, and $i$ always wins a request $x=1$ since it then has no competitor; that is, $\nu\MatchKernel_0=\nu$ and $\nu\MatchKernel_1=0$.

For a bounded measurable function $f$ on ages, write $\nu f\defeq\int f\,\d\nu$, which is the expected value of $f$ at the current age, counting only outcomes in which $i$ has lost every comparison so far.
Then, $(\nu\MatchKernel_x)f$ is the same quantity after the request.
We also let $\MatchKernel_x$ act on functions, so that $\MatchKernel_xf$ maps each initial age to this expected value.
Formally,
\begin{equation}
    (\MatchKernel_xf)(a)\defeq\E\bigl[f(A(x))\,\bigl(1-e^{-(1-x)Y}\bigr)\bigm|A(0)=a\bigr].
    \label{eqn:matching-kernel-definition}
\end{equation}
By Fubini's theorem, $(\nu\MatchKernel_x)f=\nu(\MatchKernel_xf)$, so we may write $\nu\MatchKernel_xf$ without ambiguity.

In matrix terms, $\MatchKernel_x$ is a substochastic transition matrix, measures are row vectors, and functions are column vectors.
Applying kernels in sequence corresponds to consecutive requests.

\paragraph{Reachable measures.}
For any offline vertex $i$, applying the kernels of its requests in arrival order, starting from $\pi$, yields $\nu_i\defeq\pi\prod_{j:(i,j)\in E}\MatchKernel_{x_{ij}}$.
More generally, we call a measure \emph{reachable} if it can be obtained by consecutively applying matching kernels to the stationary distribution $\pi$.

The total mass $\nu_i\mathbf1$ is the probability that $i$ loses every comparison, which bounds the probability that $i$ is unmatched by the reduction above:
\begin{equation}
    \Pr[i\text{ is unmatched}]\le\nu_i\mathbf1.
    \label{eqn:matching-unmatched-bound}
\end{equation}
Consequently, the expected matching size is at least $\sum_i(1-\nu_i\mathbf1)$.

\subsection{Potential of Reachable Measures}
\label{sec:matching-potential}

We construct a \emph{potential} function $\Gamma$ on ages.
Our online primal-dual analysis tracks each offline vertex $i$ through the potential $\nu_i\Gamma$ of its current measure $\nu_i$, with the potential $\Gamma^\star=\pi\Gamma$ of the stationary distribution as the target competitive ratio.

Intuitively, the worst case for matching consists of a sequence of infinitesimal requests, so we derive $\Gamma$ from the limit of such requests.
We define the \emph{infinitesimal generator} of the matching kernel by setting, for each bounded and continuously differentiable function $f$ with bounded derivative,
\begin{equation*}
    (\InfGenerator_{\MatchKernel}f)(a)
    \defeq\lim_{x\downarrow0}
    \frac{(\MatchKernel_xf)(a)-f(a)}{x}.
\end{equation*}
As derived in \Cref{app:matching-gap-kernel-generator}, we have the following lemma.

\begin{lemma}
    \label{lem:matching-infinitesimal-generator}
    For every bounded continuously differentiable function $f$ with bounded derivative,
    $$
        (\InfGenerator_{\MatchKernel}f)(a)
        =f'(a)-(\lambda+e^{-\lambda})f(a)
        +\lambda\bigl(1-e^{-\lambda e^{-\lambda a}}\bigr)f(0).
    $$
    The derivative at zero is the right derivative.
\end{lemma}

Let $\MatchKernel_x^\star$ describe the evolution over total allocation $x$ in the infinitesimal-request limit, i.e. $\MatchKernel_x^\star = \lim_{\epsilon \downarrow 0} (\MatchKernel_\epsilon)^{x/\epsilon}$.
This can be made formal using the exponential operator $\MatchKernel_x^\star := \exp(x\InfGenerator_{\MatchKernel})$, or equivalently by the solution to
$$
    \frac{\d}{\d x}(\MatchKernel_x^\star f)
    =\InfGenerator_{\MatchKernel}(\MatchKernel_x^\star f),
    \qquad
    \MatchKernel_0^\star f=f,
$$
for the test functions $f$ specified above.

To describe the \emph{infinitesimal-request age process} explicitly, consider a small request $x$ at age $a$, whose competing clock has rate $1-x\approx1$.
With probability about $\lambda x$, the request's interval contains an arrival with age about $a$.
By \Cref{lem:first-age-cdf}, its clock is then $Y\approx\lambda e^{-\lambda a}$, so $i$ wins with probability about $e^{-\lambda e^{-\lambda a}}$, and otherwise its age restarts near zero.
Without an arrival, the clock is randomized over the atom of $F_x$ at zero, and $i$ wins with probability $xe^{-\lambda(1-x)}\approx xe^{-\lambda}$, i.e., at rate $e^{-\lambda}$ per unit allocation.
The limiting age process therefore increases at unit rate between Poisson arrivals, resets to zero after a losing arrival, and stops at the first win.

OCS-based analyses of matching~\cite{BlancCharikar21,GaoEtAl21} translate a bound $g(x)$ on the probability of remaining unmatched after total allocation $x$ into a competitive ratio of $1-\int_0^\infty e^{-x}g(x)\,\d x$.
In the infinitesimal-request age process started at age $a$, $g(x)=(\MatchKernel_x^\star\mathbf1)(a)$, which leads to the following potential that intuitively captures the conditional competitive ratio given the initial age $a$:
\begin{equation}
    \Gamma(a)
    \defeq1-\int_0^\infty e^{-x}(\MatchKernel_x^\star\mathbf1)(a)\,\d x.
    \label{eqn:matching-potential-definition}
\end{equation}
The potential is increasing in the age, since an older age makes $i$ more likely to win at the next arrival, after which the age resets to zero unless $i$ wins.
Accordingly, the potential of a reachable measure $\nu$ is $\nu\Gamma$.
The target competitive ratio is
\begin{equation}
    \Gamma^\star
    =\pi\Gamma
    =1-\int_0^\infty e^{-t}\,\pi\MatchKernel_t^\star\mathbf1\,\d t.
    \label{eqn:gamma-star}
\end{equation}

\Cref{app:gamma-star} shows numerically that $\Gamma^\star>0.556$.

\subsection{Online Primal-Dual Algorithm and Analysis}
\label{sec:matching-primal-dual}

The algorithm maintains dual variables $\alpha_i$ and $\beta_j$ alongside the rounding.
To show that the algorithm is $\Gamma^\star$-competitive, it suffices to establish two properties:
\begin{enumerate}[label=(\arabic*)]
    \item \emph{Reverse weak duality}: the expected matching size is at least $\sum_i\alpha_i+\sum_j\beta_j$.
    \item \emph{Approximate dual feasibility}: $\alpha,\beta\ge0$, and $\alpha_i+\beta_j\ge\Gamma^\star$ for every edge $(i,j)$.
\end{enumerate}
Then, $(\alpha/\Gamma^\star,\beta/\Gamma^\star)$ is dual feasible, so by weak duality, the optimal matching size is at most $(\sum_i\alpha_i+\sum_j\beta_j)/\Gamma^\star$, and thus at most $1/\Gamma^\star$ times the expected matching size.

The offline dual variable $\alpha_i\defeq\Gamma^\star-\nu_i\Gamma$ records the potential spent by $i$ from its initial value $\Gamma^\star=\pi\Gamma$.
For an arriving online vertex $j$, let $\nu_i$ denote neighbor $i$'s measure before the allocation.
After allocating mass $x_{ij}$ to $i$, its offline dual becomes $\alpha_i=\Gamma^\star-\nu_i\MatchKernel_{x_{ij}}\Gamma$.
Thus, approximate dual feasibility requires $\beta_j \ge \nu_i\MatchKernel_{x_{ij}}\Gamma$ for every neighbor $i$.
This motivates a water-filling procedure.
We refer to $\nu_i\MatchKernel_{x_{ij}}\Gamma$ as $i$'s offer to $j$.
Starting from $x_{ij}=0$, we continuously allocate $j$ to neighbors with the largest current offers, splitting allocation among tied neighbors to keep their offers equal.
All neighbors receiving positive allocation then have the same maximum final offer.
After water filling, each neighbor $i$ contributes $x_{ij}\nu_i\MatchKernel_{x_{ij}}\Gamma$ to $\beta_j$, using its final offer.
Since the allocated masses sum to $1$, $\beta_j$ equals the maximum final offer and satisfies the required inequalities.

\begin{algorithm}{Online Primal-Dual Allocation with First-Age Poisson OCEC (Matching)}
    \textbf{Internal states for each offline vertex $i$:}
    \begin{itemize}
        \item An independent First-Age Poisson generator $\Generator_i$.
        \item A measure $\nu_i$, initially $\pi$, and the corresponding offline dual $\alpha_i=\Gamma^\star-\nu_i\Gamma$.
    \end{itemize}

    \medskip

    \textbf{Upon the arrival of an online vertex $j$:}
    \begin{enumerate}
        \item \textbf{Allocate:~} Starting from $x_{ij}=0$, continuously allocate to the neighbors $i$ with the largest current offer $\nu_i\MatchKernel_{x_{ij}}\Gamma$ until $\sum_{i:(i,j)\in E}x_{ij}=1$.
        \item \textbf{Update duals:~} Set $\beta_j \gets \max_{i:(i,j)\in E} \nu_i\MatchKernel_{x_{ij}}\Gamma$. For every neighbor $i$, set $\nu_i\gets\nu_i\MatchKernel_{x_{ij}}$ and update $\alpha_i$ accordingly.
        \item \textbf{Round:~} If an unmatched neighbor $i$ wins the OCEC rounding, match $j$ to $i$.
    \end{enumerate}
\end{algorithm}

We need three lemmas about the reachable measures, the potential, and how they interact with the matching kernel.
We give only proof sketches for two of them below, since all three follow from the stronger Display Ads lemmas stated in \Cref{sec:display-ads}, which we prove in \Cref{app:matching-display}.

\begin{lemma}
    \label{lem:matching-measures-bounds}
    For every reachable measure $\nu$, $e^{\lambda a}\nu([a,\infty))$ is nondecreasing in $a\ge0$ and at most one.
    In particular, for $a\geq 0$, we have $(\nu\mathbf1)e^{-\lambda a}\le\nu([a,\infty))\le e^{-\lambda a}.$
\end{lemma}

\begin{proof}[Proof Sketch]
    The tail bound says that conditioning on losing every comparison favors older ages relative to $\pi$.
    Poisson arrivals reset the age and increase the probability of winning, so conditioning on losses favors paths without recent arrivals.
    More precisely, $e^{\lambda a}\nu([a,\infty))$ is the probability of losing every comparison conditional on no arrivals in the last $a$ time units, which is at most one and cannot decrease as this arrival-free interval grows.
\end{proof}

\begin{lemma}\label{lem:matching-request-monotonicity}
    For every reachable measure $\nu$, $\nu\MatchKernel_x\Gamma$ is continuous and nonincreasing in $x$ on $[0,1]$.
\end{lemma}

\begin{lemma}\label{lem:matching-potential-inequalities}
    For every reachable measure $\nu$ and every request $x\in[0,1]$, the probability of a first win on the request covers the resulting dual increase:
    \begin{equation}
        \underbrace{\nu\mathbf1-\nu\MatchKernel_{x}\mathbf1\vphantom{\MatchKernel_{x}\Gamma\mathbf1}}_{\substack{\textnormal{first-win}\\\textnormal{probability}}}
        ~~ \ge ~~
        \underbrace{\nu\Gamma-\nu\MatchKernel_{x}\Gamma\vphantom{\MatchKernel_{x}\Gamma\mathbf1}}_{\substack{\textnormal{potential}\\\textnormal{spent}}}
        ~~+ \underbrace{x\,\nu\MatchKernel_{x}\Gamma\vphantom{\MatchKernel_{x}\Gamma\mathbf1}}_{\substack{\textnormal{$x$ times}\\\textnormal{remaining potential}}}.
        \label{eqn:finite-potential-inequal}
    \end{equation}
\end{lemma}

\begin{proof}[Proof Sketch]
    Collecting terms, it suffices to prove $\nu H_x\ge0$, where $H_x\defeq\mathbf1-\Gamma-\MatchKernel_x\mathbf1+(1-x)\MatchKernel_x\Gamma$.
    For the nondecreasing lower envelope $\bar H_x(a)=\inf_{b\ge a}H_x(b)$, we have $\nu H_x\ge\nu\bar H_x$.
    Moreover, \Cref{lem:matching-measures-bounds} implies $\nu\bar H_x\ge(\nu\mathbf1)\pi\bar H_x$, so it suffices to verify $\pi\bar H_x\ge0$ for the stationary distribution $\pi$.
    Finally, we combine this reduction with discretization to certify the inequality for all $x\in[0,1]$.
\end{proof}

\begin{proof}[Proof of \Cref{thm:matching}]
    We verify the two required properties of online primal-dual analysis.

    \noindent
    \emph{Reverse weak duality.~}
    Consider the arrival of $j$, and let $\nu_i$ be the measure of each neighbor $i$ before it.
    The request $x_{ij}$ increases $\alpha_i$ by $\nu_i\Gamma-\nu_i\MatchKernel_{x_{ij}}\Gamma$ and contributes $x_{ij}\nu_i\MatchKernel_{x_{ij}}\Gamma$ to $\beta_j$, while it decreases $\nu_i\mathbf1$ by $\nu_i\mathbf1-\nu_i\MatchKernel_{x_{ij}}\mathbf1$.
    By \Cref{lem:matching-potential-inequalities}, the decrease is at least the dual increase.
    Summing over all requests, with each $\nu_i$ starting at $\pi$ and each $\alpha_i$ at zero, the final values satisfy $\sum_i(1-\nu_i\mathbf1)\ge\sum_i\alpha_i+\sum_j\beta_j$, and \eqref{eqn:matching-unmatched-bound} bounds the left side by the expected matching size.

    \noindent
    \emph{Approximate dual feasibility.~}
    By \Cref{lem:matching-request-monotonicity}, $\nu_i\MatchKernel_x\Gamma$ is continuous and nonincreasing in $x$, so water filling is well defined and each $\alpha_i$ is nondecreasing from $0$.
    Moreover, $\beta_j\ge0$ since $\Gamma\ge0$.
    After water filling and the dual update for $j$, every neighbor with $x_{ij}>0$ attains the smallest offline dual among the neighbors, so its remaining potential is $\Gamma^\star-\min_{i':(i',j)\in E}\alpha_{i'}$.
    Since the allocations sum to $1$, $\beta_j=\Gamma^\star-\min_{i':(i',j)\in E}\alpha_{i'}$.
    Hence $\alpha_i+\beta_j\ge\Gamma^\star > 0.556$ for every $(i,j) \in E$, and this persists since $\alpha_i$ never decreases.
\end{proof}

\section{Display Ads}
\label{sec:display-ads}

We next extend the analysis to Display Ads, or edge-weighted matching with free disposal.
In this problem, advertisers $i\in\OfflineObj$ are given upfront, and impressions $j\in\OnlineObj$ arrive one at a time.
On the arrival of $j$, the algorithm observes the non-negative weights $(w_{ij})_{i\in\OfflineObj}$ and immediately allocates $j$ to at most one advertiser.
Advertisers have free disposal, so the value of advertiser $i$ is the maximum weight $\max_{j\in S_i}w_{ij}$ among the impressions $S_i$ assigned to it, and zero if $S_i$ is empty.
The objective is to maximize the total value of all advertisers.
The offline optimum is the weight of a maximum-weight matching.
Hence, the standard fractional benchmark and dual are
\begin{align*}
    \max_{y\ge0}\quad
     & \textstyle\sum_{i\in\OfflineObj}\sum_{j\in\OnlineObj}w_{ij}y_{ij}
     &
    \min_{\alpha,\beta\ge0}\quad
     & \textstyle\sum_{i\in\OfflineObj}\alpha_i+\sum_{j\in\OnlineObj}\beta_j \\
    \text{s.t.}\quad
     & \textstyle\sum_{j\in\OnlineObj}y_{ij}\le1
    \quad\forall i\in\OfflineObj
     &
    \text{s.t.}\quad
     & \alpha_i+\beta_j\ge w_{ij}
    \quad\forall i\in\OfflineObj,\ j\in\OnlineObj                            \\
     & \textstyle\sum_{i\in\OfflineObj}y_{ij}\le1
    \quad\forall j\in\OnlineObj.
\end{align*}

We use the same First-Age Poisson generators, matching kernel, and potential function as in \Cref{sec:matching}.
We now track a separate age measure for each advertiser $i$ at each weight level $w\ge0$.
Although this requires additional ideas, we retain the warm-up's competitive ratio.

\begin{theorem}
    \label{thm:display-ads}
    With the parameter $\lambda=4.5$, Online Primal-Dual Allocation with First-Age Poisson OCEC in \Cref{sec:display-ads-algorithm} is $0.556$-competitive for Display Ads.
\end{theorem}

\subsection{Analysis via Weight Levels}
\label{sec:display-ads-levels}

Fix an advertiser $i$ and a weight level $w \geq 0$.
Let $\OnlineObj_i(w) \defeq \{j\in\OnlineObj:w_{ij}\ge w\}$ denote the set of online impressions with weight at least $w$ for advertiser $i$.
We say that level $w$ is covered if $i$ receives at least one impression in $\OnlineObj_i(w)$.
An advertiser whose best impression has weight $v$ covers exactly the levels $0<w\le v$, so
\begin{equation}
    \max_{j\in S_i}w_{ij}
    =\int_0^\infty\Ind\{S_i\cap\OnlineObj_i(w)\ne\varnothing\}\,\d w.
    \label{eqn:display-levels}
\end{equation}
Thus, at each fixed weight level, we seek to lower-bound the probability that the level is covered.

\paragraph{Age measures by weight levels.}
We extend $\nu_i$ to a family of age measures $\nu_{i,w}$, one for each weight level $w\ge0$.
Fix fractional allocations rounded by OCEC with the First-Age Poisson generator, and consider the fresh competitor process in \Cref{sec:matching}.
For each weight level $w$, the function $S\mapsto\Ind\{\exists j\in S:w_{ij}\ge w\}$ is submodular.
Thus, \Cref{lem:ocec-dominance} bounds the probability that advertiser $i$ receives no impression of weight at least $w$ by the probability of losing every comparison with $w_{ij}\ge w$ against fresh clocks $Z_j\sim\Exp(1-x_{ij})$.
We again track this probability through each $\nu_{i,w}$ one
request at a time, keeping $i$'s generator and removing mass only for outcomes in which $i$ wins an impression with $w_{ij}\ge w$.

\subsection{Gap Kernel}
\label{sec:display-ads-gap}

An impression with $w_{ij}<w$ cannot cover level $w$, so its request removes no mass from $\nu_{i,w}$.
Nevertheless, the request advances the generator's internal time by $x_{ij}$ and evolves the age measure.
We describe this evolution by a gap kernel.

The age $A(t)$ increases at unit rate and resets to zero at each rate-$\lambda$ Poisson arrival.
For any bounded measurable function $f$, define the \emph{gap kernel} by
$$
    (\GapKernel_t f)(a)
    \defeq\E\!\left[f(A(t))\,\middle|\,A(0)=a\right],
    \qquad t\ge0.
$$
We have $\GapKernel_0 = \IdKernel$.
It does not remove mass along the process, i.e., $\GapKernel_t\mathbf1=\mathbf1$, and the stationary age distribution is invariant, i.e., $\pi\GapKernel_t=\pi$.
The age dynamics give the following infinitesimal generator.

\begin{lemma}
    \label{lem:gap-infinitesimal-generator}
    For every bounded continuously differentiable function $f$ with bounded derivative,
    $$
        (\InfGenerator_{\GapKernel}f)(a)
        =f'(a)-\lambda f(a)+\lambda f(0).
    $$
    The derivative at zero is the right derivative.
\end{lemma}
\begin{proof}
    Without arrivals, the age increases at unit rate, so the drift contributes $f'(a)$.
    With arrivals, which happen at rate $\lambda$, the age resets to zero, changing the function value by $f(0)-f(a)$.
    Combining these contributions gives the claimed generator.
\end{proof}

\subsection{Potential of Reachable Measures}

We use the same matching kernel $\MatchKernel_x$ and potential function $\Gamma$ as in \Cref{sec:matching}.
Recall that the potential of a reachable measure $\nu$ is $\nu\Gamma$.

\paragraph{Reachable measures at each weight level.}
We extend the warm-up's definition of a reachable measure by allowing a finite sequence of both matching kernels $\MatchKernel_x$, for $x\in[0,1]$, and gap kernels $\GapKernel_t$, for $t\ge0$, starting from $\pi$.
For advertiser $i$ and level $w$, initialize $\nu_{i,w}=\pi$.
After choosing the allocation $x_{ij}$ for impression $j$, update
\begin{equation}
    \nu_{i,w}\gets
    \begin{cases}
        \nu_{i,w}\MatchKernel_{x_{ij}}, & w\le w_{ij}, \\[.5ex]
        \nu_{i,w}\GapKernel_{x_{ij}},   & w>w_{ij}.
    \end{cases}
    \label{eqn:level-update}
\end{equation}
For any measurable set $H$, $\nu_{i,w}(H)$ is the probability that advertiser $i$ has not yet won an impression of weight at least $w$ against fresh clocks and that its current age lies in $H$.
Its total mass is therefore the probability that level $w$ remains uncovered in this comparison process.
Using negative association, we can compare this to the actual probability that level $w$ remains uncovered as follows.

\begin{lemma}
    \label{lem:level-no-arrival}
    For every advertiser $i$ and level $w>0$, the final measure satisfies
    \begin{equation}
        \Pr[S_i\cap\OnlineObj_i(w)=\varnothing]
        \le\nu_{i,w}\mathbf1.
        \label{eqn:level-bound}
    \end{equation}
\end{lemma}
\begin{proof}
    Apply \Cref{lem:ocec-dominance} to the monotone submodular function $\Ind\{S\cap\OnlineObj_i(w)\ne\varnothing\}$.
    It bounds the probability of no relevant win by the corresponding probability against fresh competitors.
    Composing the updates in~\eqref{eqn:level-update} gives exactly this latter probability, just as in~\eqref{eqn:matching-unmatched-bound}.
\end{proof}

The following lemmas generalize and strengthen \Crefrange{lem:matching-measures-bounds}{lem:matching-potential-inequalities}.
In particular, they consider reachable measures that also allow gap updates.

\begin{lemma}
    \label{lem:display-measures-bounds}
    For every reachable measure $\nu$, $e^{\lambda a}\nu([a,\infty))$ is nondecreasing in $a\ge0$ and at most one.
    In particular, for $a\geq 0$ we have
    $
        (\nu\mathbf1)e^{-\lambda a}
        \le\nu([a,\infty))
        \le e^{-\lambda a}.
    $
\end{lemma}

\begin{lemma}
    \label{lem:display-request-monotonicity}
    For every reachable measure $\nu$, $\nu\MatchKernel_x\Gamma$ and $\nu\GapKernel_x\Gamma$
    are continuous and nonincreasing  in $x$ on $[0,1]$.
\end{lemma}

\begin{lemma}\label{lem:display-potential-inequalities}
    For every reachable measure $\nu$ and every request $x\in[0,1]$,
    \begin{equation*}
        \nu\mathbf1-\nu\MatchKernel_x\mathbf1
        \ge
        \nu\Gamma-\nu\MatchKernel_x\Gamma
        +x\nu\MatchKernel_x\Gamma.
    \end{equation*}
\end{lemma}

Besides the above extensions, we only need one new lemma.
A gap can weaken the correlation between past losses and future wins.
This can decrease the remaining potential without covering the weight level.
The following lemma bounds how much potential a small gap can consume: the rate of potential decrease during a gap is at most the potential already spent.
\begin{lemma}
    \label{lem:gap-evolution}
    For every reachable measure $\nu$ and every $x\ge0$,
    $$
        0\leq -\frac{\d}{\d x}\nu \GapKernel_x \Gamma \leq \Gamma^\star - \nu\GapKernel_x\Gamma,
    $$
    with a right derivative at $x=0$.
\end{lemma}

The proofs of these four lemmas are deferred to \Cref{app:matching-display}.

\subsection{Online Primal-Dual Algorithm and Analysis}
\label{sec:display-ads-algorithm}

The algorithm maintains dual variables $\alpha_i$ and $\beta_j$ alongside the rounding.
As in matching, $\Gamma^\star$-competitiveness follows by weak duality from two properties:
\begin{enumerate}[label=(\arabic*)]
    \item \emph{Reverse weak duality}: the expected total value is
          at least $\sum_i\alpha_i+\sum_j\beta_j$.
    \item \emph{Approximate dual feasibility}: $\alpha,\beta\ge0$
          and $\alpha_i+\beta_j\ge\Gamma^\star w_{ij}$ for every edge $(i,j)$.
\end{enumerate}

The offline duals record spent potential across all weight levels:
\begin{equation}
    \alpha_{i,w}\defeq\Gamma^\star-\nu_{i,w}\Gamma,
    \qquad
    \alpha_i\defeq\int_0^\infty\alpha_{i,w}\,\d w.
    \label{eqn:display-alpha}
\end{equation}
They start at zero and never decrease after a request.
Above all weights seen for $i$, stationarity gives
$\nu_{i,w}=\pi$ and $\alpha_{i,w}=0$.
The measures are constant between consecutive observed weights, so all level integrals reduce to finite sums, including the current weight as a breakpoint.

Let $\nu_{i,w}$ be the measures when $j$ arrives.
For a candidate mass $x\in[0,1]$, let $\alpha_i(x)$ denote advertiser $i$'s offline dual after applying \eqref{eqn:level-update} once with mass $x$.
Let $\alpha_{i,w}(x)$ denote the corresponding offline dual at weight level $w$.
Approximate dual feasibility requires $\beta_j\ge\Gamma^\star w_{ij}-\alpha_i(x_{ij})$ for every advertiser $i$.
We therefore define advertiser $i$'s offer per unit of allocation to $\beta_j$ as
\begin{equation}
    \Gamma^\star w_{ij}-\alpha_i(x)
    = \int_0^{w_{ij}}\nu_{i,w}\MatchKernel_x\Gamma\,\d w
    -\int_{w_{ij}}^\infty \bigl(\Gamma^\star-\nu_{i,w}\GapKernel_x\Gamma\bigr)\,\d w.
    \label{eqn:display-return-levels}
\end{equation}
The first term sums the remaining potential at weight levels lower than $w_{ij}$;
the second accounts for the offline dual at higher weight levels.
When all weights are one, this reduces to the offer $\nu_i\MatchKernel_x\Gamma$ for matching.
Starting from $x_{ij}=0$, we continuously allocate $j$ to advertisers with the largest positive offers, splitting allocation among tied advertisers to keep their offers equal, until $j$ is fully allocated or no positive offer remains.
By \Cref{lem:display-request-monotonicity}, each offer is continuous and nonincreasing in the allocation, so the water-filling procedure is well defined.
Each advertiser receiving positive allocation has the maximum final offer.
If $j$ is not fully allocated, all final offers are nonpositive, and those of advertisers receiving positive allocation are zero.
Thus, setting $\beta_j$ to the maximum final offer or zero, whichever is larger, gives
$$
    \beta_j=\sum_i x_{ij}\bigl(\Gamma^\star w_{ij}-\alpha_i(x_{ij})\bigr).
$$

\begin{algorithm}{Online Primal-Dual Allocation with First-Age Poisson OCEC (Display Ads)}
    \textbf{Internal states for each advertiser $i$:}
    \begin{itemize}
        \item An independent First-Age Poisson generator $\Generator_i$.
        \item A measure $\nu_{i,w}$ for every level $w>0$, initially $\pi$, and the corresponding offline duals $\alpha_{i,w}$ and $\alpha_i$ defined in~\eqref{eqn:display-alpha}.
    \end{itemize}

    \medskip

    \textbf{Upon the arrival of an impression $j$:}
    \begin{enumerate}
        \item \textbf{Allocate:~} Starting from $x_{ij}=0$, continuously allocate to advertisers $i$ with the largest positive offer defined in \eqref{eqn:display-return-levels}, until $\sum_i x_{ij}=1$ or no positive offer remains.

        \item \textbf{Update duals:~} Set $\beta_j\gets\max\bigl\{0,\max_i\bigl(\Gamma^\star w_{ij}-\alpha_i(x_{ij})\bigr)\bigr\}$.
              For each advertiser $i$, update each $\nu_{i,w}$ by~\eqref{eqn:level-update} and update $\alpha_{i,w}$ and $\alpha_i$ accordingly.

        \item \textbf{Round:~} If advertiser $i$ wins the OCEC rounding, allocate $j$ to $i$; otherwise discard $j$.
    \end{enumerate}
\end{algorithm}

\begin{proof}[Proof of \Cref{thm:display-ads}]
    We verify the two required properties.

    \noindent
    \emph{Reverse weak duality.~}
    Consider the arrival of $j$, and let $\nu_{i,w}$ denote advertiser $i$'s measure at weight level $w$ before the allocation.
    Write $\Delta\alpha_{i,w}=\alpha_{i,w}(x_{ij})-\alpha_{i,w}$ and $\Delta\alpha_i=\alpha_i(x_{ij})-\alpha_i$ for the increases in the offline duals.
    At each level $w\le w_{ij}$, \Cref{lem:display-potential-inequalities} bounds the credited gain $\nu_{i,w}\mathbf1 -\nu_{i,w}\MatchKernel_{x_{ij}}\mathbf1$
    below by $\Delta\alpha_{i,w}
        +x_{ij}(\Gamma^\star-\alpha_{i,w}(x_{ij}))$.
    At each level $w>w_{ij}$, by \Cref{lem:gap-evolution} we have
    $$
        \Delta\alpha_{i,w}
        = \int_0^{x_{ij}}-\frac{\d}{\d x}(\nu_{i, w}\GapKernel_x\Gamma)\d x
        \leq
        \int_0^{x_{ij}}
        \bigl(\Gamma^\star-\nu_{i,w}\GapKernel_x\Gamma\bigr)
        \,\d x
        \le x_{ij}\alpha_{i,w}(x_{ij}).
    $$
    Integrating over both ranges and summing over advertisers yields
    \begin{align*}
        \sum_i\int_0^{w_{ij}} \bigl(\nu_{i,w}\mathbf1 -\nu_{i,w}\MatchKernel_{x_{ij}}\mathbf1\bigr)\,\d w
         & \geq
        \sum_i\Biggl(\int_0^{w_{ij}}
        \Bigl[\Delta\alpha_{i,w}+x_{ij}\bigl(\Gamma^\star-\alpha_{i,w}(x_{ij})\bigr)\Bigr]\,\d w \\
         & \qquad\qquad\quad
        +\int_{w_{ij}}^\infty
        \Bigl[\Delta\alpha_{i,w}-x_{ij}\alpha_{i,w}(x_{ij})\Bigr]\,\d w\Biggr)                   \\
         & \geq
        \sum_i\left(\Delta\alpha_i
        +x_{ij}\bigl(\Gamma^\star w_{ij}-\alpha_i(x_{ij})\bigr)\right)
        = \sum_i \Delta\alpha_i + \beta_j.
    \end{align*}
    Summing over arrivals, the gains telescope.
    Since the measures start at $\pi$ and the offline duals at
    zero, \eqref{eqn:display-levels} and \Cref{lem:level-no-arrival} give
    $$
        \E[\textnormal{total value}]
        \ge
        \sum_i\int_0^\infty(1-\nu_{i,w}\mathbf1)\,\d w
        \ge
        \sum_i\alpha_i+\sum_j\beta_j,
    $$
    where the measures and dual variables are their final values.

    \noindent
    \emph{Approximate dual feasibility.~}
    By \Cref{lem:display-request-monotonicity}, each $\alpha_i$ is nondecreasing from zero.
    The definition of $\beta_j$ ensures $\beta_j\ge0$ and $\alpha_i(x_{ij})+\beta_j\ge\Gamma^\star w_{ij}$ for every $i$ immediately after processing $j$.
    These inequalities persist since $\alpha_i$ never decreases.
\end{proof}

\section{AdWords}
\label{sec:adwords}

In the AdWords problem, advertisers $i\in\OfflineObj$ have budgets $B_i>0$, and impressions $j\in\OnlineObj$ arrive online. On the arrival of impression $j\in\OnlineObj$, the algorithm observes bids $(b_{ij})_{i\in\OfflineObj}$ and immediately allocates $j$ to at most one advertiser.
Recall that $S_i$ is the set of impressions allocated to $i$.
The revenue from advertiser $i$ is
$\min\{B_i,\sum_{j\in S_i}b_{ij}\}$.
Thus, if a bid exceeding the remaining budget is allocated, it only contributes the remaining budget in revenue.
Hence, we may assume $0\le b_{ij}\le B_i$.

The optimal revenue is at most the optimal objectives of the following LP and its dual:
\begin{equation}
    \begin{aligned}
        \max_{y\ge0}\quad
         & \textstyle\sum_{i\in\OfflineObj}\sum_{j\in\OnlineObj}b_{ij}y_{ij}
         & \qquad
        \min_{\alpha,\beta\ge0}\quad
         & \textstyle\sum_{i\in\OfflineObj}B_i\alpha_i
        +\sum_{j\in\OnlineObj}\beta_j                                        \\
        \text{s.t.}\quad
         & \textstyle\sum_{j\in\OnlineObj}b_{ij}y_{ij}\le B_i
        \quad\forall i\in\OfflineObj
         & \qquad
        \text{s.t.}\quad
         & b_{ij}\alpha_i+\beta_j\ge b_{ij}
        \quad\forall i,j,                                                    \\
         & \textstyle\sum_{i\in\OfflineObj}y_{ij}\le1
        \quad\forall j\in\OnlineObj.
    \end{aligned}
    \label{eqn:adwords-lp}
\end{equation}
Write $\mathrm{LP}$ for the optimum of this relaxation.
As in the preceding sections, $y_{ij}$ denotes a benchmark variable and $x_{ij}$ denotes the fractional allocation passed to rounding.

Unlike in the matching setting, allocating an impression to an advertiser may not fully saturate their budget.
Hence, the state space becomes more complex than just the age distribution. To manage this, we will separately track ``small bids,'' which are less than half the advertiser's budget, and ``large bids,'' which are at least half.

For each pair $(i,j)$ with a small bid, advertiser $i$ competes for impression $j$ using a fresh exponential clock.
Because these bids are small, this use of independent clocks already gives an improvement over the $1/2$-competitive baseline analysis.
For pairs with large bids, we obtain the clock from the advertiser's First-Age Poisson generator.
Winning two large bids exhausts an advertiser’s budget, so only histories with zero or one win on a large bid can contribute further revenue.
This simplifies our state space, allowing us to conduct an analysis similar to \Cref{sec:matching}. Combining the guarantees for both large and small bids, we prove the following theorem in the rest of this section.

\begin{theorem}
    \label{thm:adwords}
    With $\lambda=4.5$, Online Primal-Dual Allocation with First-Age Poisson OCEC in Section~\ref{sec:adwords-algorithm} has expected revenue at least
    $0.530\,\mathrm{LP}$.
\end{theorem}

\subsection{Age Measures}
\label{sec:adwords-state}

\paragraph{Fresh competitor process.}
Fix an advertiser $i$ and fractional allocations independent of the rounding randomness, and suppress $i$ from the notation.
As in previous sections, we consider the evolution of its age process against fresh competing clocks.
For the purpose of this process, we will always consider arriving bids normalized by the advertiser's budget.
Write $\NormalizedBid\in[0,1]$ for a \emph{normalized} bid (i.e. the ratio $\frac{b_{ij}}{B_i}$ of the bid to the advertiser's budget) and
$x\in[0,1]$ for its allocation mass.
A bid is \emph{large} if $\NormalizedBid\ge1/2$ and
\emph{small} otherwise.
For a large bid, we obtain a clock by sending mass $x$ to the advertiser's First-Age Poisson generator.
For a small bid, we draw a fresh $\Exp(x)$ clock without advancing the generator's internal time.

By Lemma~\ref{lem:na-closure}, independent unions preserve negative association, so the First-Age clocks for large bids and the independent clocks for small bids together form an OCEC generator.
And the budget-additive revenue $S \mapsto\min\{1,\sum_{j\in S}\NormalizedBid_j\}$ is submodular.
Thus as before, Lemma~\ref{lem:ocec-dominance} reduces the analysis to competition against fresh independent clocks $Z_j\sim\Exp(1-x_j)$.
It suffices to lower-bound revenue, or equivalently upper-bound unused budget, in this comparison process.

\paragraph{The contribution of large bids.}
Let $A$ be the current age, $N$ the number of large bids won so far, and $R$ the normalized remaining budget after accounting for large bids.
Since every large bid is at least $1/2$, we have $R=0$ whenever $N\ge2$. Thus, we only need to track age distributions for the states $N = 0$ and $N = 1$, which we will do using measures $\nu_0$ and $\nu_1$ respectively.

For a measurable age set $H$, define
\[
    \nu_0(H):=\Pr[A\in H,\ N=0],
    \qquad
    \nu_1(H):=\E\!\left[R\, \Ind\{A\in H,\ N=1\}\right].
\]
The first measure is the surviving age measure of the matching problem.
The second records the measure of trajectories with one win on a large bid, \textit{weighted by the budget still available}.
For example, a first win of bid $0.7$ contributes weight $0.3$ to $\nu_1$; a second win on a large bid removes that trajectory entirely.
This weighting records the effect of different large bids without keeping a separate measure for every remainder.
Initially $(\nu_0,\nu_1)=(\pi,0)$, and at any point in the process, the expected remaining budget after accounting for large bids is
$(\nu_0+\nu_1)\mathbf1=\E[R]$.

Recall the matching and gap kernels $\MatchKernel_x$ and $\GapKernel_x$ from the previous sections.
Define the \emph{win kernel}
$\WinKernel_x:=\GapKernel_x-\MatchKernel_x$.
An impression with large bid $\NormalizedBid$ and mass $x$ updates the state $(\nu_0, \nu_1)$ to $(\nu'_0, \nu'_1)$ as follows:
\begin{equation}
    \nu_0'=\nu_0\MatchKernel_x,
    \qquad
    \nu_1'=\nu_1\MatchKernel_x+(1-\NormalizedBid)\nu_0\WinKernel_x.
    \label{eqn:adwords-update}
\end{equation}
Indeed, we know that $\nu_0$ updates according to the standard matching kernel, just as in the warm-up.
For $\nu_1$, existing mass updates according to $\MatchKernel_x$ since another large bid win would exhaust the budget, while winning mass from $\nu_0$ enters with weight $1-\NormalizedBid$.

\paragraph{The contribution of small bids.}
Impressions with small bids leave both age measures fixed.
We record only their normalized fractional load with the scalar quantity $s$:
\[
    s:=\sum_{j:\,\NormalizedBid_j<1/2}\NormalizedBid_jx_j.
\]
We seek to upper-bound the expected available budget as a function of $\nu_0$, $\nu_1$, and $s$.
Intuitively, the worst case for small bids comes from the Poissonized limit with many bids of size $\NormalizedBid_j \approx 1/2$ with mass $x_j$ close to 0.
Assuming small bids contribute $\Poisson(2s)$-many wins of size $1/2$, we get the unused-budget bound
\begin{equation}
    U(\nu_0,\nu_1,s)
    :=e^{-2s}\bigl((1+s)\nu_0\mathbf1+\nu_1\mathbf1\bigr).
    \label{eqn:adwords-residual}
\end{equation}
This expression comes from two cases.
If no large bid is won, then $\Poisson(2s)$ wins on bids of size $1/2$ leave an expected remaining budget of $e^{-2s}(1+s)$, and this happens with probability $\nu_0 \1$.
If exactly one large bid is won, the Poisson comparison leaves budget $R$ when there are no Poisson wins, and zero otherwise.
Since no Poisson win occurs with probability $e^{-2s}$, these histories contribute $e^{-2s}\nu_1\mathbf1$ to the expected remaining budget.
We make this comparison precise in the following lemma.

\begin{lemma}\label{lem:adwords-residual}
    At every input prefix, the advertiser's expected normalized unused budget under the rounding is at most $U(\nu_0,\nu_1,s)$.
\end{lemma}
\begin{proof}
    By the OCEC comparison above, it suffices to consider fresh competitors.
    Let $X$ be the sum of the small bids won in that process.
    Its summands are independent $\NormalizedBid_j\Bernoulli(x_j)$ variables, independent also of $R$.
    Replacing each summand by $\tfrac12\Poisson(2\NormalizedBid_jx_j)$ can only increase the expectation of a convex function of their sum.
    To see this, first replace it by $\tfrac12\Bernoulli(2\NormalizedBid_jx_j)$ using the chord inequality on $[0,1/2]$.
    Then replace the Bernoulli variable by a Poisson variable of the same mean: at every nonnegative integer, a convex function lies above the line through its values at zero and one.
    Conditioning on the other summands justifies these replacements one at a time.

    We write $(z)_+$ for $\max\{z,0\}$.
    Thus, for $K\sim\Poisson(2s)$ independent of $R$,
    \[
        \E[(R-X)_+\mid R]
        \le\E[(R-\tfrac12K)_+\mid R]
        =e^{-2s}\bigl(R+s\mathbf1\{R=1\}\bigr).
    \]
    The equality uses $R\in\{1\}\cup[0,1/2]$: only $K=0$ can leave a positive remainder when $R\le1/2$, while $K=1$ leaves $1/2$ when $R=1$.
    Averaging and using $\E[R]=(\nu_0+\nu_1)\mathbf1$ and $\Pr[R=1]=\nu_0\mathbf1$ gives~\eqref{eqn:adwords-residual}.
    Impressions with large or small bids may be interleaved; their comparison outcomes are independent across the two groups, and the final remaining budget depends only on the sum of their winning bids.
\end{proof}

\subsection{Potential of Reachable States}
\label{sec:adwords-potential}

As in previous sections, we use a potential $\Gamma$ on the state $(\nu_0,\nu_1,s)$ to choose fractional allocations and dual variables.
The competitive analysis requires nonnegativity, continuity, monotonicity, and an accounting inequality for every reachable state.

We assign an age-value function to each of the two measures and define
\begin{equation}
    \Gamma(\nu_0,\nu_1,s)
    :=\nu_0\Gamma_{0,s}+\nu_1\Gamma_{1,s},
    \qquad
    \AdwordsRatio:=\Gamma(\pi,0,0).
    \label{eqn:adwords-potential}
\end{equation}
Here $\Gamma_{0,s}$ assigns potential to histories with no large bid won, while $\Gamma_{1,s}$ assigns potential per unit of normalized budget remaining after one large bid is won.
Both depend on $s$, since small-bid load reduces the opportunity for future revenue.
They replace the single age-value function $\Gamma$ used in matching.
Our primal-dual analysis certifies competitive ratio $\AdwordsRatio$.

A state is \emph{reachable} if it is obtained from $(\pi,0,0)$ by applying~\eqref{eqn:adwords-update} to impressions with large bids and $s\gets s+\NormalizedBid x$ to impressions with small bids.
For one such update, abbreviate the before and after values by $U, \Gamma$ and $U', \Gamma'$.
The required accounting for our primal-dual analysis is
\begin{equation}
    \underbrace{\vphantom{\big|} U-U'}_{\text{credited primal gain}}
    \ge
    \underbrace{\vphantom{\big|} \Gamma-\Gamma'}_{\text{offline dual increase}}
    +\underbrace{\vphantom{\big|} \NormalizedBid x\,\Gamma'}_{\text{online dual contribution}}.
    \label{eqn:adwords-accounting}
\end{equation}
This is the warm-up's inequality with the online contribution scaled by the normalized bid $\NormalizedBid$.

The following lemma summarizes the properties needed to prove \Cref{thm:adwords}.
\Cref{sec:adwords-optimization} describes the finite LP used to find this potential, and \Cref{app:adwords} proves the lemma.

\begin{lemma}\label{lem:adwords-potential}
    With $\lambda=4.5$, there exists a potential $\Gamma$ in the form \eqref{eqn:adwords-potential} with the following properties:

    \begin{enumerate}
        \item $\Gamma_{0,s}(a)$ and $\Gamma_{1,s}(a)$ are jointly continuous in $(a,s)$, continuously differentiable and bounded in $a$ for each fixed $s$, and locally Lipschitz in $s$ uniformly over $a$ (hence locally absolutely continuous in $s$).
        \item $\Gamma(\pi,0,0)=\pi\Gamma_{0,0}=\AdwordsRatio\ge0.530$.
        \item Fix any reachable state $(\nu_0,\nu_1,s)$ and the normalized bid $\bar{b} \in [0,1]$ of an impression.
              Write $U,\Gamma$ and $U',\Gamma'$ for the values before and after assigning a fraction $x\in[0,1]$ of the impression to the advertiser.
              Then:
              \begin{enumerate}
                  \item (Nonnegativity) $\Gamma \geq 0$.
                  \item (Monotonicity) $\Gamma'$ is continuous and nonincreasing in $x$; in particular, $\Gamma'\le\Gamma$.
                  \item (Accounting) $U - U' \geq \Gamma - \Gamma' + \NormalizedBid x\Gamma'$.
              \end{enumerate}
    \end{enumerate}
\end{lemma}

\subsection{Online Primal-Dual Algorithm and Analysis}
\label{sec:adwords-algorithm}

The algorithm maintains dual variables $\alpha_i$ and $\beta_j$ alongside the rounding.
As in the previous sections, it suffices to prove:
\begin{enumerate}[label=(\arabic*)]
    \item \emph{Reverse weak duality}: the expected revenue is at least
          $\sum_i B_i\alpha_i+\sum_j\beta_j$.
    \item \emph{Approximate dual feasibility}: $\alpha,\beta\ge0$, and
          $b_{ij}\alpha_i+\beta_j
              \ge\Gamma_{\mathrm{AW}}^\star b_{ij}$
          for every pair $(i, j)$.
\end{enumerate}
Then, $(\alpha/\Gamma_{\mathrm{AW}}^\star,\beta/\Gamma_{\mathrm{AW}}^\star)$ is dual feasible.
Together with the first property, weak duality then implies that the expected revenue is at least $\Gamma_{\mathrm{AW}}^\star$ times the LP optimum.

The dual variables follow the potential.
For each advertiser $i$, write
\[
    \Gamma_i:=\Gamma(\nu_{i,0},\nu_{i,1},s_i),
    \qquad
    \alpha_i:=\Gamma_{\mathrm{AW}}^\star-\Gamma_i.
\]
Thus the offline dual $\alpha_i$ records the potential spent from its initial value, and its contribution to the dual objective is $B_i\alpha_i$.

For an arriving impression $j$, let $\NormalizedBid_{ij}=b_{ij}/B_i$ and hold each advertiser's incoming state fixed.
Write $\Gamma_i^{(j)}(t)$ for its potential after a single request of normalized bid $\NormalizedBid_{ij}$ and mass $t\in[0,1]$.
For a large bid, apply \eqref{eqn:adwords-update} while keeping $s_i$ fixed.
For a small bid, keep the age measures fixed and replace $s_i$ by $s_i+\NormalizedBid_{ij}t$.
After allocating a fraction $x_{ij}$ of impression $j$ to advertiser $i$, its offline dual becomes $\alpha_i=\Gamma_{\mathrm{AW}}^\star-\Gamma_i^{(j)}(x_{ij})$.
Thus, approximate dual feasibility requires
$$
    \beta_j\ge b_{ij}\bigl(\Gamma_{\mathrm{AW}}^\star-\alpha_i\bigr)
    =b_{ij}\Gamma_i^{(j)}(x_{ij})
$$
for every advertiser $i$.
We therefore define advertiser $i$'s offer per unit of allocation to $\beta_j$ as $b_{ij}\Gamma_i^{(j)}(t)$.
Starting from $x_{ij}=0$, we continuously allocate $j$ to advertisers with the largest current offers, splitting allocation among tied advertisers to keep their offers equal, until $\sum_i x_{ij}=1$.
We set $\beta_j$ to the maximum final offer to satisfy the required inequalities.
Every advertiser receiving positive allocation has this offer.
Since the allocated masses sum to $1$, we have
$$
    \beta_j = \max_i b_{ij}\Gamma_i^{(j)}(x_{ij}) = \sum_i x_{ij}b_{ij}\Gamma_i^{(j)}(x_{ij}).
$$

\begin{algorithm}{Online Primal-Dual Allocation with First-Age Poisson OCEC (AdWords)}
    \textbf{Internal states for each advertiser $i$:}
    \begin{itemize}
        \item An independent First-Age Poisson generator $\Generator_i$.
        \item A state $(\nu_{i,0},\nu_{i,1},s_i)$, initially $(\pi,0,0)$, and its potential $\Gamma_i=\Gamma(\nu_{i,0},\nu_{i,1},s_i)$ and offline dual $\alpha_i=\Gamma_{\mathrm{AW}}^\star-\Gamma_i$.
    \end{itemize}

    \medskip

    \textbf{Upon the arrival of an impression $j$:}
    \begin{enumerate}
        \item \textbf{Allocate:~}
              Starting from $x_{ij}=0$, continuously allocate to
              advertisers with the largest current offers
              $b_{ij}\Gamma_i^{(j)}(x_{ij})$
              until $\sum_i x_{ij}=1$.
              Distribute allocation among tied advertisers to maintain equal offers.
        \item \textbf{Update duals:~}
              Set
              $
                  \beta_j\gets
                  \max_i b_{ij}\Gamma_i^{(j)}(x_{ij}).
              $

              For each advertiser $i$, update its state according to \eqref{eqn:adwords-update} if $\NormalizedBid_{ij}\geq 1/2$, otherwise set $ s_i\gets s_i+\NormalizedBid_{ij}x_{ij}$.
              Then set $\Gamma_i\gets\Gamma(\nu_{i,0},\nu_{i,1},s_i)$ and $\alpha_i\gets\Gamma_{\mathrm{AW}}^\star-\Gamma_i$.

        \item \textbf{Round:~}
              For each advertiser with $x_{ij}>0$, send request
              $x_{ij}$ to $\Generator_i$ if $\NormalizedBid_{ij}\ge1/2$;
              otherwise, draw a fresh $\operatorname{Exp}(x_{ij})$
              clock.
              Allocate $j$ to the advertiser with the smallest clock.
    \end{enumerate}
\end{algorithm}

The incoming states remain fixed throughout water filling; only the final chosen requests update them.
Lemma~\ref{lem:adwords-potential} supplies both the accounting inequality
and the continuity and monotonicity of the offers required for water filling.

\begin{proof}[Proof of \Cref{thm:adwords}]
    We verify the two required properties of online primal-dual analysis.

    \noindent
    \emph{Reverse weak duality.~}
    Consider the arrival of impression $j$.
    For advertiser $i$, let $U_i$ and $\Gamma_i$ denote the upper bound on its normalized unused budget and its remaining potential before the request, and let $U_i'$ and $\Gamma_i'$ denote the corresponding values afterward.
    The request $x_{ij}$ increases $\alpha_i$ by $\Gamma_i - \Gamma_i'$ and contributes $x_{ij}b_{ij}\Gamma_i' = x_{ij}b_{ij}\Gamma_i^{(j)}(x_{ij})$ to $\beta_j$, while it decreases $B_iU_i$ by $B_i(U_i - U_i')$.
    Multiplying the accounting inequality in Lemma~\ref{lem:adwords-potential} by $B_i$ shows that the decrease covers the weighted dual increase.
    Summing over all requests, with each $U_i$ starting at $1$
    and each $\alpha_i$ at $0$, the final state satisfies
    $
        \sum_i B_i(1-U_i)
        \geq
        \sum_i B_i\alpha_i+\sum_j\beta_j.
    $
    By \Cref{lem:adwords-residual}, the expected revenue is at least the left-hand side.

    \noindent
    \emph{Approximate dual feasibility.~}
    By \Cref{lem:adwords-potential}, the offers $b_{ij}\Gamma_i^{(j)}(x_{ij})$ are
    continuous and nonincreasing in the allocation, so water
    filling is well defined and each $\alpha_i$ is nondecreasing
    from $0$.
    Moreover, $\beta_j\ge0$ since the potentials are nonnegative.
    By definition, immediately after processing impression $j$,
    $
        \beta_j
        =\max_i b_{ij}\Gamma_i'
        =\max_i b_{ij}
        \bigl(\Gamma_{\mathrm{AW}}^\star-\alpha_i\bigr).
    $
    Hence $b_{ij}\alpha_i+\beta_j
        \ge\Gamma_{\mathrm{AW}}^\star b_{ij}$ for every $i$,
    and these inequalities persist since $\alpha_i$ never decreases.
\end{proof}

\subsection{Choosing the Potential in Lemma~\protect{\ref{lem:adwords-potential}}}
\label{sec:adwords-optimization}
Here, we outline how we find the explicit potential function in \Cref{lem:adwords-potential} while optimizing for the competitive ratio.
\paragraph{The optimization problem.}
Fix $\lambda=9/2$ and an advertiser with budget one. We seek age-value
functions $\Gamma_{0,s}$ and $\Gamma_{1,s}$ defining
\[
    \Gamma(\nu_0,\nu_1,s)=\nu_0\Gamma_{0,s}+\nu_1\Gamma_{1,s}.
\]
Their initial value $\rho=\Gamma(\pi,0,0)$ is the competitive ratio in
the primal--dual analysis. We therefore leave $\rho$ free and ask for
the largest value compatible with the other properties of
Lemma~\ref{lem:adwords-potential}:
\begin{equation}
    \begin{aligned}
        \text{maximize}\quad & \rho=\pi\Gamma_{0,0}                           \\
        \text{subject to}\quad
                             & \Gamma\ge0,\qquad \Gamma'\le\Gamma,            \\
                             & U-U'\ge\Gamma-\Gamma'+\NormalizedBid x\Gamma'.
    \end{aligned}
    \label{eqn:adwords-design-accounting}
\end{equation}
The constraints range over every reachable state $(\nu_0, \nu_1, s)$ and every normalized request $(\NormalizedBid,x)$, where $\Gamma'$ and $U'$ denote the values after that request.
We also require the regularity and monotonicity in request mass stated in the lemma.
Even if we omit the regularity conditions, the above formulation already poses three challenges for numerical optimization:
\begin{enumerate}
    \item Our solution space is infinite dimensional (namely, functions of age and small-bid load).
    \item Constraints are given by reachable states $(\nu_0, \nu_1, s)$, which are implicitly defined.
    \item There are infinitely many constraints given by each reachable state and each request.
\end{enumerate}
We will address each of these challenges in turn, reducing our optimization problem to a finite LP whose candidate solutions can then be verified over all states and requests.

\paragraph{Reducing to finite variables.}
We first simplify the optimization space by restricting the form of our functions.
Since unused-budget $U$ already contains the factor $e^{-2s}$, we will factor it out of our unknown functions.
Let $z=e^{-\lambda a}$ and write
\begin{equation}
    \begin{gathered}
        \Gamma_{0,s}(a)=e^{-2s}F_s(z),\qquad
        \Gamma_{1,s}(a)=e^{-2s}G_s(z),\\
        \Gamma(\nu_0,\nu_1,s)=e^{-2s}(\nu_0F_s+\nu_1G_s).
    \end{gathered}
    \label{eqn:adwords-design-prices}
\end{equation}
The change of variables from age $a$ to $z$ gives a bounded domain: $z=1$ is age zero, while $z=0$ is the limit of very old ages.
We use the same notation for an age measure and its image under this change of coordinates.
In particular, $\pi=\Exp(\lambda)$ becomes the uniform measure on $[0,1]$,
so the objective is simply $\rho=\int_0^1F_0(z)\,\d z$.

We choose a polynomial degree $d$ for our functions and set ``load knots'' $0=s_0<\cdots<s_K=S$ to discretize the space of the small-bid load $s$.
We use the Bernstein basis $\mathcal B_j^d(z)=\binom djz^j(1-z)^{d-j}$ and write
\begin{equation}
    F_{s_k}(z)=\sum_{j=0}^d f_{kj}\mathcal B_j^d(z),
    \qquad
    G_{s_k}(z)=\sum_{j=0}^d g_{kj}\mathcal B_j^d(z).
    \label{eqn:adwords-design-bernstein}
\end{equation}
This basis provides simple linear sufficient conditions for nonnegativity and monotonicity, as well as uniform bounds for verification.
For example, nonnegative, nonincreasing Bernstein coefficients yield a nonnegative, nonincreasing polynomial on $[0,1]$, allowing us to enforce these regularity conditions through linear inequalities.
The objective also has a simple linear form
$$
    \rho=\int_0^1F_0(z)\,\d z=\frac1{d+1}\sum_{j=0}^d f_{0j}.
$$
Between knots, interpolate the coefficients linearly. Beyond the last
knot, use
\begin{equation}
    G_s=G_S,\qquad F_s=F_S+(s-S)h\quad(s\ge S),
    \label{eqn:adwords-design-continuation}
\end{equation}
where $h(z)=\sum_{j=0}^d h_j\mathcal B_j^d(z)$ is another degree-$d$
polynomial. In this way, we cover all possible small-bid load values $s$.

We will use $\theta$ to represent the coefficient vector which consists of all $f_{kj},g_{kj}$ and the coefficients of $h$.
Both the potential and the objective are linear in $\theta$, so our optimization problem reduces to an LP in $\theta$.
However, this LP still has infinitely many linear constraints, which we address next.

\paragraph{Relaxing reachable measures to explicit inequalities.}
Before we can discretize our constraints, we must make the space of constraints explicit by relaxing the set of reachable states to those satisfying specific inequalities. These include the tail bounds introduced in \Cref{sec:matching}, along with a signed-score inequality.

In addition to $\nu_0$ and $\nu_1$, we define the measure $\sigma=\nu_0+2\nu_1$. Both $\nu_0$ and $\sigma$ attach
weights to histories that can only decrease when another large bid is
won, giving the tail condition
\[
    a\longmapsto e^{\lambda a}\eta([a,\infty))
    \quad\text{is nondecreasing for }\eta\in\{\nu_0,\sigma\}.
\]
Let $\AgeCone$ denote the cone of finite nonnegative absolutely continuous measures with the rescaled-tail property above.

However, these tail restrictions alone still allow adverse states.
For example, for sufficiently large $T$, the pair $\nu_0=\pi|_{[T,\infty)}$, $\nu_1=0$ satisfies the tail condition even though all surviving mass is very old.
As it turns out, such states are problematic for our optimization, despite not arising as reachable measures.\footnote{
    Poisson arrivals can reset the age without producing a win on a large bid, so any nonzero reachable $\nu_0$ retains young surviving histories.
    Hence, the truncated stationary distribution $\pi|_{[T,\infty)}$ is unreachable.
    Our potential assigns more value to older histories, reflecting better opportunities to win future large bids.
    However, small-bid rounding ignores age.
    Concentrating surviving mass at old ages can therefore increase the dual charges $\Gamma-\Gamma'+\NormalizedBid x\Gamma'$ without a corresponding increase in the revenue gain $U-U'$ from small bids, making accounting more restrictive.
}
To fix this, we further introduce a \emph{signed score} function $H$, for which $H(e^{-\lambda a})$ is positive for small ages $a$ and negative for sufficiently large $a$.
We prove that all reachable measures $\nu_0$ have
\(
\nu_0H\ge0,
\)
ruling out these truncated tails.

Ultimately, the class of measures we consider is as follows.
We will enforce our accounting inequality constraint on all pairs $(\nu_0, \nu_1)$ satisfying
\begin{equation}
    \nu_0,\nu_1\ge0,\qquad
    \nu_0,\ \nu_0+2\nu_1\in\AgeCone,\qquad \nu_0H\ge0.
    \label{eqn:adwords-design-cone}
\end{equation}
Every reachable state belongs to this class, so this only strengthens the constraints of our program.

\paragraph{Discretizing constraints.}
Finally, we reduce our optimization problem to one with finitely many constraints. Nonnegativity and monotonicity follow from simple linear inequalities on the polynomial coefficients; the accounting constraint requires more work.

For fixed measures $(\nu_0,\nu_1)$, polynomial bounds for $\NormalizedBid$ and interpolation between knots for $s$ reduce accounting to finitely many test families.
The LP imposes these tests at finitely many values of $x$; uniform bounds between test points then certify all request masses during verification.\footnote{
    As in matching, small request masses require special care because the accounting slack vanishes at $x=0$.
    For large bids, we divide the slack by $x$ and evaluate its limit at zero using the infinitesimal generator.
    For small bids, we verify a differential accounting inequality in the load $s$ and integrate it over the increment $\NormalizedBid x$.
}
The resulting LP tests have the form $\nu_0C_0+\nu_1C_1\ge0$, where the costs
$C_0,C_1$ depend affinely on $\theta$.

The signed-score inequality also gives additional flexibility in each test. For any $\tau\ge0$,
it suffices to show
$\nu_0(C_0-\tau H)+\nu_1C_1\ge0$, since $\nu_0H\ge0$.
We can choose a different $\tau$ for each test. After this adjustment,
we need to satisfy these tests for all pairs $(\nu_0, \nu_1)$ satisfying the tail bounds.

To do this, we replace each measure with a finite-dimensional sketch. Choose a partition $0=t_0<\cdots<t_N=1$ in the transformed coordinate
$z=e^{-\lambda a}$. We abuse notation slightly and use $\nu_r$, for $r\in\{0,1\}$, to denote the image of the original age measure under this transformation. Replace this transformed measure by a \emph{sketch} recording just its mass in each bucket:
\[
    y_{ri}=\nu_r((t_{i-1},t_i]),\qquad r\in\{0,1\},\quad 1\le i\le N.
\]
To evaluate a test from this sketch, we bound $C_0-\tau H$ and $C_1$ below by constants $v_{0i}$ and $v_{1i}$ throughout bucket $i$.
Then $v^\top y$ is a lower bound on $\nu_0(C_0-\tau H)+\nu_1C_1$, regardless of how the mass is distributed inside each bucket.
For measures with $\nu_0H\ge0$, it therefore also lower-bounds the original accounting cost $\nu_0C_0+\nu_1C_1$.
Finer buckets can reduce this approximation loss.

The tail bounds from \eqref{eqn:adwords-design-cone} translate into linear inequalities on cumulative
bucket masses, which apply to both $\nu_0$ and $\nu_0+2\nu_1$.
Because the accounting inequality is homogeneous in the measures, we can normalize their
total mass to one, and the sketches allowed by these tail constraints form a polytope
\[
    \mathcal Y=\{y\ge0:Ay\le0,\ \1^\top y=1\},
\]
where $A$ has finitely many rows.
Thus, to certify a test for all pairs satisfying~\eqref{eqn:adwords-design-cone}, it suffices to require $\min_{y\in\mathcal Y}v^\top y\ge0$.

By LP duality, the requirement $v^\top y\ge0$ for every $y\in\mathcal Y$ is equivalent to the existence of multipliers $u$ satisfying
$$
    u\ge0,\qquad v+A^\top u\ge0.
$$
These finite linear inequalities certify the test once the bucket bounds $v_{ri}$ are valid.

\section{Load Balancing}
\label{sec:load-balancing}

In this problem, machines $i\in\OfflineObj$ are given upfront, and jobs $j\in\OnlineObj$ arrive online one at a time.
On the arrival of job $j$, the algorithm observes the job size $p_{ij} > 0$ on each machine $i$; $p_{ij} = \infty$ if it is infeasible to process $j$ on machine $i$.
We assume every job has at least one feasible machine and use the convention $\infty\cdot0=0$.
The algorithm then immediately assigns job $j$ to a machine.
The objective is to minimize the total squared load
$$
  \sum_{i\in\OfflineObj}L_i^2,\quad \text{where } L_i=\sum_{j\in S_i}p_{ij}
  ~,
$$
recalling that $S_i$ is the set of jobs assigned to machine $i$.

Write $\operatorname{ALG}=\sum_i\E[L_i^2]$ for the algorithm's expected objective, and $\operatorname{OPT}$ for the minimum objective of an offline integral assignment.
Write $X_{ij}=\1\{j\in S_i\}$ for the allocation indicator.

\begin{theorem}\label{thm:load-balancing}
  With the common rate $\lambda=4.5$, the algorithm in Section~\ref{sec:load-balancing-algorithm} is $\Gamma$-competitive for minimizing squared loads on unrelated machines, where $\Gamma=4.69$.
\end{theorem}

\subsection{Convex Program Duality}

We use the convex relaxation of \cite{FengHL26}, specialized to squared loads, and its dual.
Let $x_{ij}\ge0$ be the fraction of job $j$ assigned to machine $i$, and write $x_i=(x_{ij})_j$.
We set $x_{ij}=0$ whenever $p_{ij}=\infty$.
For a job size $p>0$ and a threshold $\bar p\ge0$, define
$$
  \begin{gathered}
    \Phi(\bar p;p) = \int_0^p2\max\{t,\bar p\}\,dt =\begin{cases}
      p^2+\bar p^2 & \bar p\le p \\
      2p\bar p     & \bar p> p
    \end{cases}, \qquad
    \mathcal C_i(x_i)=\max_{\bar p_i\ge0}\biggl\{\sum_jx_{ij}\Phi(\bar p_i;p_{ij})-\bar p_i^2\biggr\}.
  \end{gathered}
$$

Each $\mathcal C_i$ is convex as a maximum of affine functions of $x_i$.
For integral allocations, $\mathcal C_i(x_i)$ equals the squared load on machine $i$, as shown in Lemma~\ref{lem:load-relaxation}.
If $\sum_jx_{ij}>1$, the unique maximizer $\bar p_i>0$ satisfies the first-order condition
$$
  \sum_j x_{ij} \min \bigl\{ p_{ij}, \bar{p}_i \bigr\} = \bar{p}_i.
$$
It is a normalization threshold: capping job sizes by $\bar p_i$ makes the total fractional load on machine $i$ equal to exactly $\bar{p}_i$.
If $\sum_jx_{ij}\le1$, we choose the maximizer $\bar p_i=0$.

The convex program and its dual are
\begin{align*}
  \min_{x\ge0} \quad            & \sum_i\mathcal C_i(x_i)
                                &
  \max_{\alpha,\beta\ge0} \quad & \sum_j\beta_j-\sum_i \alpha_i^2                                                                           \\
  \text{s.t.}    \quad          & \sum_i x_{ij}\ge1 \quad \forall j\in\OnlineObj
                                & \text{s.t.}  \quad                             & \beta_j\le \Phi(\alpha_i;p_{ij})    \quad \forall (i,j).
\end{align*}
Writing $\gamma_i=\alpha_i^2$ gives the convex form of the Fenchel dual, since $\gamma\mapsto\Phi(\sqrt\gamma;p)$ is concave.
We use the square-root variables $\alpha_i=\sqrt{\gamma_i}$ to simplify notation in the competitive analysis.

Our dual is a restriction of the SDP dual used by \cite{BorstKashaev25}.
The dual assignment in their analysis satisfies this restriction.\footnote{
  In their notation, restrict $v_{ij}=q_{ij}e_i$ to $0\le q_{ij}\le\sqrt{2}p_{ij}$, where $e_i$ is the unit vector for machine $i$, as their analysis does.
  These bounds ensure all pairwise vector constraints.
  For fixed $\nu=\sqrt{2}\alpha$, maximizing each remaining constraint's right-hand side gives $q_{ij}=\sqrt{2}\min\{p_{ij},\alpha_i\}$ and $y_j\le\Phi(\alpha_i;p_{ij})$.
  Identifying $y_j=\beta_j$ preserves the dual objective $\sum_j\beta_j-\sum_i\alpha_i^2$.
}
The improved competitive ratio comes from the First-Age Poisson rounding and its amortized analysis, rather than from a stronger relaxation.

\begin{lemma}[Relaxation and Weak Duality]\label{lem:load-relaxation}
  The primal program is a relaxation of load balancing.
  Consequently, every feasible dual solution satisfies
  $$
    \sum_j\beta_j-\sum_i\alpha_i^2\le\operatorname{OPT}.
  $$
\end{lemma}

\begin{proof}
  For any integral assignment and any fixed $\bar p$, $\Phi(\bar p;p)$ is convex in $p$ and vanishes at $p=0$, so
  $$
    \sum_{j\in S_i}\Phi(\bar p_i;p_{ij})-\bar p_i^2
    \le\Phi(\bar p_i;L_i)-\bar p_i^2
    =L_i^2-(\bar p_i-L_i)_+^2\le L_i^2.
  $$
  Equality holds at $\bar p_i=L_i$, where $\Phi(L_i;p_{ij})=2L_ip_{ij}$, proving $\mathcal C_i(x_i)=L_i^2$.
  Since every integral assignment is feasible for the primal program, the program is a relaxation.
  Finally, for any feasible dual solution $(\alpha,\beta)$, the definition of $\mathcal C_i$ gives $\mathcal C_i(x_i)\ge\sum_jx_{ij}\beta_j-\alpha_i^2$.
  Summing over machines in an optimal integral assignment proves the dual bound.
\end{proof}

\subsection{Covariance Savings and Cost Accounting}

Fix any machine $i$.
Write $j\sim i$ if entry $(i,j)$ is passed to machine $i$'s persistent First-Age Poisson generator.
Let $t_i(j)$ denote the internal time of machine $i$'s correlated timeline immediately before job $j$.
For $j\sim i$, its request interval is $I_{ij}=[t_i(j),t_i(j)+x_{ij})$.
An independent entry uses a fresh $\Exp(x_{ij})$ clock and does not advance that timeline.
The designation of an entry, as well as its mass $x_{ij}$, is independent of the sampled clocks.
This section bounds the covariance savings from the entries correlated via OCEC.

\begin{lemma}\label{lem:load-marginals}
  For every $i\in\OfflineObj$ and $j\in\OnlineObj$, we have $\E[X_{ij}]=x_{ij}$.
  For every $i\in\OfflineObj$ and distinct $j,k\in\OnlineObj$,
  $$
    \E[X_{ij}X_{ik}]
    \le\E\bigl[e^{-(1-x_{ij})Y_{ij}-(1-x_{ik})Y_{ik}}\bigr]
    \le x_{ij}x_{ik}.
  $$
\end{lemma}
\begin{proof}
  The marginal identity follows by the exponential marginals, the independence of different offline objects' OCEC generators, and the fact that $\sum_{i'\in\OfflineObj}x_{i'j}=1$.
  The pairwise bounds are immediate if either mass is zero or one.
  Otherwise, conditional on $Y_{ij}=y$ and $Y_{ik}=z$, independence across offline objects and negative association on each other object give
  $$
    \begin{aligned}
      \Pr[X_{ij}=X_{ik}=1\mid Y_{ij}=y,Y_{ik}=z]
       & \le\prod_{i'\ne i}\Pr[Y_{i'j}>y]\Pr[Y_{i'k}>z] \\
       & =e^{-(1-x_{ij})y-(1-x_{ik})z}.
    \end{aligned}
  $$
  Taking expectations gives the first inequality; negative association gives the second:
  $$
    \E\bigl[e^{-(1-x_{ij})Y_{ij}-(1-x_{ik})Y_{ik}}\bigr]
    \le\E\bigl[e^{-(1-x_{ij})Y_{ij}}\bigr]\E\bigl[e^{-(1-x_{ik})Y_{ik}}\bigr]
    =x_{ij}x_{ik}.
  $$
\end{proof}

Our algorithm in \Cref{sec:load-balancing-algorithm} ensures $x_{ij}\le\overline{x}=0.15$ for every entry $(i,j)$ passed to OCEC.
We now quantify the pairwise covariance savings of such jobs on a machine.
\begin{lemma}[Assignment Covariance]\label{lem:load-rounding}
  Let $\lambda=4.5$ and $\overline{x}=0.15$, and let the covariance saving function be
  $$
    \delta(s)=\frac{43}{50}\exp\!\left(-s-\frac{25}{3}s^2\right),\qquad s\ge0.
  $$
  If jobs $j<k$ satisfy $j\sim i$ and $k\sim i$, with masses at most $\overline{x}$, then
  \begin{equation}\label{eqn:load-pair}
    \operatorname{Cov}(X_{ij},X_{ik})\le
    -\int_{I_{ij}}\int_{I_{ik}} \delta(v-u)\,dv\,du.
  \end{equation}
\end{lemma}

\Cref{lem:load-rounding} simplifies the more precise covariance bound in \Cref{lem:load-covariance} below.
The latter bound holds with equality when the competing machines use fresh clocks for these two jobs.

\begin{lemma}\label{lem:load-covariance}
  Suppose jobs $j<k$ satisfy $j\sim i$, $k\sim i$, and $0<x_{ij},x_{ik}<1$.
  Recall from \eqref{eqn:matching-win-probability} that $D_x(a)$ is the probability of winning a request $x$ against a fresh competing clock of rate $1-x$, conditional on first age $a$.
  Let $dD_x$ denote its positive Stieltjes measure on $[0,\infty)$, including its right jump at zero.
  Then
  \begin{equation}\label{eqn:load-covariance}
    \operatorname{Cov}(X_{ij},X_{ik})
    \le -\lambda^2\int_{I_{ij}}\int_{I_{ik}}
    \biggl(\int_{[0,\infty)}e^{-\lambda\max\{a,t-t_i(j)\}}\,dD_{x_{ij}}(a)\biggr)\!
    \biggl(\int_{(t'-t,\infty)}e^{-\lambda b}\,dD_{x_{ik}}(b)\biggr)\,dt'\,dt.
  \end{equation}
\end{lemma}
\begin{proof}
  By Lemma~\ref{lem:load-marginals}, averaging over the private uniforms gives
  $$
    \operatorname{Cov}(X_{ij},X_{ik})
    \le\E[D_{x_{ij}}(\FirstAge(I_{ij}))D_{x_{ik}}(\FirstAge(I_{ik}))]-x_{ij}x_{ik}.
  $$
  Write $A_j=\FirstAge(I_{ij})$ and $A_k=\FirstAge(I_{ik})$.
  Since $D_x$ is nondecreasing, for any $A\ge0$ we have
  $$
    D_x(A)=D_x(0)+\int_{[0,\infty)}\1\{A>a\}\,dD_x(a).
  $$
  The constant terms disappear from the covariance, so
  $$
    \operatorname{Cov}\bigl(D_{x_{ij}}(A_j),D_{x_{ik}}(A_k)\bigr) =\int_{[0,\infty)}\int_{[0,\infty)}
    \operatorname{Cov}\bigl(\1\{A_j>a\},\1\{A_k>b\}\bigr)
    \,dD_{x_{ik}}(b)\,dD_{x_{ij}}(a).
  $$
  Here $a$ and $b$ are thresholds for the first ages of jobs $j$ and $k$ on machine $i$, respectively.
  We evaluate these indicator covariances by examining the first arrivals in the two intervals.

  Fix thresholds $a,b \ge 0$.
  An arrival at $t\in I_{ij}$ is first in $I_{ij}$ and has age greater than $a$ precisely when its preceding interval of length $\max\{a,t-t_i(j)\}$ is empty.
  The corresponding condition at $t'\in I_{ik}$ requires an empty interval of length $\max\{b,t'-t_i(k)\}$.
  These events are incompatible if $t'-t\le b$; otherwise their empty intervals are disjoint and their Poisson void probabilities multiply.
  On incompatible pairs, the later empty interval has length $b$.
  Integrating the product densities over these pairs therefore gives
  $$
    \begin{aligned}
       & \Pr[\FirstAge(I_{ij})>a]\Pr[\FirstAge(I_{ik})>b]-\Pr[\FirstAge(I_{ij})>a,\FirstAge(I_{ik})>b] \\
       & \qquad=\lambda^2\int_{I_{ij}}\int_{I_{ik}}e^{-\lambda(\max\{a,t-t_i(j)\}+b)}
      \1\{t'-t\le b\}\,dt'\,dt.
    \end{aligned}
  $$
  Integrating the preceding identity against $dD_{x_{ij}}(a)dD_{x_{ik}}(b)$ gives the covariance saving for the conditional win probabilities.
  Using $\E[D_{x_{ij}}(\FirstAge(I_{ij}))]=x_{ij}$ and $\E[D_{x_{ik}}(\FirstAge(I_{ik}))]=x_{ik}$ and applying Tonelli's theorem proves \eqref{eqn:load-covariance}.
\end{proof}

\begin{figure}[htbp]
  \centering
  \includegraphics[width=\textwidth]{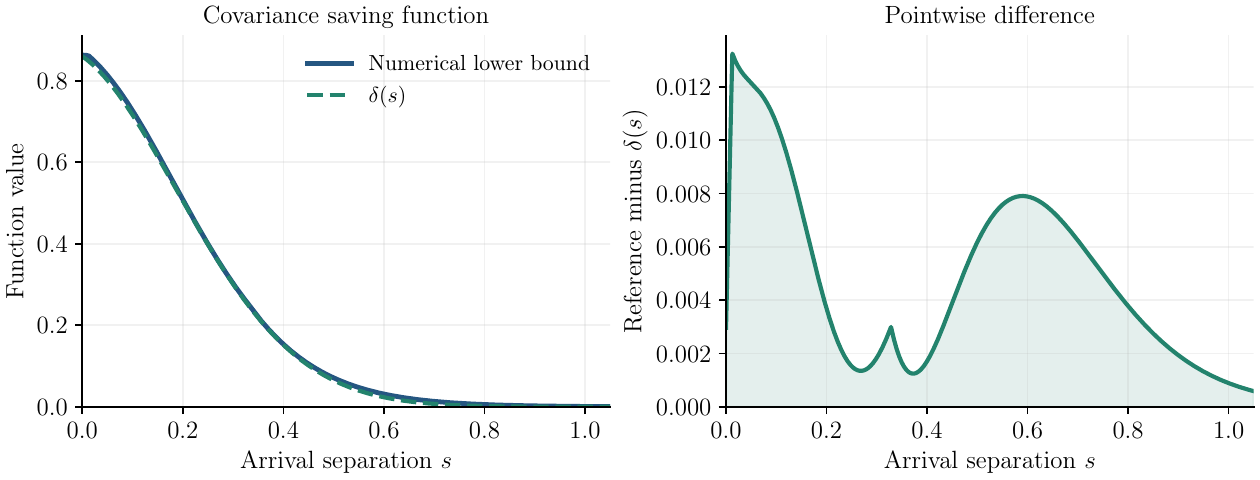}
  \caption{The closed-form covariance saving function $\delta(s)=(43/50)e^{-s-(25/3)s^2}$ and a numerical lower bound on the infimum in the proof sketch of Lemma~\ref{lem:load-rounding}, at $\lambda=4.5$ and $\overline{x}=0.15$.
    The right panel shows their pointwise difference.}
  \label{fig:load-kernel-comparison}
\end{figure}

\begin{proof}[Proof Sketch of Lemma~\ref{lem:load-rounding}]
  For a fixed separation $s=t'-t>0$, the covariance saving integrand in Lemma~\ref{lem:load-covariance} depends on the two request masses $x,x'$ and the earlier arrival's position $r=t-t_i(j)\in[0,x]$.
  To obtain a bound depending only on $s$, it suffices to choose $\delta(s)$ below the infimum over these choices:
  $$
    \delta(s)\le
    \inf_{\substack{0<x,x'\le\overline{x}\\0\le r\le x}}
    \lambda^2
    \left(\int_{[0,\infty)}e^{-\lambda\max\{a,r\}}\,dD_x(a)\right)
    \left(\int_{(s,\infty)}e^{-\lambda b}\,dD_{x'}(b)\right).
  $$
  The first integral is minimized over $r$ at $r=x$, and this minimum is nonincreasing in $x$.
  Thus the infimum reduces to setting $r = x=\overline{x}$ and minimizing over the later mass $x'$.
  Figure~\ref{fig:load-kernel-comparison} shows that the proposed closed-form $\delta$ closely follows a numerical lower bound on this infimum.
  Section~\ref{sec:load-balancing-kernel} certifies the displayed inequality for every $s>0$.
  Substituting it into Lemma~\ref{lem:load-covariance} and integrating over the two request intervals gives \eqref{eqn:load-pair}.
\end{proof}

We now aggregate these pairwise savings into a bound on the expected squared loads.
For each job $j$, let $\ell_i(j)=\sum_{k<j}p_{ik}x_{ik}$ denote the expected load on machine $i$ immediately before $j$.
When discussing a fixed job $j$, we omit the time argument and use a superscript $+$ for the state after processing it.
For example, $\ell_i=\ell_i(j)$ and $\ell_i^+=\ell_i(j+1)$.
Immediately before job $j$, define the covariance credit per unit of job size for a trial mass $x\ge0$ by
$$
  \Delta_{ij}(x)=2\sum_{\substack{k<j\\k\sim i}}p_{ik}
  \int_{I_{ik}}\int_{t_i}^{t_i+x}\delta(v-u)\,dv\,du.
$$
By Lemma~\ref{lem:load-rounding}, $p_{ij}\Delta_{ij}(x_{ij})$ is a guaranteed saving in the squared-load cost bound from allocating $x_{ij}$ of job $j$ to machine $i$ via OCEC, relative to independent rounding.

\begin{lemma}[Cost Accounting]\label{lem:load-cost}
  If $x_{ij}\le\overline{x}$ for all jobs $j\sim i$, we have
  $$
    \E \bigl[ L_i^2 \bigr]\le\sum_j\bigl(p_{ij}^2x_{ij}+2p_{ij}\ell_i(j)x_{ij}\bigr)
    -\sum_{j\sim i}p_{ij}\Delta_{ij}(x_{ij}).
  $$
\end{lemma}

\begin{proof}Expand the square and apply Lemmas~\ref{lem:load-marginals} and~\ref{lem:load-rounding}, charging each correlated pair to its later job:
  $$
    \begin{aligned}
      \E[L_i^2]
       & =\sum_j\bigl(p_{ij}^2x_{ij}+2p_{ij}\ell_i(j)x_{ij}\bigr)
      +2\sum_{k<j}p_{ik}p_{ij}\operatorname{Cov}(X_{ik},X_{ij})     \\
       & \le\sum_j\bigl(p_{ij}^2x_{ij}+2p_{ij}\ell_i(j)x_{ij}\bigr)
      -\sum_{j\sim i}p_{ij}\Delta_{ij}(x_{ij}).
    \end{aligned}
  $$
\end{proof}

\begin{corollary}\label{cor:load-cost}
  Define
  $$
    \overline{\operatorname{ALG}}
    =\sum_{i,j}\bigl(p_{ij}^2x_{ij}+2p_{ij}\ell_i(j)x_{ij}\bigr)
    -\sum_{i,j:\,j\sim i}p_{ij}\Delta_{ij}(x_{ij}).
  $$
  If $x_{ij}\le\overline{x}$ for every $j\sim i$, then $\operatorname{ALG}\le\overline{\operatorname{ALG}}$.
\end{corollary}
\begin{proof}
  Sum the bound in Lemma~\ref{lem:load-cost} over all machines.
\end{proof}

\subsection{Online Primal-Dual Algorithm}\label{sec:load-balancing-algorithm}

Instead of setting $\alpha_i$ to be the maximizing threshold $\bar{p}_i$, we maintain it heuristically as half the expected load.
That is, we let
$$
  \alpha_i=\frac{\ell_i}{2}
  ~,
$$
where $\ell_i$ is the expected load of machine $i$ from jobs seen so far.

Correspondingly, we use $\Phi(\ell_i/2;p_{ij})$ as a surrogate marginal price.
Upon the arrival of job $j$, choose the fractional allocation by water filling with these prices:
\begin{equation}\label{eqn:load-water-filling}
  (x_{ij})_i\in\operatorname*{arg\,min}_{x_i\ge0,\ \sum_i x_i=1}
  \sum_i\int_0^{x_i}\Phi\!\left(\frac{\ell_i+p_{ij}u}{2};p_{ij}\right)\,du.
\end{equation}
After allocating job $j$, we have $\ell_i^+=\ell_i+p_{ij}x_{ij}$.

Each feasible machine's price is continuous and strictly increasing, so the objective is strictly convex and the fractional allocation is unique.
At a minimizer, positive coordinates have a common endpoint price, and zero coordinates have no smaller price.
Set $\beta_j$ to this common price, equivalently
$$
  \beta_j=\min_i \Phi\!\left(\frac{\ell_i^+}{2};p_{ij}\right) = \sum_i x_{ij}\Phi\!\left(\frac{\ell_i^+}{2};p_{ij}\right).
$$
This choice satisfies the dual constraints for job $j$, and monotonicity of the expected loads preserves all previous dual constraints.

\paragraph{Choosing the Rounding Rule.}
For a positive $x_{ij} > 0$, independent rounding increases $\overline{\operatorname{ALG}}$ by $p_{ij}^2x_{ij}+2p_{ij}\ell_ix_{ij}$, and increases the dual objective by $x_{ij}\Phi(\ell_i^+/2;p_{ij})-((\ell_i^+)^2-\ell_i^2)/4$.
We use independent rounding if the former is at most $\Gamma_0$ times the latter, that is, if
\begin{equation}\label{eqn:load-independent}
  p_{ij}^2x_{ij}+2p_{ij}\ell_ix_{ij} \le \Gamma_0 \biggl( x_{ij}\Phi\!\left(\frac{\ell_i^+}{2};p_{ij}\right)-\frac{(\ell_i^+)^2-\ell_i^2}{4} \biggr),
\end{equation}
and use OCEC otherwise.
We choose
\begin{equation}\label{eqn:load-parameters}
  \Gamma_0=4.573,
\end{equation}
slightly smaller than the final competitive ratio $\Gamma=4.69$, to account for the loss from an auxiliary potential in the final competitive analysis.

The test depends only on the expected load, the job size, and the allocated mass.
The following lemma ensures that correlated entries satisfy the mass bound required by Lemma~\ref{lem:load-cost}.

\begin{lemma}[Correlated Mass Bound]\label{lem:load-mass}
  Every entry with $j\sim i$ satisfies $x_{ij}<\overline{x}$.
\end{lemma}
\begin{proof}
  Fix a positive entry $(i,j)$ and let $b=\frac12+\frac2{\Gamma_0}\in(0,1)$.
  The formula for $\Phi$ gives
  $$
    \Phi\!\left(\frac{\ell_i^+}2;p_{ij}\right)
    \ge p_{ij}^2(1-b^2)+bp_{ij}\ell_i^+.
  $$
  Subtracting the right side of \eqref{eqn:load-independent} from the left and dividing by $\Gamma_0p_{ij}^2x_{ij}$ therefore gives at most
  $$
    \frac1{\Gamma_0}-1+b^2-\left(b-\frac14\right)x_{ij}
    \le\frac1{\Gamma_0}-1+b^2-\left(b-\frac14\right)\overline{x}<0
  $$
  whenever $x_{ij}\ge\overline{x}$.
  Thus every such entry uses independent rounding.
\end{proof}

\begin{algorithm}{Online Primal-Dual Allocation with First-Age Poisson OCEC (Load Balancing)}
  \textbf{Internal states for each machine $i$:}
  \begin{itemize}
    \item An independent First-Age Poisson generator $\Generator_i$.
    \item The expected load $\ell_i$, internal time $t_i$, and corresponding offline dual $\alpha_i=\ell_i/2$, all initially zero.
  \end{itemize}

  \medskip

  \textbf{Upon the arrival of a job $j$:}
  \begin{enumerate}
    \item \textbf{Allocate:~} Starting from $x_{ij}=0$, continuously allocate $j$ to feasible machines $i$ with the smallest current price $\Phi((\ell_i+p_{ij}x_{ij})/2;p_{ij})$ until $\sum_i x_{ij}=1$.
          For each feasible machine, set $\ell_i^+\gets\ell_i+p_{ij}x_{ij}$.
    \item \textbf{Update duals:~} For each feasible machine, set $\ell_i\gets\ell_i^+$ and $\alpha_i\gets\ell_i/2$.
          Set $\beta_j\gets\min_i\Phi(\alpha_i;p_{ij})$.
    \item \textbf{Round:~} For each machine with $x_{ij}>0$, if \eqref{eqn:load-independent} fails (using the pre-arrival value of $\ell_i$), send request $x_{ij}$ to $\Generator_i$ and set $t_i\gets t_i+x_{ij}$; otherwise, draw a fresh independent $\Exp(x_{ij})$ clock.
          Assign $j$ to the machine with the smallest clock.
  \end{enumerate}
\end{algorithm}

\subsection{Competitive Analysis}\label{sec:load-balancing-amortized}

A correlated entry can incur a deficit before it generates covariance savings: the first correlated entry on a machine has $\Delta_{ij}(x_{ij})=0$, but fails \eqref{eqn:load-independent}.
Savings from its interactions with later entries can compensate for this deficit, so we amortize the comparison across arrivals.
\cite{BorstKashaev25} likewise aggregate covariance savings over groups of jobs.

\begin{lemma}[Amortized Cost Bound]\label{lem:load-amortized}
  There is a potential $\Psi$ with the following properties:
  \begin{enumerate}
    \item Initially, $\Psi=0$.
    \item At every prefix,
          $$
            \Psi\le\nu\sum_i\ell_i^2,\qquad \nu=0.00539.
          $$
    \item For each arriving job $j$, write $D$ for the current dual objective.
          Then
          \begin{equation}\label{eqn:load-amortized}
            \overline{\operatorname{ALG}}^{+}-\overline{\operatorname{ALG}}
            \le\Gamma_0\bigl(D^+-D+\Psi^+-\Psi\bigr).
          \end{equation}
  \end{enumerate}
\end{lemma}

Lemma~\ref{lem:load-amortized} immediately implies Theorem~\ref{thm:load-balancing}, and finally we give a proof of Lemma \ref{lem:load-amortized}.

\begin{proof}[Proof of Theorem~\ref{thm:load-balancing}]
  Sum \eqref{eqn:load-amortized} over jobs and apply Lemma~\ref{lem:load-mass} and Corollary~\ref{cor:load-cost} to obtain
  $$
    \operatorname{ALG}
    \le\overline{\operatorname{ALG}}
    \le\Gamma_0(D+\Psi)
    \le\Gamma_0(\operatorname{OPT}+\Psi),
  $$
  where the last inequality follows from dual feasibility and weak duality.
  Moreover, $\ell_i=\E[L_i]$, so Lemma~\ref{lem:load-amortized} and Jensen's inequality give
  $$
    \Psi\le\nu\sum_i\ell_i^2\le\nu\operatorname{ALG}.
  $$
  Hence
  $$
    \operatorname{ALG}
    \le\frac{\Gamma_0}{1-\Gamma_0\nu}\,\operatorname{OPT}
    <4.69\,\operatorname{OPT}=\Gamma\,\operatorname{OPT}.
  $$
\end{proof}

\begin{proof}[Proof Sketch of Lemma~\ref{lem:load-amortized}]
  For auxiliary values $0\le\widetilde{\alpha}_i\le\alpha_i$ defined below, set
  $$
    \Psi=\sum_i(\alpha_i^2-\widetilde{\alpha}_i^2).
  $$
  Then, we have $D+\Psi=\sum_j\beta_j-\sum_i\widetilde{\alpha}_i^2$, so the right-hand side of \eqref{eqn:load-amortized} is $\Gamma_0$ times the increase in this auxiliary objective.
  Moreover, $\alpha_i+\widetilde{\alpha}_i\le2\alpha_i=\ell_i$, so
  $$
    \Psi
    =\sum_i(\alpha_i+\widetilde{\alpha}_i)(\alpha_i-\widetilde{\alpha}_i)
    \le\sum_i\ell_i(\alpha_i-\widetilde{\alpha}_i).
  $$
  Thus, property 2 follows if $\alpha_i-\widetilde{\alpha}_i\le\nu\ell_i$ for every machine at every prefix.

  Recall that $\Delta_{ij}(x)$ is the covariance credit per unit of job size for a trial mass $x$, right before processing job $j$.
  Hence, $\Delta_{ij}(\infty)$ captures the total remaining covariance credit from interactions between the existing correlated intervals and the future timeline, per unit of future job size.
  Here the current history stays fixed, and only the trial interval extends to infinity.
  We use a scaled, capped version of this remaining credit to define the gap $\alpha_i-\widetilde{\alpha}_i$.
  Cap each job size at half the expected load immediately before its arrival:
  $$
    \tilde{p}_{ij} = \min\{p_{ij},\alpha_i(j)\} = \min\Bigl\{p_{ij},\frac{\ell_i(j)}2\Bigr\}.
  $$
  Each capped size is fixed upon arrival and remains unchanged thereafter.
  Replacing each earlier job size by its capped value gives
  $$
    \widetilde{\Delta}_{ij}(x)=2\sum_{\substack{k<j\\k\sim i}}\tilde p_{ik}
    \int_{I_{ik}}\int_{t_i}^{t_i+x}\delta(v-u)\,dv\,du.
  $$
  Thus $0\le\widetilde{\Delta}_{ij}(x)\le\Delta_{ij}(x)$.
  Set
  $$
    \widetilde{\alpha}_i=\alpha_i-0.15\,\widetilde{\Delta}_{ij}(\infty).
  $$

  Capping the job sizes plays two roles.
  First, it keeps the remaining credit small relative to the current load, which bounds the potential and establishes the second property of the lemma.
  The high-level plan is to use both $\tilde p_{ik}\le\alpha_i(k)$ and the increase $p_{ik}x_{ik}/2\ge\tilde p_{ik}x_{ik}/2$ in $\alpha_i$ caused by each allocation to constrain the history of capped sizes.
  These constraints yield a differential inequality for the cumulative capped load in the recent history.
  Integrating its solution against the covariance saving function bounds $\alpha_i-\widetilde{\alpha}_i$ by a constant times the current load.
  Section~\ref{sec:load-balancing-potential-proof} certifies that this constant is below $\nu$ by numerically bounding a single scalar integral.

  Second, capping controls the cost of releasing earlier credit in the amortized comparison.
  When a correlated entry $(i,j)$ advances the timeline, interactions between earlier intervals and $I_{ij}$ no longer belong to the future timeline.
  Their removal decreases $\alpha_i-\widetilde{\alpha}_i$ by $0.15\,\widetilde{\Delta}_{ij}(x_{ij})$.
  We call this decrease \emph{releasing credit}; it raises $\widetilde{\alpha}_i$, reducing the increase in the auxiliary objective.

  For example, preload $50$ machines to load $2$ using jobs feasible on only one machine.
  Then present $30$ jobs with size $3\alpha_i/2$ on each machine $i$, followed by one with size $5\alpha_i/8$, evaluating $\alpha_i$ before each arrival.
  Water filling assigns mass $1/50$ per machine, and all these entries fail \eqref{eqn:load-independent}.
  With the uncapped reserve, $\widetilde{\alpha}_i=\alpha_i-0.15\Delta_{ij}(\infty)$, the amortized comparison \eqref{eqn:load-amortized} fails on the final job.
  Here the cap reduces the earlier deposits by one third while leaving the final job's deposit unchanged.
  Lowering the cap also reduces the new credit available to cover the current deficit.
  The local comparison keeps this deposit exactly and bounds the released credit jointly with the reserve before the arrival.
  Both come from the same capped history; the bounds $\tilde p_{ik}\le\alpha_i(k)\le\alpha_i$ yield a lower bound on the reserve in terms of the released credit.
  A larger reserve makes $\widetilde{\alpha}_i$ smaller, reducing its quadratic charge for a given increase.

  Section~\ref{sec:load-balancing-potential-proof} derives this relation and verifies the resulting numerical comparison.
\end{proof}

\section*{Declaration of AI Usage}
The authors developed this work with the help of extensive back-and-forth interactions with GPT Astra.
The concept of OCEC generators and a framework for constructing them from Poisson arrivals was proposed by the authors, who then guided the model's exploration of mapping Poisson arrivals to exponential clocks, potential functions, and proof strategies. The authors selected and refined the suggested approaches to optimize the presentation of results in the paper.
The model also assisted with the formulation of finite optimization problems, the implementation and debugging of numerical verification programs, and the organization and polishing of the manuscript.
The authors verified the correctness of the results and take full responsibility for the algorithms, proofs, numerical certificates, and all other content of the paper.

\bibliographystyle{alphaurl}
\bibliography{ref}

\appendix
\crefalias{section}{appendix}
\crefalias{subsection}{appendix}
\crefalias{subsubsection}{appendix}

\section{Deferred Proof for Online Correlated Exponential Clocks}
\label{app:ocec}

\begin{proof}[Proof of \Cref{lem:first-age-cdf}]
    By stationarity, consider $I=[0,x)$.
    If $T_1<x$, the first arrival in $I$ is $T_1$, with age $T_1-T_0$; otherwise, $I$ contains no arrivals and $\FirstAge(I)=0$.
    Since $-T_0$ and $T_1$ independently follow $\Exp(\lambda)$, conditioning on $T_1=t$ gives, for every $a\ge0$,
    \begin{align*}
        F_x(a)
         & = \Pr[T_1\ge x]+\int_0^{\min\{a,x\}}\Pr[-T_0\le a-t]\,\lambda e^{-\lambda t}\,dt               \\
         & = e^{-\lambda x}+\int_0^{\min\{a,x\}}\bigl(1-e^{-\lambda(a-t)}\bigr)\lambda e^{-\lambda t}\,dt \\[1ex]
         & = 1+e^{-\lambda x}-e^{-\lambda\min\{a,x\}}-\lambda\min\{a,x\}e^{-\lambda a}.
    \end{align*}
    Separating the cases $a<x$ and $a\ge x$ yields~\eqref{eqn:first-age-cdf}.
\end{proof}

\section{Deferred Proofs for Matching and Display Ads}
\label{app:matching-display}

\subsection{Additional Notation and Properties}

For a request $0<x<1$, let $D_x(a)$ be the probability that $i$ wins against a fresh competing clock of rate $1-x$, conditional on first age $a$.

\begin{lemma}
    \label{lem:matching-win-probability}
    For every request $0<x<1$,
    \begin{equation}
        D_x(a)=
        \begin{cases}
            c_x\defeq xe^{-\lambda(1-x)}, & a=0, \\[1ex]
            F_x(a)^{(1-x)/x},             & a>0.
        \end{cases}
        \label{eqn:matching-win-probability}
    \end{equation}
    In particular, $D_x$ is nondecreasing in $a$, with minimum $D_x(0)=c_x$.
\end{lemma}

\begin{proof}
    Given $U$ in \eqref{eqn:first-age-uniform}, the clock $Y=-\log U/x$ wins with probability $e^{-(1-x)Y}=U^{(1-x)/x}$.
    Equation \eqref{eqn:matching-win-probability} follows by substituting $U=F_x(a)$ for $a>0$, and averaging over $U = e^{-\lambda x} V$ with $V\sim\mathrm{Unif}[0,1]$ for $a=0$.
    Monotonicity follows since $F_x$ is nondecreasing and $F_x(a) \ge F_x(0)=e^{-\lambda x}$, so $D_x(a)\ge e^{-\lambda(1-x)}>c_x$ for $a>0$.
\end{proof}

Next, we derive an explicit formula for $\Gamma$ in terms of its value at age zero.

\begin{lemma}\label{lem:potential-first-event}
    For every age $a\ge0$,
    \begin{equation}
        \Gamma(a)
        =\int_0^\infty e^{-(1+\lambda+e^{-\lambda})t}
        \bigl[e^{-\lambda}+\lambda\Gamma(0)+\lambda(1-\Gamma(0))e^{-\lambda e^{-\lambda(a+t)}}\bigr]\,\d t.
        \label{eqn:matching-first-event-potential}
    \end{equation}
    In particular, $\Gamma$ is nondecreasing.
\end{lemma}

\begin{proof}
Consider the infinitesimal-request age process and fix $x\ge0$.
The quantity $(\MatchKernel_x^\star\mathbf1)(a)$ is the probability that $i$ has not won by time $x$, starting from age $a$.
Consider the first event, either a Poisson arrival or an independent win.
With probability $e^{-(\lambda+e^{-\lambda})x}$, neither occurs before $x$, so $i$ remains unmatched.
Otherwise, remaining unmatched requires the first event to be a Poisson arrival that $i$ loses.
The joint density that the first event occurs at time $t\in(0,x)$ and is a Poisson arrival is $\lambda e^{-(\lambda+e^{-\lambda})t}$.
The pre-arrival age is then $a+t$, so $i$ loses with probability $1-e^{-\lambda e^{-\lambda(a+t)}}$.
Upon losing, the process restarts from age zero, and the probability of remaining unmatched until time $x$ is
$(\MatchKernel_{x-t}^\star\mathbf1)(0)$.
Hence,
$$
    (\MatchKernel_x^\star\mathbf1)(a)
    =
    e^{-(\lambda+e^{-\lambda})x}
    +\int_0^x
        \lambda e^{-(\lambda+e^{-\lambda})t}
        \bigl(1-e^{-\lambda e^{-\lambda(a+t)}}\bigr)
        (\MatchKernel_{x-t}^\star\mathbf1)(0)\,\d t.
$$
Multiplying by $e^{-x}$ and integrating over $x\ge0$,
Tonelli's theorem and \eqref{eqn:matching-potential-definition} give
$$
    1-\Gamma(a)
    =
    \frac{1}{1+\lambda+e^{-\lambda}}
    +\lambda(1-\Gamma(0))
        \int_0^\infty
        e^{-(1+\lambda+e^{-\lambda})t}
        \bigl(1-e^{-\lambda e^{-\lambda(a+t)}}\bigr)\,\d t.
$$
Rearranging yields
\eqref{eqn:matching-first-event-potential}.

By \Cref{eqn:matching-potential-definition}, $\Gamma(0)\le1$.
Since $e^{-\lambda e^{-\lambda(a+t)}}$ is increasing in $a$ and its coefficient $\lambda(1-\Gamma(0))$ is nonnegative, \eqref{eqn:matching-first-event-potential} implies that $\Gamma$ is nondecreasing.
\end{proof}

As a corollary, we obtain a differential equation for $\Gamma$.

\begin{lemma}
    \label{lem:potential-derivative}
    The potential $\Gamma$ is continuously differentiable with
    \begin{equation}
        \Gamma'(a)=(1+\lambda+e^{-\lambda})\Gamma(a)-e^{-\lambda}-\lambda\Gamma(0)-\lambda(1-\Gamma(0))e^{-\lambda e^{-\lambda a}}.
        \label{eqn:potential-derivative}
    \end{equation}
\end{lemma}

\begin{proof}
    Substituting $u = a+t$ in \eqref{eqn:matching-first-event-potential} gives
    $$
        \Gamma(a)=e^{(1+\lambda+e^{-\lambda})a}\int_a^\infty e^{-(1+\lambda+e^{-\lambda})u}
        \bigl[e^{-\lambda}+\lambda\Gamma(0)+\lambda(1-\Gamma(0))e^{-\lambda e^{-\lambda u}}\bigr]\,\d u.
    $$
    The integrand is continuous, so differentiating proves the lemma.
\end{proof}

\begin{lemma}
\label{lem:matching-kernel-formula}
For every bounded measurable function $f$, every $a\ge0$,
and $0<x<1$,
\begin{equation*}
    \begin{split}
        (\MatchKernel_xf)(a)
        ={}&
        \bigl(e^{-\lambda x}-xe^{-\lambda}\bigr)f(a+x)\\
        &+\int_0^x\lambda e^{-\lambda s}
        \bigl(1-D_x(a+s)\bigr)
        \left[
            e^{-\lambda(x-s)}f(x-s)
            +\int_0^{x-s}\lambda e^{-\lambda v}f(v)\,dv
        \right] \d s.
    \end{split}
\end{equation*}
\end{lemma}
\begin{proof}
If no Poisson arrival occurs during the request, the final age is $a+x$ and $i$ loses with probability $1-c_x$.
This contributes $e^{-\lambda x}(1-c_x)f(a+x) =(e^{-\lambda x}-xe^{-\lambda})f(a+x)$.

Otherwise, the first arrival occurs at time $s\in(0,x)$ with density $\lambda e^{-\lambda s}$.
Its first age is $a+s$, so $i$ loses with probability
$1-D_x(a+s)$.
After this arrival, the age resets to zero.
The final age equals $x-s$ if no further arrival occurs, and otherwise has density $\lambda e^{-\lambda v}$ on $(0,x-s)$.
Thus, the bracketed expression is the conditional expected
value of $f$ at the final age.
Integrating over $s$ proves it.
\end{proof}

\subsection{Numerical Value of \texorpdfstring{$\Gamma^\star$}{Gamma*}}
\label{app:gamma-star}
Taking expectation over $a \sim \pi$ in \eqref{eqn:potential-derivative} gives
$$
    \pi\Gamma'=(1 + \lambda + e^{-\lambda})\Gamma^\star-e^{-\lambda}-\lambda\Gamma(0)-(1-\Gamma(0))(1-e^{-\lambda}),
$$
where the last term uses $\int_0^\infty\lambda e^{-\lambda a}\cdot\lambda e^{-\lambda e^{-\lambda a}}\,\d a=1-e^{-\lambda}$.
On the other hand, we also have
\begin{align*}
    \pi\Gamma'
     =\int_0^\infty\lambda e^{-\lambda a}\Gamma'(a)\,\d a
     = - \lambda \Gamma(0)
    +\lambda\int_0^\infty\lambda e^{-\lambda a}\Gamma(a)\,\d a
     =\lambda(\Gamma^\star-\Gamma(0)),
\end{align*}
where the second equality holds by integration by parts.
Equating the two gives
\begin{equation*}
    \Gamma^\star=\frac{1-(1-e^{-\lambda})\Gamma(0)}{1+e^{-\lambda}}. 
\end{equation*}

Write
$
I\defeq\int_0^\infty e^{-(1 + \lambda + e^{-\lambda})t}e^{-\lambda e^{-\lambda t}}\,\d t
$
.
Setting $a=0$ in \eqref{eqn:matching-first-event-potential} and solving for $\Gamma(0)$ gives
$$
    \Gamma(0)=\frac{e^{-\lambda}+(1 + \lambda + e^{-\lambda})\lambda I}{1+e^{-\lambda}+(1 + \lambda + e^{-\lambda})\lambda I}.
$$
Using $\lambda = 4.5$, we can substitute and calculate each of $I$, $\Gamma(0)$, and $\Gamma^\star$ up to high precision, yielding the values:
\begin{align*}
    I &= 0.0315488474142158237512103586\dots,\\\Gamma(0) &=0.4424372533127501602229664934\dots,\\
    \Gamma^\star &=0.5562978694962254765482117855\dots.
\end{align*}
\qed

\subsection{Proof of Lemma~\ref{lem:matching-infinitesimal-generator}}
\label{app:matching-gap-kernel-generator}

Fix $a\ge0$ and apply \Cref{lem:matching-kernel-formula}.
By Taylor expansion, the first term satisfies
$$
    \bigl(e^{-\lambda x}-xe^{-\lambda}\bigr)f(a+x)
    =
    f(a)+x\bigl[f'(a)-(\lambda+e^{-\lambda})f(a)\bigr]+o(x).
$$

By \Cref{lem:first-age-cdf,lem:matching-win-probability},
uniformly over $0<s<x$ as $x\downarrow0$, we have
$F_x(a+s)=1-\lambda e^{-\lambda a}x+O(x^2)$ and $D_x(a+s)=e^{-\lambda e^{-\lambda a}}+o(1)$.
Continuity of $f$ also gives, uniformly over the same range,
$
    e^{-\lambda(x-s)}f(x-s)
    +\int_0^{x-s}\lambda e^{-\lambda v}f(v)\,\d v
    =f(0)+o(1).
$
Since $e^{-\lambda s}=1+o(1)$ uniformly, the integral term in
\Cref{lem:matching-kernel-formula} equals
$
    \lambda x\bigl(1-e^{-\lambda e^{-\lambda a}}\bigr)f(0)+o(x).
$

Combining the two terms yields
$$
    \begin{aligned}
        (\MatchKernel_xf)(a)
        ={}&f(a)
        +x\Bigl[
            f'(a)-(\lambda+e^{-\lambda})f(a)
            +\lambda\bigl(1-e^{-\lambda e^{-\lambda a}}\bigr)f(0)
        \Bigr]
        +o(x),
    \end{aligned}
$$
which proves the claimed infinitesimal generator formula.
\qed

\subsection{Proof of Lemma~\ref{lem:matching-measures-bounds} and Lemma~\ref{lem:display-measures-bounds}}
\label{app:matching-display-measures-bounds}

It suffices to prove Lemma~\ref{lem:display-measures-bounds}.
Let $\nu$ be obtained from $\pi$ by a finite sequence of matching and gap updates, and let $x_j$ denote the mass of matching request $j$.
We may assume $0<x_j<1$, since $\MatchKernel_0$ is the identity and $\MatchKernel_1$ makes the final measure zero.
Realize the ages by a stationary rate-$\lambda$ Poisson process on $(-\infty,t]$, where $t$ is the total length of all updates and the matching requests and gaps occupy consecutive intervals of $[0,t]$ in the order of the updates.
Given the process, $i$ loses matching request $j$ independently with probability $1-D_{x_j}(b_j)$, where $b_j$ is its first age.
No loss condition is imposed during gaps.

For $r\le t$, let $P(r)$ denote the probability that $i$ loses every matching request after the arrivals in $(r,t]$ are removed.
An empty request interval remains empty.
For a nonempty matching request $j$, if its first arrival is
after $r$, the removal deletes every arrival in that interval
and sets $b_j$ to zero; otherwise, it keeps that arrival and
all earlier ones, so $b_j$ is unchanged.
Since $D_x(b)\ge D_x(0)$ by \Cref{lem:matching-win-probability}, removing more arrivals can only increase the probability of losing every matching request.
Hence, $P$ is nonincreasing in $r$.

The event $A(t)\ge a$ means no arrival in $(t-a,t]$.
It has probability $e^{-\lambda a}$ and, by independent increments, conditioning on it does not affect the process before $t-a$.
Thus, $\nu([a,\infty))=e^{-\lambda a}P(t-a)$.
Consequently, $e^{\lambda a}\nu([a,\infty))=P(t-a)$ is nondecreasing in $a$ and at most one.
The stated tail bounds follow from its value $\nu\mathbf1$ at $a=0$.
\qed

\subsection{Proof of Lemma~\ref{lem:matching-request-monotonicity} and Lemma~\ref{lem:display-request-monotonicity}}
\label{app:matching-display-request-monotonicity}

It suffices to prove Lemma~\ref{lem:display-request-monotonicity}.
The following statement immediately proves Lemma~\ref{lem:display-request-monotonicity} since $\Gamma$ is nonnegative, bounded, and nondecreasing.
\begin{lemma}
\label{lem:request-tail-monotonicity}
Let $\nu$ be obtained from $\pi$ by a finite sequence of matching and gap updates.
For every $a\ge0$, both $\nu\MatchKernel_x([a,\infty))$ and $\nu\GapKernel_x([a,\infty))$ are continuous and nonincreasing in $x$ on $[0,1]$.
\end{lemma}

\begin{proof}
We first consider the matching kernel.
Fix a threshold $a\ge0$, and let $0<x<1$.
If $x\le a$, ending with age at least $a$ requires an empty request and an initial age at least $a-x$, so
\begin{equation}
    (\nu\MatchKernel_x)([a,\infty))
    =(1-c_x)e^{-\lambda a}
        \cdot e^{\lambda(a-x)}\nu([a-x,\infty)).
    \label{eqn:matching-request-tail-empty}
\end{equation}
If $x>a$, ending with age at least $a$ requires no arrival in the last $a$ units of the request, an event of probability $e^{-\lambda a}$.
Given this event and initial age $b$, $i$ wins with probability
\begin{equation}
    W_x(b)\defeq
    e^{-\lambda(x-a)}c_x
    +\int_0^{x-a}\lambda e^{-\lambda s}D_x(b+s)\,\d s,
    \label{eqn:poisson-arrival-win}
\end{equation}

where the two terms correspond to no arrival and a first
arrival at time $s$.
Hence,
\begin{equation}
    (\nu\MatchKernel_x)([a,\infty))
    =
    e^{-\lambda a}\int\bigl(1-W_x(b)\bigr)\,\nu(\d b).
    \label{eqn:matching-request-tail-arrival}
\end{equation}

\paragraph{Continuity.}
Since the underlying age process is stationary and $\nu$ counts only outcomes in which $i$ loses every matching request, with no loss condition imposed during gaps, we have $\nu\leq\pi$.
In particular, $\nu$ has no atoms since $\pi$ has none, so \eqref{eqn:matching-request-tail-empty} is continuous in $x$.
Moreover, \eqref{eqn:matching-request-tail-arrival} is continuous in $x$ since $W_x(b)$ is, and the two expressions agree at $x=a$ since $W_a(b)=c_a$.
Finally, the tail $(\nu\MatchKernel_x)([a,\infty))$ tends to $\nu([a,\infty))$ as $x\downarrow0$ and to zero as $x\uparrow1$, the latter since $D_x\ge c_x\to1$.
These limits agree with $\nu\MatchKernel_0=\nu$ and $\nu\MatchKernel_1=0$.
Hence the tail is continuous on $[0,1]$.

\paragraph{Monotonicity.}
The right side of \eqref{eqn:matching-request-tail-empty} is nonincreasing in $x$, since $e^{\lambda(a-x)}\nu([a-x,\infty))$ is nonincreasing by Lemma~\ref{lem:display-measures-bounds}, and $c_x$ is increasing.
For \eqref{eqn:matching-request-tail-arrival}, it suffices to show that $W_x(b)$ is increasing in $x$.

We first establish the following bound for $0<x<1$ and $b>0$:
\begin{equation}
    \partial_x\log D_x(b)
    \ge-\frac{\lambda}{e^{\lambda x}-1}.
    \label{eqn:matching-win-mass-derivative}
\end{equation}
Recall from \Cref{lem:matching-win-probability} that $\log D_x(b)=\frac{1-x}{x}\log F_x(b)$ for $b>0$; we treat the two cases of \Cref{lem:first-age-cdf} separately.

For $b\ge x$, we have $F_x(b)=1-vx$ with $v\defeq\lambda e^{-\lambda b}$, so
$$
    \partial_x\log D_x(b)
    =-\frac{\log(1-vx)}{x^2}
     -\frac{(1-x)v}{x(1-vx)},
    \qquad
    \partial_b\partial_x\log D_x(b)
    =-\frac{\lambda v(1-v)}{(1-vx)^2}.
$$
As $v$ decreases in $b$, the mixed derivative changes sign at most once, from positive to negative.
Hence, the infimum over $b\ge x$ is attained at $b=x$ or approached as $b\to\infty$.
The limit as $b\to\infty$ equals $0$ and satisfies the bound.
At $b=x$, the bound $-\log(1-vx)\ge vx$ gives $\partial_x\log D_x(b)\ge v(1-v)/(1-vx)$.
With $z=\lambda x$, it therefore suffices to show
$$
    (\lambda+1+z)e^{-z}-\lambda e^{-2z}\le2.
$$
Using $z-\log2\le e^z/2-1$, we obtain
\begin{align*}
    (\lambda+1+z)e^{-z}-\lambda e^{-2z}
    &\le
    \frac12+(\lambda+\log2)e^{-z}-\lambda e^{-2z}\\
    &\le
    \frac12+\frac{(\lambda+\log2)^2}{4\lambda}
    <2,
\end{align*}
where the last inequality uses $\lambda=4.5$.

For $0<b<x$, we have
$\partial_xF_x(b)=-\lambda e^{-\lambda x}$ and
$\partial_bF_x(b)=\lambda^2be^{-\lambda b}$, so
$$
    \partial_x\log D_x(b)
    =-\frac{\log F_x(b)}{x^2}
     -\frac{\lambda(1-x)e^{-\lambda x}}{xF_x(b)},
$$
and
$$
    \partial_b\partial_x\log D_x(b)
    =
    \frac{\lambda x(1-x)e^{-\lambda x}-F_x(b)}
         {x^2F_x(b)^2}
    \,\lambda^2be^{-\lambda b}.
$$
As $F_x(b)$ increases in $b$, the mixed derivative changes sign at most once, from positive to negative.
Hence, the infimum over $0<b<x$ is approached as $b\downarrow0$ or $b\uparrow x$.
Both limits satisfy the bound: the former is $\lambda$, since $F_x(b)\to e^{-\lambda x}$, and the latter equals the value at $b=x$ verified above.
This proves \eqref{eqn:matching-win-mass-derivative}.

Returning to $W_x(b)$, for $b>0$, differentiating in $x$ gives
$$
    \frac{\d}{\d x}W_x(b)
    =
    e^{-\lambda(x-a)}
        \bigl(c_x'-\lambda c_x+\lambda D_x(b+x-a)\bigr)
    +\int_0^{x-a}
        \lambda e^{-\lambda s}\partial_xD_x(b+s)\,\d s.
$$
Since $\partial_xD_x=D_x\,\partial_x\log D_x$, \eqref{eqn:matching-win-mass-derivative} and $D_x(b+s)\le D_x(b+x-a)$ by
\Cref{lem:matching-win-probability} bound the integral from below by
$$
    -\frac{\lambda}{e^{\lambda x}-1}
        \int_0^{x-a}\lambda e^{-\lambda s}D_x(b+s)\,\d s
    \ge
    -\frac{1-e^{-\lambda(x-a)}}{e^{\lambda x}-1}
        \,\lambda D_x(b+x-a).
$$
Substituting this bound and $c_x'-\lambda c_x=e^{-\lambda(1-x)}$ gives
\begin{align*}
    \frac{\d}{\d x}W_x(b)
    &\ge
    e^{-\lambda(1-a)}
    +\lambda D_x(b+x-a)
        \left(
            e^{-\lambda(x-a)}
            -\frac{1-e^{-\lambda(x-a)}}{e^{\lambda x}-1}
        \right)\\
    &=
    e^{-\lambda(1-a)}
    +\lambda D_x(b+x-a)
        \frac{e^{\lambda a}-1}{e^{\lambda x}-1}
    >0.
\end{align*}
Hence \eqref{eqn:matching-request-tail-arrival} is nonincreasing in $x$ for $x>a$.
Together with \eqref{eqn:matching-request-tail-empty} and continuity, this proves that the tail is nonincreasing on $[0,1]$.

Then, consider the gap kernel. 
The same age argument gives
\begin{equation}
(\nu\GapKernel_x)([a,\infty))
=
\begin{cases}
    e^{-\lambda a}\cdot
    e^{\lambda(a-x)}\nu([a-x,\infty)),
    & x\le a,\\[2pt]
    e^{-\lambda a}\nu\mathbf1,
    & x>a.
\end{cases}
\label{eqn:gap-kernel-tail-formula}
\end{equation}
Indeed, when $x\le a$, the gap must be empty and the initial
age must be at least $a-x$; when $x>a$, only the last $a$
units of the gap must be empty.
The first expression is continuous since $\nu$ has no atoms,
and nonincreasing in $x$ by
Lemma~\ref{lem:display-measures-bounds}.
The second is constant in $x$, and the two expressions agree
at $x=a$.
Thus every tail of $\nu\GapKernel_x$ is also continuous and
nonincreasing on $[0,1]$.
\end{proof}

This completes the proof for Lemma~\ref{lem:display-request-monotonicity} and Lemma~\ref{lem:matching-request-monotonicity}.
\qed

\subsection{Proof of Lemma~\ref{lem:matching-potential-inequalities} and Lemma~\ref{lem:display-potential-inequalities}}
\label{app:matching-display-potential-inequalities}

It suffices to prove Lemma~\ref{lem:display-potential-inequalities}.
Let $\nu$ be obtained from $\pi$ by a finite sequence of matching and gap updates.
Define
$
    R_x\defeq
    \mathbf1-\Gamma-\MatchKernel_x\mathbf1
    +(1-x)\MatchKernel_x\Gamma.
$
It remains to prove $\nu R_x\ge0$ for every $x\in[0,1]$.
We reduce this inequality
to a finite numerical verification.

\paragraph{Reduction to Finite Checks.}
We partition $[0,1]$ into intervals and associate each interval $I$ with a bounded function $f_I$ that has a limit at infinity.
As shown below, verifying $\nu f_I\ge0$ for all these intervals establishes $\nu R_x\ge0$ throughout $[0,1]$.
Fix an interval $I$ and define
$$
    \bar f_I(a)\defeq\inf_{b\ge a}f_I(b).
$$
Then $\bar f_I\le f_I$, and $\bar f_I$ is nondecreasing.
Applying the lower tail bound in
\Cref{lem:display-measures-bounds} gives
$$
    \nu f_I
    \ge \nu\bar f_I
    \ge (\nu\mathbf1)\pi\bar f_I.
$$
Thus, it suffices to show that $\pi\bar f_I\ge0$.
With the change of variables $z=e^{-\lambda a}$, we have
$$
    \pi\bar f_I
    =\int_0^\infty
        \lambda e^{-\lambda a}\bar f_I(a)\,\d a
    =\int_0^1
        \bar f_I\left(-\frac{\log z}{\lambda}\right)\,\d z.
$$
Since the integrand is nonincreasing in $z$, its right-endpoint sum gives
$$
    \pi\bar f_I
    \ge B_N(f_I)
    \defeq
    \frac1N\sum_{j=1}^N
        \bar f_I\left(-\frac{\log(j/N)}{\lambda}\right).
$$
It therefore suffices to verify that $B_N(f_I)\ge0$.
To obtain a computable lower bound on $B_N(f_I)$, we partition $[0,1]$ into $N$ equal intervals in $z$ and use interval arithmetic to obtain numbers $L_j$ satisfying
$$
    f_I(-\log z/\lambda)\geq L_j,~
    \text{for every }
    z\in\left[\frac{j-1}{N},\frac{j}{N}\right],
    ~j=1,\ldots,N,
$$
where the value at $z=0$ is interpreted by its limit.
Then, we have
$$
    \bar f_I\left(-\frac{\log(j/N)}{\lambda}\right)
    \ge\min_{1\le j'\le j}L_{j'}.
$$
Consequently, it suffices to verify
\begin{equation}
    \frac1N\sum_{j = 1}^N\min_{1\le j'\le j}L_{j'} \geq 0.
    \label{eqn:display-ads-verify-require}
\end{equation}
The program verifies \eqref{eqn:display-ads-verify-require} for every interval $I$.
We carry out this verification separately for large and
small requests.

\paragraph{Requests $x\in[\frac{1}{10},1]$.}
For each interval
$
    I=[\ell,r]
    =\left[\frac{16}{k+1},\frac{16}{k}\right],
    ~k=16,\ldots,159,
$
take
$$
    f_I
    \defeq C_{\ell,r}
    =\mathbf1-\Gamma-\MatchKernel_\ell\mathbf1+(1-r)\MatchKernel_r\Gamma.
$$
By \Cref{lem:display-request-monotonicity,lem:request-tail-monotonicity},
both $\nu\MatchKernel_x\Gamma$ and
$\nu\MatchKernel_x\mathbf1$ are nonincreasing in $x$.
Hence $\nu R_x\ge\nu f_I$ for every $x\in I$.
Fix one such interval $I$.
We use $N=512$ and compute a lower bound $L_j$ valid throughout
each $z$-interval $[z_-,z_+]$, where
$z_-=(j-1)/512$ and $z_+=j/512$.
Substituting $v=e^{-\lambda s}$ in
\Cref{lem:matching-kernel-formula} and using the monotonicity
of $\Gamma$ and $D_y$ in age gives, for every $z\in[z_-,z_+]$,
\begin{align*}
    f_I\left(-\frac{\log z}{\lambda}\right)
    \geq{}&
    \ell e^{-\lambda}
    -\Gamma\left(-\frac{\log z_-}{\lambda}\right)\\
    &+\int_{e^{-\lambda\ell}}^1
        D_\ell\left(-\frac{\log(z_+v)}{\lambda}\right)
        \,\d v\\
    &+(1-r)\bigl(e^{-\lambda r}-re^{-\lambda}\bigr)
        \Gamma\left(r-\frac{\log z_+}{\lambda}\right)\\
    &+(1-r)\int_{e^{-\lambda r}}^1
        \left[
            1-D_r\left(-\frac{\log(z_-v)}{\lambda}\right)
        \right]
        (\GapKernel_{r+\log v/\lambda}\Gamma)(0)
        \,\d v.
\end{align*}
Values at $z_-=0$ are interpreted by their limits.

The two displayed integrals are bounded using finite interval sums over $256$ equal subintervals.
To enclose the potential values, we transform the integral defining $\Gamma$ to $[0,1]$ and bound it using weighted endpoint sums over $8192$ equal subintervals.
Using upper bounds for negative terms and lower bounds for positive terms yields a certified lower bound $L_j$ on the right-hand side.
For each of the $144$ request intervals, the program verifies \eqref{eqn:display-ads-verify-require} for $N = 512$.

\paragraph{Requests $x\in[0,\frac{1}{10}]$.}
For $0\le x\le1/10$ and $b\ge0$, let $v=\lambda e^{-\lambda b}$ and define
$$
    \underline D_x(b)\defeq
    \exp\!\left(
        -(1-x)\left(
            v+\frac{xv^2}{2}+\frac{20x^2v^3}{33}
        \right)
    \right).
$$
For $x>0$ and $b>0$, $\underline D_x(b)\leq (1-xv)^{(1-x)/x}$. Thus, it is a smooth lower bound on $D_x(b)$.
Replace $D_x$ by $\underline D_x$ only in the integral of \Cref{lem:matching-kernel-formula}, and denote the resulting kernel and residual by $\underline{\MatchKernel}_x$ and $\underline R_x$.
Since $\mathbf1-(1-x)\Gamma\ge0$, we have $R_x\ge\underline R_x$.
Moreover, $\underline D_0(b)=e^{-\lambda e^{-\lambda b}}$, so \Cref{lem:matching-infinitesimal-generator,lem:potential-derivative} give $\underline R_0=\partial_x\underline R_0=0$.

Set $T_x\defeq\partial_x^2\underline R_x$, with right derivatives
at zero.
Taylor's integral formula gives
\[
    \nu R_x
    \ge\nu\underline R_x
    =\int_0^x(x-t)\nu T_t\,\d t.
\]
Thus, it suffices to prove $\nu T_x\ge0$ for every
$x\in[0,1/10]$.

For each interval
$
    I=\left[\frac{k}{1280},\frac{k+1}{1280}\right],~
    k=0,\ldots,127,
$
take
$$
    f_I(a)\defeq\inf_{x\in I}T_x(a).
$$
Then $\nu T_x\ge\nu f_I$ for every $x\in I$,
so it suffices to verify $B_{1024}(f_I)\ge0$ for each interval.

Fix one such interval $I$.
We use $N=1024$ and compute a lower bound $L_j$ valid throughout each $z$-interval $[z_-,z_+]$, where $z_-=(j-1)/1024$ and $z_+=j/1024$.
Substituting $s=xu$ in the kernel formula and differentiating twice in $x$ gives
\begin{align*}
    T_x(a)
    ={}&
    -\partial_x^2\left[
        \bigl(e^{-\lambda x}-xe^{-\lambda}\bigr)
        \bigl(1-(1-x)\Gamma(a+x)\bigr)
    \right]\\
    &-\int_0^1
        \partial_x^2\left[
            \lambda x e^{-\lambda xu}
            \bigl(1-\underline D_x(a+xu)\bigr)
            \bigl(1-(1-x)(\GapKernel_{x(1-u)}\Gamma)(0)\bigr)
        \right]\,\d u.
\end{align*}

We bound the integral by finite interval sums over $256$ equal subintervals of $[0,1]$.
On each subinterval, interval arithmetic bounds the integrand uniformly over $x\in I$ and $z\in[z_-,z_+]$.
The required potential and derivative values are enclosed
using the same finite-sum construction with $8192$ integration
subintervals.
Combining these bounds with the nonintegral term gives
a certified lower bound $L_j$ satisfying
$$
    T_x\left(-\frac{\log z}{\lambda}\right)\geq L_j,~
    \text{for all }x\in I,\ z\in[z_-,z_+],
$$
where the value at $z=0$ is interpreted by its limit.
This also bounds $f_I(-\log z/\lambda)$ below by $L_j$.

For each of the $128$ request intervals, the program verifies \eqref{eqn:display-ads-verify-require} for $N = 1024$.

\qed

\subsection{Proof of Lemma~\ref{lem:gap-evolution}}
\label{app:display-gap-evolution}

For any finite measure $\nu$, the Kolmogorov equation gives
$$
    \frac{\d}{\d x}\nu\GapKernel_x\Gamma
    =\nu\GapKernel_x\InfGenerator_{\GapKernel}\Gamma,
$$
with a right derivative at $x=0$.
Thus, it suffices to prove that for any reachable measure $\nu$,
\begin{equation}
    -\Gamma^\star+\nu\Gamma \leq \nu\InfGenerator_\GapKernel\Gamma \leq 0.
    \label{eqn:gap-generator-bounds}
\end{equation}
Moreover, stationarity gives
$
    \pi\InfGenerator_{\GapKernel}\Gamma
    =0.
$

For the upper bound, \Cref{lem:display-request-monotonicity} and $\GapKernel_0=\IdKernel$ imply $\nu\GapKernel_x\Gamma\le\nu\Gamma$ for every $x\in[0,1]$.
Then,
$$
    \nu\InfGenerator_{\GapKernel}\Gamma
    =\lim_{x\downarrow0}
        \frac{\nu\GapKernel_x\Gamma-\nu\Gamma}{x}
    \le0.
$$

For the lower bound, we show that the function $(\Gamma - \InfGenerator_\GapKernel\Gamma)(a)$ is nonnegative and nondecreasing.
Together with the upper tail bound in Lemma~\ref{lem:display-measures-bounds}, i.e., $\nu([a, \infty))\leq \pi([a, \infty))$ for any $a\geq0$, it gives 
$$
\nu(\Gamma-\InfGenerator_\GapKernel\Gamma) \leq \pi (\Gamma-\InfGenerator_\GapKernel\Gamma) = \pi\Gamma = \Gamma^\star,
$$
which proves the lower bound by rearranging.
By \Cref{lem:potential-derivative,lem:gap-infinitesimal-generator},
\begin{align*}
    (\Gamma-\InfGenerator_{\GapKernel}\Gamma)(a)
    =(1+\lambda)\Gamma(a)-\Gamma'(a)-\lambda\Gamma(0) =e^{-\lambda}(1-\Gamma(a))
      +\lambda(1-\Gamma(0))e^{-\lambda e^{-\lambda a}}.
\end{align*}
This is nonnegative since $0<\Gamma<1$.
Let $q:=1+\lambda+e^{-\lambda}$ to simplify the notation.
Differentiating \eqref{eqn:potential-derivative} gives
$$
    \frac{\d}{\d a}\bigl(e^{-qa}\Gamma'(a)\bigr)
    =
    -\lambda^3(1-\Gamma(0))
    e^{-(q+\lambda)a}e^{-\lambda e^{-\lambda a}}.
$$
\Cref{eqn:potential-derivative} and the boundedness of $\Gamma$ imply that $\Gamma'$ is bounded.
Integrating from $a$ to infinity therefore yields
\begin{align*}
    \Gamma'(a)
    =
    \lambda^3(1-\Gamma(0))e^{-\lambda a}
    \int_0^\infty
        e^{-(q+\lambda)t}
        e^{-\lambda e^{-\lambda(a+t)}}\,\d t
    \leq
    \frac{\lambda^3(1-\Gamma(0))e^{-\lambda a}}
         {q+\lambda}.
\end{align*}
Consequently,
\begin{align*}
    \frac{\d}{\d a}
        (\Gamma-\InfGenerator_{\GapKernel}\Gamma)(a)
    &=
    -e^{-\lambda}\Gamma'(a)
    +\lambda^3(1-\Gamma(0))
        e^{-\lambda a}e^{-\lambda e^{-\lambda a}}\\
    &\ge
    \lambda^3(1-\Gamma(0))e^{-\lambda a}
    \left(
        e^{-\lambda e^{-\lambda a}}
        -\frac{e^{-\lambda}}{q+\lambda}
    \right)\\
    &\ge
    \lambda^3(1-\Gamma(0))e^{-\lambda(a+1)}
    \left(1-\frac{1}{q+\lambda}\right)
    >0,
\end{align*}
where the second inequality uses
$e^{-\lambda e^{-\lambda a}}\ge e^{-\lambda}$.

\qed

\section{Verification of the AdWords Potential}
\label{app:adwords}

We complete the construction in Section~\ref{sec:adwords-optimization}
and prove Lemma~\ref{lem:adwords-potential}. We use the same notation
$z=e^{-\lambda a}$, $F_s,G_s,h$, and $\theta$, with measures understood
in $z$ coordinates when integrated against these functions. The
certificate consists of rational price coefficients, interval-cost
bounds, and LP dual witnesses. The checks below establish the required
inequalities for every reachable state and every request.

\subsection{The Candidate and the Explicit LP}
\label{app:adwords-certificate}
\label{app:aw-alt-finite-lp}

\begin{figure}[htbp]
 \centering
 \includegraphics[width=.98\linewidth]{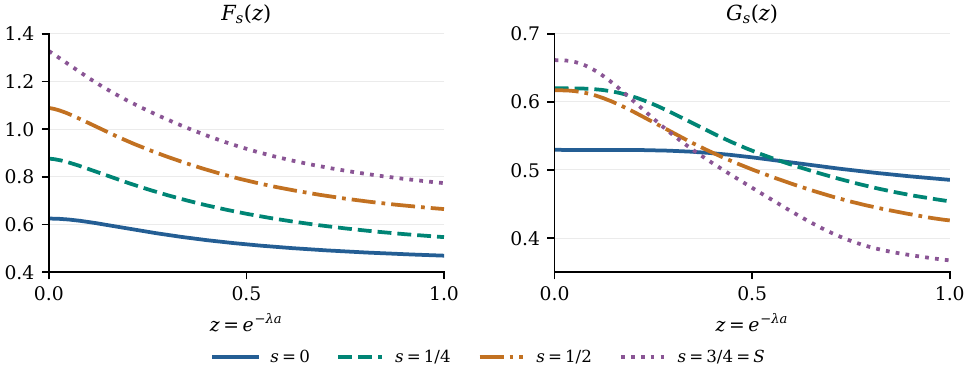}
 \caption{The certified degree-$12$ functions $F_s$ and $G_s$ before
 restoring the factor $e^{-2s}$. The coordinate $z=e^{-\lambda a}$ runs
 from the limit of very old ages to age zero; panels have different
 vertical scales. Beyond $S=3/4$, $G_s=G_S$ and $F_s=F_S+(s-S)h$.}
 \label{fig:aw-alt-functions}
\end{figure}

The candidate uses $\lambda=9/2$, degree $d=12$, and the load knots
\begin{equation}
 0,\ \tfrac1{20},\ \tfrac1{10},\ \tfrac3{20},\ \tfrac15,
 \ \tfrac14,\ \tfrac3{10},\ \tfrac25,\ \tfrac12,
 \ \tfrac58,\ S=\tfrac34.
 \label{eqn:aw-cert-knots}
\end{equation}
All $299$ coefficients are rational and are supplied in
the program files. With the parametrization
\eqref{eqn:adwords-design-prices}--\eqref{eqn:adwords-design-continuation},
their exact initial value is
\begin{equation}
 \Gamma(\pi,0,0)=\int_0^1F_0(z)\,\d z
     =\frac1{13}\sum_{j=0}^{12}f_{0j}=\frac{53}{100}.
 \label{eqn:aw-cert-initialization}
\end{equation}
We verify the initial value and all shape and load constraints in \eqref{eqn:aw-alt-design-lp} by program using exact rational arithmetic.
Figure~\ref{fig:aw-alt-functions} shows the resulting functions.

\paragraph{Accounting tests.}
\label{app:aw-alt-families}
For a request with a large bid, substituting~\eqref{eqn:adwords-update} gives
\begin{equation}
\begin{aligned}
 e^{2s}[U-U'-\Gamma+(1-\NormalizedBid x)\Gamma']&=\nu_0C_0+\nu_1C_1,\\
 C_0&=(s+\NormalizedBid)\WinKernel_x\1-F_s
       +(1-\NormalizedBid x)[\MatchKernel_xF_s+(1-\NormalizedBid)\WinKernel_xG_s],\\
 C_1&=\WinKernel_x\1-G_s+(1-\NormalizedBid x)\MatchKernel_xG_s.
\end{aligned}
\label{eqn:aw-cert-large-costs}
\end{equation}
Their degree-two Bernstein coefficients in $\NormalizedBid\in[1/2,1]$ are the
endpoint costs and
\begin{equation}
 C_{0,\mathrm{mid}}=\tfrac12(C_{0,1/2}+C_{0,1})
                   -\tfrac{x}{8}\WinKernel_xG_s,
 \qquad
 C_{1,\mathrm{mid}}=\tfrac12(C_{1,1/2}+C_{1,1}).
 \label{eqn:aw-cert-bid-coefficients}
\end{equation}
Nonnegativity of the three integrated coefficients suffices for all
bids. The costs are affine in $s$ between knots. Beyond $S$, their
integrated slope is
$\nu_0[\WinKernel_x\1-h+(1-\NormalizedBid x)\MatchKernel_xh]$, which decreases
with $\NormalizedBid$ when $h\ge0$.
Thus the three tests at each knot and the
continuation slope at $\NormalizedBid=1$ give $3\cdot11+1=34$ families, each to
be checked for all $x$.

For requests with small bids, the differential accounting condition is
\begin{equation}
 \nu_0(1+2s-3F_s+\partial_sF_s)
 +\nu_1(2-3G_s+\partial_sG_s)\ge0.
 \label{eqn:aw-cert-small-cost}
\end{equation}
It is affine on each load interval. Checking both endpoints with that
interval's derivative gives twenty tests. The continuation value at
$S$ and slope $\nu_0(2-3h)$ give two more. These $22$ tests suffice
almost everywhere in load, including across knots by integration.

\paragraph{The state constraints.}
\label{app:aw-interval-projection}
For a partition $0=t_0<\cdots<t_N=1$, the interval masses of a measure
in $\AgeCone$ satisfy
\begin{equation}
 t_{i-1}\sum_{j\le i}y_j\le t_i\sum_{j<i}y_j
 \qquad(1\le i\le N).
 \label{eqn:aw-interval-tail}
\end{equation}
Indeed, the tail condition in $z$ coordinates says that
$P(z)/z$ is nonincreasing, where $P(z)=\eta([0,z])$.
Conversely, assign density $c_i=y_i/(t_i-t_{i-1})$ on interval $i$.
Then $P(z)-zP'(z)=\sum_{j<i}y_j-t_{i-1}c_i\ge0$ by the displayed
inequality. Constructing both measures this way proves that imposing
these rows on $y_0$ and $y_0+2y_1$, together with $y_0,y_1\ge0$,
describes their possible interval masses exactly. For nonzero total
mass, normalize and write
\begin{equation}
 \mathcal Y=\{y\ge0:Ay\le0,\ \1^\top y=1\}.
 \label{eqn:aw-interval-state-lp}
\end{equation}
For a scalar continuation test, retain only $\nu_0$ and its tail rows,
and normalize $\nu_0\1=1$. Tests with zero normalization mass have
zero cost and hold automatically.

\paragraph{Finite affine cost bounds.}
\label{app:aw-draft-linear-enclosures}
The fixed score is
\begin{equation}
 H(z)=\frac{103}{100}-\int_z^1\frac{1-e^{-\lambda t}}t\,\d t,
 \qquad \lambda=\frac92.
 \label{eqn:aw-cert-moment}
\end{equation}
The integrand extends continuously to zero, and
Lemma~\ref{lem:aw-alt-score-invariant} proves $\nu_0H\ge0$ for reachable
states.
For each accounting test, put $D_0=C_0-\tau H$, $D_1=C_1$, with
$\tau\ge0$.
Since $\nu_0H\ge0$, lower bounds on $\nu_0D_0+\nu_1D_1$ also bound
the original accounting cost. Write
$C_r^\theta=a_r+\sum_j\theta_j b_{rj}$, where $a_r,b_{rj}$ are fixed
functions. On each interval, choose polynomial enclosures in a common
local Bernstein basis:
\[
 a_r\ge\sum_q a^-_{riq}\mathcal B_q,\qquad
 b_{rj}\ge\sum_q b^-_{rijq}\mathcal B_q,\qquad
 H\le\sum_q H^+_{iq}\mathcal B_q.
\]
The coefficient constraints below imply $\theta_j\ge0$. Hence
\begin{equation}
 D_r(z)\ge\sum_q L_{riq}(\theta,\tau)\mathcal B_q(z),\qquad
 L_{riq}=a^-_{riq}+\sum_j\theta_j b^-_{rijq}
                         -\1\{r=0\}\tau H^+_{iq}.
 \label{eqn:aw-interval-affine-enclosure}
\end{equation}
All coefficients with superscripts $-$ or $+$ are fixed enclosures
computed before solving the LP. A uniform error bound is subtracted
from each Bernstein coefficient for a lower enclosure and added for
an upper enclosure; degree elevation gives a common basis. Since the
basis is nonnegative and sums to one, the finite inequalities
$v_{ri}\le L_{riq}$ ensure $v_{ri}\le D_r(z)$ throughout the interval.
Consequently, accounting follows from $v^\top y\ge0$ for all
$y\in\mathcal Y$, equivalently from $v+A^\top u\ge0$ for some
$u\ge0$. This is LP duality; sufficiency also follows directly from
$v^\top y\ge-u^\top Ay\ge0$.

\paragraph{The linear program.}
Let $\mathcal T$ contain the $22$ tests for small bids and the $34$
test families for large bids at chosen masses.
At positive masses use
$C/x$, and at zero use its limit $\partial_xC|_{x=0}$.
Each test $\ell$ has a partition, tail matrix $A_\ell$, and affine cost bounds $L_{\ell,riq}$ as above. Put
$\Delta s_k=s_{k+1}-s_k$ and $\mathcal I=\{(k,j,e):0\le k<K,\ 0\le j\le d,\ e\in\{k,k+1\}\}$.
The finite LP is
\begin{equation}
\boxed{\begin{alignedat}{2}
 \underset{\theta,(\tau_\ell,v_\ell,u_\ell)_{\ell\in\mathcal T}}
       {\text{maximize}}\quad&\frac1{d+1}\sum_{j=0}^d f_{0j}\\[1ex]
 \text{subject to}\quad&g_{k0}\ge\cdots\ge g_{kd}\ge0
       &&\quad(0\le k\le K),\\
 &f_{k0}-\tfrac12g_{k0}\ge\cdots\ge f_{kd}-\tfrac12g_{kd}\ge0
       &&\quad(0\le k\le K),\\
 &h_0\ge\cdots\ge h_d\ge0,\\
 &f_{k+1,j}-f_{kj}\le2\Delta s_k f_{ej}
       &&\quad((k,j,e)\in\mathcal I),\\
 &g_{k+1,j}-g_{kj}\le2\Delta s_k g_{ej}
       &&\quad((k,j,e)\in\mathcal I),\\
 &h_j\le2f_{Kj}&&\quad(0\le j\le d),\\
 &\tau_\ell\ge0,\quad u_\ell\ge0,\quad
           v_\ell+A_\ell^\top u_\ell\ge0&&\quad(\ell\in\mathcal T),\\
 &v_{\ell,ri}\le L_{\ell,riq}(\theta,\tau_\ell)
       &&\quad(\ell,r,i,q).
\end{alignedat}}
\label{eqn:aw-alt-design-lp}
\end{equation}
The first six rows impose the shape and load conditions; the last two
certify accounting at the selected tests. For the reported certificate,
we fix $\theta$ and solve the associated state LPs using direct
enclosures of the combined costs. These can be sharper than the
fixed-basis bounds in the design LP. Partitions are refined as needed,
and the uniform bounds below establish coverage of all request masses.

\subsection{Reachability and Monotonicity}
\label{app:adwords-structure}

\begin{lemma}[Tail restrictions]
\label{lem:adwords-tail}
For every reachable state, $\nu_0,\nu_1$ are absolutely continuous and,
with $\sigma=\nu_0+2\nu_1$, both functions
$a\mapsto e^{\lambda a}\eta([a,\infty))$, $\eta\in\{\nu_0,\sigma\}$,
are nondecreasing with values in $[0,1]$.
\end{lemma}
\begin{proof}
Fix the requests $(\NormalizedBid_j,x_j)$ with large bids so far, omitting zero masses, and
put $t=\sum_jx_j$.
Realize the stationary age process by a rate-$\lambda$ Poisson process
on the real line. Conditional on this path, request $j$ is won with
probability $D_{x_j}(\FirstAge_j)$ from
\eqref{eqn:matching-win-probability}, independently across requests;
a request of mass one is won surely.
For $r\in[0,t]$, suppress arrivals after $r$, retaining each request's
full mass, and use the same comparison uniforms. Let $S(r)$ be the
resulting set of wins. Increasing $r$ changes a request's first age
from zero to its original positive value when its first arrival is
restored. Since $D_x(a)\ge D_x(0)$ for $a>0$, the sets $S(r)$ increase.

The weights defining the two measures are
\[
 q_0(S)=\1\{S=\varnothing\},\qquad
 q_\sigma(S)=
 \begin{cases}
  1,&S=\varnothing,\\
  2(1-\NormalizedBid_j),&S=\{j\},\\
  0,&|S|\ge2.
 \end{cases}
\]
Both lie in $[0,1]$ and decrease when wins are added, since $\NormalizedBid_j\ge1/2$.
Thus $V_\eta(r):=\E[q_\eta(S(r))]$ is nonincreasing.
Conditioning on no arrival in $(t-a,t]$ and using independence of
Poisson increments gives
\[
 \eta([a,\infty))=e^{-\lambda a}V_\eta((t-a)_+).
\]
For $a>t$, all requests are empty and their comparison probabilities
are independent of the initial age, giving the same formula with
$V_\eta(0)$. This proves the tail claim. The weights are at most one,
so $\nu_0,\sigma\le\pi$ as measures; hence both, and also $\nu_1$,
are absolutely continuous.
Requests with small bids leave these measures fixed.
\end{proof}

\paragraph{Coefficient conditions.}
The shape constraints in~\eqref{eqn:aw-alt-design-lp} make $G_s$ and
$F_s-G_s/2$ nonnegative and nonincreasing in $z$. Indeed, nonnegative
Bernstein coefficients imply nonnegativity, while
$p'(z)=d\sum_{j=0}^{d-1}(p_{j+1}-p_j)\mathcal B_j^{d-1}(z)$ shows
that nonincreasing coefficients imply $p'\le0$.
Interpolation between knots and the continuation
\eqref{eqn:adwords-design-continuation} preserve these properties.
Thus the corresponding age functions are nonnegative and nondecreasing,
and
\begin{equation}
 \Gamma(\nu_0,\nu_1,s)=\tfrac12\sigma\Gamma_{1,s}
      +\nu_0(\Gamma_{0,s}-\tfrac12\Gamma_{1,s})\ge0.
 \label{eqn:aw-alt-monotone-decomposition}
\end{equation}
The load constraints impose coefficientwise
\begin{equation}
 \partial_sF_s\le2F_s,\qquad \partial_sG_s\le2G_s,
 \label{eqn:aw-alt-load-shape}
\end{equation}
so both prices are nonincreasing in $s$. Checks at the two endpoints
of each load interval suffice, using its one-sided derivative.
For $s\ge S$, the constraints reduce to $h\le2F_S$ because
$h,G_S\ge0$. The polynomial parametrization gives prices bounded in
age for each fixed load, joint continuity, continuous age derivatives,
and local Lipschitz continuity in load
uniformly over ages; hence these almost-everywhere load inequalities
imply monotonicity for every load change.

\paragraph{Continuity and monotonicity in request mass.}
\label{app:adwords-request-monotonicity}
Fix a reachable incoming state and normalized bid $\NormalizedBid$.
Let $f$ be bounded,
continuous, nonnegative, and nondecreasing in age. For any finite
nonnegative absolutely continuous measure $\eta$ with nondecreasing
rescaled tails, Appendix~\ref{app:matching-display-request-monotonicity} proves
that $x\mapsto\eta\MatchKernel_xf$ is continuous and nonincreasing;
its proof uses precisely these properties of the incoming measure.
The same holds for $\eta\GapKernel_xf$, since, writing
$h_\eta(a)=e^{\lambda a}\eta([a,\infty))$, its tails are
\[
 (\eta\GapKernel_x)([a,\infty))
 =e^{-\lambda a}h_\eta((a-x)_+).
\]
For the win kernel, $x\mapsto\eta\WinKernel_xf$ is continuous and
nondecreasing for every finite nonnegative absolutely continuous
$\eta$. For $x\le a$, its tail is
\[
 (\eta\WinKernel_x)([a,\infty))
 =xe^{-\lambda}\eta([a-x,\infty)),
\]
which is nondecreasing in $x$. For $x>a$, the tail is
$e^{-\lambda a}\int W_x(v)\,\eta(dv)$, with $W_x$ from
\eqref{eqn:poisson-arrival-win}; the derivative argument in
Appendix~\ref{app:matching-display-request-monotonicity} proves that $W_x(v)$
increases with $x$. 
The two expressions agree at $x=a$ and are
continuous, including at $x=0,1$. Tail integration and bounded
convergence give the stated properties for $f$.

For $\NormalizedBid\ge1/2$, let $\eta_{\NormalizedBid}=\nu_1+(\NormalizedBid-\tfrac12)\nu_0\ge0$.
Using $\WinKernel_x=\GapKernel_x-\MatchKernel_x$, the potential after
one request equals
\[
 \nu_0\MatchKernel_x(\Gamma_{0,s}-\tfrac12\Gamma_{1,s})
 +\tfrac12\sigma\GapKernel_x\Gamma_{1,s}
 -\eta_{\NormalizedBid}\WinKernel_x\Gamma_{1,s}.
\]
Lemma~\ref{lem:adwords-tail} and the coefficient conditions make the
first two terms continuous and nonincreasing in $x$, and the
subtracted term continuous and nondecreasing. For a small bid, the
post-request potential is $\Gamma(\nu_0,\nu_1,s+\NormalizedBid x)$, which has the
required continuity and monotonicity by~\eqref{eqn:aw-alt-load-shape}.
At $x=0$ the state is unchanged, so in both cases $\Gamma'\le\Gamma$.

\subsection{The Signed-Score Invariant}
\label{app:aw-alt-score-verification}

\begin{lemma}[Signed-score invariant]
\label{lem:aw-alt-score-invariant}
For $H$ in~\eqref{eqn:aw-cert-moment} and all $x,z\in[0,1]$,
\begin{equation}
 \MatchKernel_xH(z)\ge\alpha(x)H(z),\qquad
 \alpha(x)=\bigl(e^{-\lambda x}-xe^{-\lambda}\bigr)
                  \left(1+\frac{x}{100}+\frac{13x^2}{10}\right).
 \label{eqn:aw-cert-moment-preservation}
\end{equation}
Consequently, every reachable state satisfies $\nu_0H\ge0$.
\end{lemma}
\begin{proof}
We first describe the rigorous numerical certificate for
\begin{equation}
 \Delta_H(x,z):=
 \frac{\MatchKernel_xH(z)-\alpha(x)H(z)}{x(1-x)}
 >5\cdot10^{-5}\qquad(0<x<1,\ 0\le z\le1).
 \label{eqn:aw-alt-score-certified-margin}
\end{equation}
For $0<x<1$, put $r=e^{-\lambda x}$, $\mu=e^{-\lambda}$, and define
the comparison functions
\begin{equation}
 \Psi_x(y)=
 \begin{cases}
  1-\lambda xy,&0\le y\le r,\\
  1+r-y+y\log y,&r\le y\le1,
 \end{cases}
 \qquad \mathcal D_x(y)=\Psi_x(y)^{(1-x)/x}.
 \label{eqn:aw-alt-comparison}
\end{equation}
Set
\[
 C_x=\frac{r-\mu x}{1-x},\qquad
 R_x(y)=\frac{1-\mathcal D_x(y)}{1-x}.
\]
These functions have continuous endpoint extensions. Convexity of
$e^{-\lambda x}$ and $r\le\Psi_x(y)\le1$ imply
$0\le C_x\le1$ and
$0\le R_x(y)\le-\log\Psi_x(y)/x\le\lambda$.
Since $zH'(z)=1-e^{-\lambda z}$, ordinary age evolution from age zero gives
\[
 (\GapKernel_tH)(1)=Q_H(e^{-\lambda t}),\qquad
 Q_H(w)=wH(w)+\int_w^1H(v)\,\d v
       =H(1)-1+w+\frac{e^{-\lambda w}-\mu}{\lambda}.
\]
Here the argument $1$ denotes age zero in $z$ coordinates. Substitution in the formula of Lemma~\ref{lem:matching-kernel-formula}, together with
$H(rz)-H(z)=-\lambda x\int_0^1(1-e^{-\lambda ze^{-\lambda xt}})\,\d t$,
yields the regular expression
\begin{equation}
\begin{split}
 \Delta_H(x,z)
 ={}&\int_0^1\lambda\Bigl[
 e^{-\lambda xt}R_x(ze^{-\lambda xt})Q_H(e^{-\lambda x(1-t)})
 -C_x\bigl(1-e^{-\lambda ze^{-\lambda xt}}\bigr)\Bigr]\,\d t\\
 &-C_x\left(\frac1{100}+\frac{13x}{10}\right)H(z).
\end{split}
\label{eqn:aw-alt-score-regular-gap}
\end{equation}
It extends to both mass endpoints and is verified on two coordinate
squares: $z=e^{-\lambda x}u$ and $z=e^{-\lambda xu}$, with $x,u\in[0,1]$.
The first covers $z\le r$, where the integrand uses one comparison
branch; the second covers $z\ge r$, where splitting at $t=1-u$
separates the branches.

On a rectangle with half-widths $h_x,h_u$, let $B_x,B_u$ bound the
absolute pure second derivatives of the chart expression. Successive
one-dimensional interpolation bounds give
\begin{equation}
 \Delta_H\ge\min_{\text{four corners}}\Delta_H
                 -\tfrac12h_x^2B_x-\tfrac12h_u^2B_u.
 \label{eqn:aw-alt-score-rectangle-bound}
\end{equation}
The verifier uses certified Arb integration at corners and interval
integration bounds for derivatives. It bisects rectangles until this
lower bound is positive and checks their exact dyadic coverage.

The implementation evaluates the same integral for
$\widetilde H(z)=c_0+\operatorname{Ein}(\lambda z)$, where
\[
 c_0=-1.05336646,\qquad
 \operatorname{Ein}(v)=\int_0^v\frac{1-e^{-t}}t\,\d t.
\]
An alternating Taylor remainder encloses the constant
$\delta=H-\widetilde H$. Replacing $\widetilde H$ by $H$ changes
\eqref{eqn:aw-alt-score-regular-gap} by
\[
 \delta\left[\int_0^1\lambda e^{-\lambda xt}
                   R_x(ze^{-\lambda xt})\,\d t
        -C_x\left(\frac1{100}+\frac{13x}{10}\right)\right],
\]
whose absolute value is at most $(\lambda^2+131/100)|\delta|$.
After deducting this allowance, less than $5.245\cdot10^{-8}$, we verify by program that every rectangle bound exceeds $5.03308\cdot10^{-5}$.
The program also checks that the $629$ first-chart and
$1181$ second-chart rectangles cover both coordinate squares,
including their boundaries.
This proves \eqref{eqn:aw-alt-score-certified-margin}
for the exact function $H$.
At $x=0,1$,~\eqref{eqn:aw-cert-moment-preservation} is an equality:
$\MatchKernel_0=\mathrm{Id}$, $\alpha(0)=1$, and
$\MatchKernel_1=0=\alpha(1)$.

Finally, uniformity of $\pi$ in $z$ and Fubini give
$\pi H=3/100+(1-e^{-\lambda})/\lambda>0$.
Also $\alpha(x)\ge0$, since $e^{-\lambda x}\ge xe^{-\lambda}$.
Thus a request with a large bid preserves the score inequality through
$\nu_0'H=\nu_0\MatchKernel_xH\ge\alpha(x)\nu_0H$;
requests with small bids leave $\nu_0$ unchanged.
Induction proves the claim.
\end{proof}

\subsection{Rigorous Accounting Certificates}
\label{app:aw-alt-certification}
\label{app:adwords-verification}

All price coefficients and LP witnesses are rational. The verifier
recomputes outward enclosures of the accounting costs, checks the LP
residuals exactly, and covers the intervals between request masses.

\paragraph{Cost enclosures.}
For each test and rational $\tau\ge0$, bound $C_0-\tau H$ and $C_1$
together with their approximation errors on each state interval.
The score has the polynomial upper enclosure
\[
 S_n(t)=\sum_{k=1}^n\frac{(-1)^{k+1}t^k}{k\,k!},\qquad
 \widehat H(z)=\frac{103}{100}-S_{80}(\lambda)
             +S_{26}(\lambda z)+\frac{\lambda^{27}}{27\cdot27!}\ge H(z).
\]
The inequality follows from the alternating remainder. Hence
$C_0-\tau\widehat H\le C_0-\tau H$.
For tests for small bids and zero-mass generator tests, the resulting functions are
$p(z)+ce^{-\lambda z}$; interval arithmetic, derivative signs, and
curvature bounds enclose their minima. Positive masses use the
following kernel enclosure.

For $0<x<1$, use $r=e^{-\lambda x}$, $\mu=e^{-\lambda}$, and the
comparison functions in~\eqref{eqn:aw-alt-comparison}.
Here $\mathcal D_x(e^{-\lambda a})$ is the win probability conditional
on positive first age $a$. For $p(z)=\sum_jp_jz^j$, let
$Q_p(w)=\overline p+\sum_{j\ge1}jp_jw^{j+1}/(j+1)$, where
$\overline p=\int_0^1p(z)\,\d z$. Ordinary age evolution from age zero
gives $Q_p(e^{-\lambda t})=(\GapKernel_t p)(1)$ in $z$ coordinates.
Thus the formula of Lemma~\ref{lem:matching-kernel-formula} becomes
\begin{equation}
 (\MatchKernel_xp)(z)=(r-\mu x)p(rz)
       +\int_r^1[1-\mathcal D_x(zv)]Q_p(r/v)\,\d v,
 \qquad \MatchKernel_0=\IdKernel,\quad\MatchKernel_1=0.
 \label{eqn:aw-alt-compact-kernel}
\end{equation}
On $0\le y\le r$, expand $\mathcal D_x(y)=\sum_{k\ge0}d_k(y/r)^k$:
\[
 d_0=1,\qquad d_k=-\frac{\lambda r(1-kx)}k\,d_{k-1}.
\]
For $k\ge42$ the absolute recurrence multiplier is at most
$\max\{\lambda/k,\lambda xe^{-\lambda x}\}\le e^{-1}$, so truncation
at degree $40$ has error at most $|d_{41}|/(1-e^{-1})$.
On each interval $[l,u]\subseteq[r,1]$, use a degree-$20$ midpoint
Taylor polynomial with error
$\sup_{[l,u]}|\mathcal D_x^{(21)}|((u-l)/2)^{21}/21!$.
All coefficients and remainders are enclosed outward, with subdivision
as needed. Substitution into~\eqref{eqn:aw-alt-compact-kernel} gives
integrable powers and logarithms; comparison error $\varepsilon_D$
contributes at most $(1-r)\varepsilon_D\|p\|_\infty$.
On each output interval, a Taylor remainder and Bernstein conversion
bound the error $\varepsilon$ of a proposed degree-$12$ polynomial.
Subtracting $\varepsilon$ from its Bernstein coefficients gives a
pointwise lower enclosure. All comparison-branch intersections are
included, and the expression on the interval touching $z=0$ is polynomial.

\paragraph{Exact LP witnesses.}
For rational interval costs $v$, multipliers $u\ge0$, and scalar $t$,
the exact residual gives
\begin{equation}
 e=v+A^\top u-t\1,\qquad
 v^\top y\ge t+\min_j e_j\quad(y\in\mathcal Y).
 \label{eqn:aw-cert-dual-residual-correction}
\end{equation}
Indeed, $v^\top y=t-u^\top Ay+e^\top y$, and $Ay\le0$, $y\ge0$,
$\1^\top y=1$. In the implementation's cumulative coordinates
$q_{ri}=\sum_{j\le i}y_{rj}$, normalization is $q_{0N}+q_{1N}=1$
(or $q_{0N}=1$ for scalar tests). A residual with coefficients $e_{ri}$
on $q_{ri}$ has coefficients $\sum_{i\ge j}e_{ri}$ on $y_{rj}$;
the correction is therefore the minimum of these suffix sums.
Using the cost enclosures above, we verify by program that all $5700$ rational LP witnesses give nonnegative corrected lower bounds.
These witnesses certify the 22 small-bid tests and all 34 large-bid families at $x=0$ and at 166 positive mass endpoints.
The remaining endpoint $x=2^{-21}$ is certified from the generator bounds using the curvature estimate below, without additional LP witnesses.

\paragraph{Coverage of every request mass.}
Fix a test family for large bids and an admissible state, and write $A(x)$
for its original integrated cost after removing $e^{-2s}$.
Let $m=(\nu_0+\nu_1)\1$, or $m=\nu_0\1$ for a scalar test.
An endpoint certificate gives $A(x)/(mx)\ge d_x$ for $x>0$.
At zero, $A(0)=0$, and the generator in
Lemma~\ref{lem:matching-infinitesimal-generator} gives $A'(0)/m\ge d_0$.
Endpoint partitions and score multipliers may differ: they bound the
same original cost, to which we apply the curvature estimates below.
If $|A''|\le mB$ on $[l,u]$ and $t=(x-l)/(u-l)$, then
\begin{equation}
 \frac{A(x)}m\ge q(x):=(1-t)l d_l+t u d_u
                   -\frac B2(u-l)^2t(1-t).
 \label{eqn:aw-cert-mass-interpolation}
\end{equation}
The difference $A/m-q$ is concave and nonnegative at both endpoints.
For $l>0$, the exact quadratic minimum gives the uniform normalized
bound $\min q/u$ whenever $\min q\ge0$. On the first interval,
Taylor's integral formula instead gives
\begin{equation}
 \frac{A(x)}{mx}\ge d_0-\frac{Bu}{2}\qquad(0<x\le u).
 \label{eqn:aw-alt-zero-mass}
\end{equation}
Using the curvature bounds below, we verify by program that \eqref{eqn:aw-cert-mass-interpolation} and \eqref{eqn:aw-alt-zero-mass} give nonnegative lower bounds for all $34$ families on each of the $167$ accepted mass intervals.
The program also checks that these intervals meet exactly and cover $[0,1]$.

\subsection{Mass-Curvature Bounds}
\label{app:aw-alt-curvature-details}

The constants $B$ above are obtained from the following uniform bounds.
For fixed physical age $v$, write
\[
 Q_x(v)=1-\int_0^x\lambda e^{-\lambda\max(v,t)}\,\d t,
 \quad w(t)=\frac{\lambda e^{-\lambda\max(v,t)}}{Q_t(v)},
 \quad L(x)=-\frac{\log Q_x(v)}x=\int_0^1w(\theta x)\,\d\theta.
\]
Thus $Q_x(v)=\Psi_x(e^{-\lambda v})$, and the comparison factor in
the loss-kernel integral is $R(x,v)=1-e^{-(1-x)L(x)}$.
We have $0\le w\le\lambda$, with $w'=w^2$ before $t=v$ and
$w'=w(w-\lambda)$ after it. Its derivative jumps by $-\lambda w(v)$.
Consequently,
\[
 -\lambda^2/8\le L'\le\lambda^2/2,\qquad
 |L''_{\rm reg}|\le2\lambda^3/3,
\]
while the moving boundary contributes
$-\lambda w(v)v^2x^{-3}\1\{v<x\}$, of magnitude at most
$\lambda^2x^{-1}\1\{v<x\}$.
On a mass interval $[l,u]$, set
\begin{align*}
 R_*&=1-e^{-\lambda(1-l)},&
 a_1&=\max\{(1-l)\lambda^2/2,\lambda+(1-l)\lambda^2/8\},\\
 a_2&=a_1^2+\lambda^2+2(1-l)\lambda^3/3,&
 c_1&=1+(1-l)\lambda,\\
 c_2&=2\lambda+2(1-l)\lambda^2,&
 d_k&=\max_{0\le t\le\lambda}t^ke^{-(1-u)t},\\
 A_1&=\min\{a_1,c_1d_1\},&
 A_2&=\min\{a_2,c_1^2d_2+c_2d_1\},\\
 E&=e^{-(1-u)\lambda e^{-\lambda u}},&A_v&=\lambda/(e-1).
\end{align*}
The maximum defining $d_k$ occurs at $\min\{\lambda,k/(1-u)\}$,
interpreted as $\lambda$ when $u=1$. The bounds
$|w'|\le\lambda w$ and $|w''_{\rm reg}|\le2\lambda^2w$ also give
$|L'|\le\lambda L$ and $|L''_{\rm reg}|\le2\lambda^2L$.
Differentiating $R$ therefore yields
\[
 0\le R\le R_*,\qquad |R_x|\le A_1,\qquad
 |R_{xx}|\le A_2+E(1-l)\lambda^2x^{-1}\1\{v<x\}.
\]
Here $v<x$ implies $L(x)\ge\lambda e^{-\lambda x}$, justifying $E$.
At the integration boundary $v=a+x\ge x$, $|R_v|\le A_v$:
putting $q=\lambda xe^{-\lambda v}\le1/e$ and $\beta=(1-x)/x$,
use $\max_{\beta\ge0}\beta(1-q)^\beta
=[e(-\log(1-q))]^{-1}$ and $-\log(1-q)\ge q$.

For a function $f$ of physical age, put
$N_j=\|f^{(j)}\|_\infty$ for $j=0,1,2$,
$f_0=|f(0)|$, $d_0=|f'(0)|$,
$g_j=\sup_{t\ge0}|\partial_t^j(\GapKernel_t f)(0)|$,
$r_l=e^{-\lambda l}$, $r_u=e^{-\lambda u}$, and $V_l=r_l-\mu l$.
Then $\|\partial_x^j\MatchKernel_xf\|_\infty\le B_j(f)$, where
\begin{align*}
 B_1(f)&=(\lambda r_l+\mu)N_0+V_lN_1+\lambda r_lR_*f_0
                    +(1-r_u)(A_1N_0+R_*g_1),\\
 B_2(f)&=\lambda^2r_lN_0+2(\lambda r_l+\mu)N_1+V_lN_2\\
 &\quad+\lambda r_l[(2A_1+A_v+\lambda R_*)f_0+R_*d_0]\\
 &\quad+(1-r_u)(A_2N_0+2A_1g_1+R_*g_2)+E(1-l)\lambda^3N_0.
\end{align*}
These follow by applying the Leibniz rule to
\[
 (\MatchKernel_xf)(a)=(e^{-\lambda x}-\mu x)f(a+x)
 +\int_0^x\lambda e^{-\lambda t}R(x,a+t)(\GapKernel_{x-t}f)(0)\,\d t.
\]
In particular, the last term of $B_2$ bounds the moving-boundary term
using $\int_0^x\lambda e^{-\lambda t}x^{-1}\1\{a+t<x\}\,\d t\le\lambda$.
The estimates hold weakly at branch crossings and extend to both mass
endpoints. All polynomial norms are bounded by local Bernstein
coefficients on $16$ equal rational intervals. The required derivatives
are obtained from $p\mapsto-\lambda zp'$,
$\InfGenerator_{\GapKernel}p=-\lambda zp'+\lambda(p(1)-p)$, and
$Q_p(e^{-\lambda t})$ above.

For the costs at a load knot and endpoint bid, put
$J=F_s-(1-\NormalizedBid)G_s$.
Differentiation gives
\begin{align*}
 C_{1,\NormalizedBid}''&=\MatchKernel_x''[(1-\NormalizedBid x)G_s-\1]-2\NormalizedBid\MatchKernel_x'G_s,\\
 C_{0,\NormalizedBid}''&=\MatchKernel_x''[(1-\NormalizedBid x)J-(s+\NormalizedBid)\1]-2\NormalizedBid\MatchKernel_x'J\\
 &\quad+(1-\NormalizedBid)\GapKernel_x[(1-\NormalizedBid x)\InfGenerator_{\GapKernel}^2G_s
                                -2\NormalizedBid\InfGenerator_{\GapKernel}G_s].
\end{align*}
Kernel derivatives here hold the test function fixed. The first terms
are bounded by the larger $B_2$ bound on the test functions at $x=l,u$;
the others use $B_1$ and the Markov contraction of $\GapKernel_x$.
Expanding the terms separately gives a second valid bound, and the
verifier takes the smaller. The middle bid coefficient adds at most
\[
 \frac18\left[2(\|\InfGenerator_{\GapKernel}G_s\|_\infty+B_1(G_s))
        +u(\|\InfGenerator_{\GapKernel}^2G_s\|_\infty+B_2(G_s))\right]
\]
to the average of the two $\nu_0$ endpoint bounds. Taking the maximum
of the two component bounds gives $B$ for a paired cost; the scalar
continuation uses the $C_{1,\NormalizedBid}''$ formula with $\NormalizedBid=1$ and $G_s=h$.

\subsection{Verification Results and Proof of the Potential Lemma}
\label{app:aw-alt-results}

The completed certificate gives the following strictly positive bounds.
Here $m=(\nu_0+\nu_1)\1$, or $m=\nu_0\1$ for scalar tests; the
accounting costs have their common factor $e^{-2s}$ removed.
\begin{center}
\begin{tabular}{lll}
\toprule
Condition & Certified domain & Lower bound\\
\midrule
Score defect $\Delta_H$ & $1810$ rectangles
& $>5.03308\cdot10^{-5}$\\
Cost for small bids divided by $m$ & $22$ tests
 & $>1.6489\cdot10^{-5}$\\
Cost for large bids divided by $xm$ & $167$ intervals, $34$ families
& $>4.6301\cdot10^{-7}$\\
\bottomrule
\end{tabular}
\end{center}
The mass intervals cover $[0,1]$ with $168$ endpoints; the first is
$[0,2^{-21}]$. All $34$ generator tests pass, and every state LP uses
at most $2763$ age intervals. Initialization and all coefficient
conditions are checked exactly. An independent audit checks $5700$
rational LP witnesses, recomputes the mass-curvature bounds, and
verifies both the mass cover and the signed-score rectangle covers.

\begin{proof}[Proof of Lemma~\ref{lem:adwords-potential}]
Use the supplied rational coefficients in
\eqref{eqn:adwords-design-prices}--\eqref{eqn:adwords-design-continuation}.
Equation~\eqref{eqn:aw-cert-initialization} gives the initial value.
Appendix~\ref{app:adwords-structure} proves the stated regularity,
nonnegativity, and continuity and monotonicity in request mass.
The uniform local Lipschitz bound in load also gives local absolute
continuity after integration against fixed finite age measures.

Lemmas~\ref{lem:adwords-tail} and~\ref{lem:aw-alt-score-invariant}
place every reachable state in~\eqref{eqn:adwords-design-cone}.
The certified interval bounds and dual witnesses therefore apply to
these states.
The $34$ test families for large bids cover every large bid and
load, and the mass cover supplies accounting for every $x\in[0,1]$.
The $22$ tests for small bids give
\begin{equation}
 -\partial_sU(\nu_0,\nu_1,s)
 \ge\Gamma(\nu_0,\nu_1,s)-\partial_s\Gamma(\nu_0,\nu_1,s)
 \quad\text{for almost every }s.
 \label{eqn:adwords-small-differential}
\end{equation}
With the age measures fixed, abbreviate $U(s)=U(\nu_0,\nu_1,s)$ and
$\Gamma(s)=\Gamma(\nu_0,\nu_1,s)$.
Integrating over a request with a small bid and load $\NormalizedBid x$, and using load monotonicity, gives
\begin{equation}
\begin{aligned}
 U(s)-U(s+\NormalizedBid x)
 &\ge\Gamma(s)-\Gamma(s+\NormalizedBid x)+\int_s^{s+\NormalizedBid x}\Gamma(t)\,\d t\\
 &\ge\Gamma(s)-\Gamma(s+\NormalizedBid x)+\NormalizedBid x\Gamma(s+\NormalizedBid x).
\end{aligned}
\label{eqn:adwords-small-accounting}
\end{equation}
This proves the remaining accounting inequality.
\end{proof}

\section{Covariance and Verification for Squared Loads}
\label{app:load-balancing}

\subsection{Proof of Lemma~\ref{lem:load-rounding}}\label{sec:load-balancing-kernel}

The claim is immediate if either request has zero mass, so assume both are positive.

\paragraph{Transforming the Inequality.}
Differentiating \eqref{eqn:matching-win-probability} and \eqref{eqn:first-age-cdf} gives the later-request factor
\begin{equation}\label{eqn:load-H-integral}
  H_x(s):=\lambda\int_{(s,\infty)}e^{-\lambda a}\,dD_x(a)
  =(1-x)\lambda^3\int_s^\infty e^{-2\lambda a}
  \min \Bigl\{ \frac{a}{x}, 1 \Bigr\} F_x(a)^{1/x-2}\,da.
\end{equation}

For the earlier-request factor, $0\le r\le x$ gives
$$
  \begin{aligned}
    \lambda\int_{[0,\infty)}e^{-\lambda\max\{a,r\}}\,dD_x(a)
     & \ge\lambda e^{-\lambda x}(D_x(x)-c_x)+H_x(x)                 \\
     & =1-(1-\lambda x e^{-\lambda x})^{1/x}-\lambda x e^{-\lambda}
    =:a_\lambda(x).
  \end{aligned}
$$

This expression is nonincreasing in $x$.
Indeed, writing $q=\lambda xe^{-\lambda x}\le\lambda x$, we have
$$
  x^2\frac{d}{dx}\log\bigl((1-q)^{1/x}\bigr)
  =-\log(1-q)-\frac{q(1-\lambda x)}{1-q}
  \ge-\log(1-q)-q\ge0.
$$

Consequently, Lemma~\ref{lem:load-covariance} reduces the claim to the two-variable inequality
\begin{equation}\label{eqn:load-minorant}
  a_\lambda(\overline{x})H_x(s)\ge\delta(s)
  \qquad(0<x\le\overline{x},\ s\ge0).
\end{equation}

\paragraph{Numerical Verification for $0\le s\le3/2$.}
Use the uniform grid
$$
  \tau=10^{-4},\qquad
  x_m=m\tau\quad(0\le m\le1500),\qquad
  u_n=n\tau\quad(0\le n\le30000).
$$

Set $F_0(a)=1$, its continuous extension at zero mass.
On the rectangle $(x,a) \in [x_m,x_{m+1}] \times [u_n,u_{n+1}]$, we claim that the integrand in \eqref{eqn:load-H-integral}, including its prefactor, is at least
$$
  g_{mn}:=(1-x_{m+1})\lambda^3e^{-2\lambda u_{n+1}}
  \min \Bigl\{ \frac{u_n}{x_{m+1}},1\Bigr\}
  \frac{F_{x_{m+1}}(u_n)^{1/x_{m+1}}}{F_{x_m}(u_{n+1})^2}.
$$

Then, a lower Riemann sum for $H_x(s)$ and the monotonicity of $\delta$ reduce the verification to
\begin{equation}\label{eqn:load-rectangle-check}
  a_\lambda(\overline{x})\,\tau\sum_{k=n+1}^{29999}g_{mk}
  >\delta(u_n)
  \qquad(0\le m<1500,\ 0\le n<15000).
\end{equation}
All checks pass using outward-rounded interval arithmetic.

To show that $g_{mn}$ lower-bounds the integrand throughout this rectangle, note that both $F_x(a)$ and $F_x(a)^{1/x}$ are nondecreasing in $a$ and nonincreasing in $x$ for $0<x\le\overline{x}$.
The assertions for $F_x(a)$ follow from \eqref{eqn:first-age-cdf}.
For the power, the derivative of $\log(F_x(a)^{1/x})$ has the sign of
$$
  x\partial_x F_x(a)- F_x(a)\log F_x(a)
  =\begin{cases}
    -1+F_x(a)- F_x(a)\log F_x(a),                  & x\le a, \\[4pt]
    -\lambda x e^{-\lambda x} - F_x(a)\log F_x(a), & a<x.
  \end{cases}
$$
The first case is nonpositive by $\log t\ge1-1/t$.
The second is nonpositive because $t\log t$ is nondecreasing on $[e^{-1},1]$ and $e^{-\lambda x}\le F_x(a)\le1$, with $\lambda x<1$.
Writing $F_x(a)^{1/x-2}=F_x(a)^{1/x}/F_x(a)^2$ and bounding each nonnegative factor separately gives $g_{mn}$.

\paragraph{The Infinite Tail.}
For $s\ge3/2>x$, substituting $a=s+u$ in \eqref{eqn:load-H-integral} gives
$$
  e^{2\lambda s}H_x(s)
  =(1-x)\lambda^3\int_0^\infty e^{-2\lambda u}
  F_x(s+u)^{1/x-2}\,du.
$$
This expression is nondecreasing in $s$, since $F_x$ is nondecreasing and $1/x-2>0$.
Meanwhile,
$$
  e^{2\lambda s}\delta(s)=\frac{43}{50}e^{8s-(25/3)s^2}
$$
is decreasing for $s\ge3/2$.
Thus \eqref{eqn:load-minorant}, already verified at $s=3/2$, holds throughout the remaining tail, completing the proof of Lemma~\ref{lem:load-rounding}.

\subsection{Proof of Lemma~\ref{lem:load-amortized}}\label{sec:load-balancing-potential-proof}

Use the capped job sizes and auxiliary variables from the proof sketch, and write
$$
  R_i=\alpha_i-\widetilde{\alpha}_i=0.15\,\widetilde{\Delta}_{ij}(\infty),\qquad
  \Psi=\sum_i(\alpha_i^2-\widetilde{\alpha}_i^2).
$$
Here $R_i$ is the reserve immediately before job $j$.
Empty histories give $\Psi=0$ initially, proving property 1.
Fix a machine $i$.
Recall that $\alpha_i=\ell_i/2$ is its current dual variable and $t_i$ is its current internal time.

\subsubsection{Bounding the Potential: Proof of Property 2}

First, if $\alpha_i = 0$, then all earlier capped sizes and the reserve are zero.
The bound holds trivially.
The remaining case is when machine $i$ has positive expected load, so $\alpha_i=\ell_i/2>0$.

For $s\ge0$, normalize the history of capped sizes by setting $f_i(s)=\tilde p_{ik}/\alpha_i$ when $t_i-s\in I_{ik}$ for an earlier correlated job $k$, and $f_i(s)=0$ otherwise.
Then $F_i(s)=\int_0^s f_i(u)\,du$ is the capped load in the most recent $s$ units of internal time, normalized by $\alpha_i$.
If $t_i-s\in I_{ik}$, then $\tilde p_{ik}\le\alpha_i(k)$.
Moreover, $\alpha_i-\alpha_i(k)$ equals half the fractional load added since immediately before job $k$.
This added load is at least the capped load on $[t_i-s,t_i]$, so we have
$$
  \alpha_i-\alpha_i(k) \ge \frac{\alpha_i F_i(s)}{2}.
$$

Dividing both sides by $\alpha_i$ and rearranging gives
$$
  1-\frac{F_i(s)}{2} \ge \frac{\alpha_i(k)}{\alpha_i} \ge \frac{\tilde{p}_{ik}}{\alpha_i} = f_i(s).
$$
For $s>t_i$, we have $f_i(s)=0$, and the same load accounting gives $F_i(s)\le2$.
Consequently,
$$
  0\le F_i'(s)=f_i(s)\le1-\frac{F_i(s)}2
  \quad\text{for almost every }s\ge0,\qquad F_i(0)=0,
$$
and hence $F_i(s)\le2(1-e^{-s/2})$.
Integration by parts now bounds the reserve:
\begin{align}
  \frac{R_i}{\alpha_i}
   & =0.3\int_0^\infty f_i(s)\int_s^\infty\delta(u)\,du\,ds
  \tag{definitions of $R_i$ and $f_i$}                      \\
   & =0.3\int_0^\infty F_i(s)\delta(s)\,ds
  \tag{integration by parts}                                \\
   & \le0.6\int_0^\infty(1-e^{-s/2})\delta(s)\,ds.
  \tag{$F_i(s)\le2(1-e^{-s/2})$}
\end{align}

Since $\ell_i=2\alpha_i$, numerical integration gives
\begin{equation}\label{eqn:load-reserve-bound}
  \frac{R_i}{\ell_i}\le 0.3 \int_0^\infty(1-e^{-s/2})\delta(s)\,ds < \nu.
\end{equation}

Thus $0\le R_i\le\nu\ell_i\le\alpha_i$, so $0\le\widetilde{\alpha}_i\le\alpha_i$.
Therefore, property 2 follows by
$$
  \Psi = \sum_i(\alpha_i+\widetilde{\alpha}_i)(\alpha_i - \widetilde{\alpha}_i) = \sum_i(\alpha_i+\widetilde{\alpha}_i)R_i
  \le\sum_i\ell_iR_i\le\nu\sum_i\ell_i^2.
$$

\subsubsection{Amortized Analysis: Proof of Property 3}\label{sec:load-balancing-local-proof}

Fix a machine $i$ and an arriving job $j$.
We prove \eqref{eqn:load-amortized} by comparing each machine's contribution to the increases in $\overline{\operatorname{ALG}}$ and $D+\Psi$.
Since $D+\Psi=\sum_k\beta_k-\sum_i\widetilde{\alpha}_i^2$ and $\beta_j=\sum_i x_{ij}\Phi(\alpha_i^+;p_{ij})$, the latter increase is
$$
  D^+-D+\Psi^+-\Psi
  =\sum_i\left[
    x_{ij}\Phi(\alpha_i^+;p_{ij})
    -\bigl((\widetilde{\alpha}_i^+)^2-\widetilde{\alpha}_i^2\bigr)
    \right].
$$

\subsubsection*{Independent Rounding}
If job $j$ is rounded independently on machine $i$, then it adds $p_{ij}^2x_{ij}+2p_{ij}\ell_i x_{ij}$ to $\overline{\operatorname{ALG}}$ and leaves $R_i$ fixed.
Thus, both $\widetilde{\alpha}_i$ and $\alpha_i$ increase by $p_{ij}x_{ij}/2$.
Since $\widetilde{\alpha}_i\le\alpha_i$, the increase in $\widetilde{\alpha}_i^2$ is at most that in $\alpha_i^2$.
The rounding rule \eqref{eqn:load-independent} therefore gives
$$
  \begin{aligned}
    p_{ij}^2x_{ij}+2p_{ij}\ell_i x_{ij}
     & \le \Gamma_0 \Bigl( x_{ij}\Phi(\alpha_i^+;p_{ij})-\bigl((\alpha_i^+)^2-\alpha_i^2\bigr) \Bigr)    \\
     & \le \Gamma_0 \Bigl( x_{ij}\Phi(\alpha_i^+;p_{ij})-\bigl((\widetilde{\alpha}_i^+)^2-\widetilde{\alpha}_i^2\bigr) \Bigr).
  \end{aligned}
$$

\subsubsection*{OCEC Rounding}
If job $j$ is rounded using the OCEC rule on machine $i$, we have $0<x_{ij}<\overline{x}$ by Lemma~\ref{lem:load-mass}.
Its contribution to the increase in $\overline{\operatorname{ALG}}$ includes the covariance saving $p_{ij}\Delta_{ij}(x_{ij})$.
We need to prove
$$
  p_{ij}^2x_{ij}+2p_{ij}\ell_i x_{ij}-p_{ij}\Delta_{ij}(x_{ij})
  \le \Gamma_0 \Bigl( x_{ij}\Phi(\alpha_i^+;p_{ij})
  -\bigl((\widetilde{\alpha}_i^+)^2-\widetilde{\alpha}_i^2\bigr) \Bigr).
$$

We will next transform the inequality into a polynomial inequality in five variables, establish some linear constraints on the variables, and finally verify it numerically.

\subsubsection*{Transforming the Inequality}

First, note that $\alpha_i>0$: otherwise $\alpha_i=\ell_i=0$, so \eqref{eqn:load-independent} would hold and the entry would be rounded independently.
Write $a=\alpha_i/p_{ij}$.
Dividing by $p_{ij}^2x_{ij}$ gives
\begin{equation}\label{eqn:load-local-energy}
  1+4a-\frac{\Delta_{ij}(x_{ij})}{p_{ij}x_{ij}}
  \le\Gamma_0\left(
  \Phi\!\left(a+\frac{x_{ij}}2;1\right)
  -\frac{(\widetilde{\alpha}_i^+)^2-\widetilde{\alpha}_i^2}{p_{ij}^2x_{ij}}
  \right).
\end{equation}

Normalize the capped credit and the remaining reserve by writing
$$
  c_{ij}=\frac{\widetilde{\Delta}_{ij}(x_{ij})}{2\alpha_i x_{ij}}
  =\frac1{x_{ij}}\int_0^\infty f_i(s)\int_0^{x_{ij}}\delta(s+u)\,du\,ds,
  \qquad r_i=\frac{R_i}{\alpha_i}.
$$

Since $\Delta_{ij}(x_{ij})\ge\widetilde{\Delta}_{ij}(x_{ij})$, the normalized covariance saving satisfies
\begin{equation}\label{eqn:load-normalized-saving}
  \frac{\Delta_{ij}(x_{ij})}{p_{ij}x_{ij}} \ge 2a c_{ij}.
\end{equation}

Further, let
$$
  \kappa_{ij} = \frac1{x_{ij}} \int_0^\infty \int_0^{x_{ij}} \delta(s+u) \,du\,ds.
$$

Advancing the correlated timeline releases $0.15\,\widetilde{\Delta}_{ij}(x_{ij})$ from the reserve, and the new interval adds $0.3\,\tilde p_{ij}x_{ij}\kappa_{ij}$.
Hence,
$$
  \widetilde{\alpha}_i^+-\widetilde{\alpha}_i
  =\frac{p_{ij}x_{ij}}2
  -0.3\,\tilde p_{ij}x_{ij}\kappa_{ij}
  +0.15\,\widetilde{\Delta}_{ij}(x_{ij}).
$$

Since $\tilde p_{ij}/p_{ij}=\min\{a,1\}$, the increase in $\widetilde{\alpha}_i$ per unit of allocated load is
\begin{equation}\label{eqn:load-normalized-increment}
  z=\frac{\widetilde{\alpha}_i^+-\widetilde{\alpha}_i}{p_{ij}x_{ij}}
  =\frac12-0.3\min\{a,1\}\kappa_{ij}+0.3a c_{ij}.
\end{equation}

Using $\widetilde{\alpha}_i=\alpha_i - R_i = \alpha_i(1-r_i)$ and expanding the square gives
\begin{equation}\label{eqn:load-normalized-charge}
  \frac{(\widetilde{\alpha}_i^+)^2-\widetilde{\alpha}_i^2}{p_{ij}^2x_{ij}}
  =2a(1-r_i)z+x_{ij}z^2.
\end{equation}

By Equations~\eqref{eqn:load-local-energy}, \eqref{eqn:load-normalized-saving}, \eqref{eqn:load-normalized-increment}, and~\eqref{eqn:load-normalized-charge}, it suffices to prove a piecewise polynomial inequality
\begin{equation}\label{eqn:load-polynomial-check}
  1+4a-2a c_{ij}
  \le\Gamma_0\left(
  \Phi\!\left(a+\frac{x_{ij}}2;1\right)
  -2a(1-r_i)z-x_{ij}z^2
  \right)
\end{equation}
in the five variables $a,x_{ij},c_{ij},\kappa_{ij},r_i$, with $z$ given by \eqref{eqn:load-normalized-increment}.

\paragraph{Constraints.}
We claim the following inequalities.
\begin{enumerate}
  \item\label{item:load-constraint-x} $0<x_{ij}<0.15$.
        \hspace*{\fill} (Lemma~\ref{lem:load-mass})
  \item\label{item:load-constraint-c} $0\le c_{ij}\le\kappa_{ij}$.
        \hspace*{\fill} (Definition of $c_{ij}$ and $\kappa_{ij}$, and $0\le f_i\le1$)
  \item\label{item:load-constraint-r} $0\le r_i\le0.01078$.
        \hspace*{\fill} (Inequality \eqref{eqn:load-reserve-bound})
  \item\label{item:load-constraint-a} $0<a<7/4$.
  \item\label{item:load-constraint-kappa} $0.219-0.43x_{ij}\le\kappa_{ij}\le0.22$.
  \item\label{item:load-constraint-credit-reserve} $r_i\ge0.07c_{ij}-0.004$.
\end{enumerate}

We next prove the last three constraints one by one.
For the fourth constraint, we already know $a>0$.
Suppose for contradiction that $a\ge7/4$.
Then $\alpha_i^+ \ge \alpha_i \ge7p_{ij}/4>p_{ij}$, so $\Phi(\alpha_i^+;p_{ij})=2p_{ij}\alpha_i^+$.
The contribution to the dual increase, divided by $p_{ij}^2x_{ij}$, is therefore
$$
  \frac{x_{ij}\Phi(\alpha_i^+;p_{ij})-\bigl((\alpha_i^+)^2-\alpha_i^2\bigr)}
  {p_{ij}^2x_{ij}}
  =2\left(a+\frac{x_{ij}}2\right)-\left(a+\frac{x_{ij}}4\right) =a+\frac{3x_{ij}}4.
$$
The left side of \eqref{eqn:load-independent}, divided by the same factor, equals $1+4a$ because $\ell_i=2\alpha_i$.
We have $\Gamma_0\bigl(a+\frac{3x_{ij}}4\bigr)\ge\Gamma_0a>1+4a$, since $(\Gamma_0-4)a\ge0.573\cdot(7/4)>1$.
Thus \eqref{eqn:load-independent} holds strictly, so entry $(i,j)$ would be rounded independently, contradicting $j\sim i$.

For the fifth constraint, Fubini's theorem gives
$$
  \kappa_{ij} = \frac1{x_{ij}}\int_0^{x_{ij}}\int_s^\infty\delta(u)\,du\,ds
  =\int_0^\infty\delta(u)\,du-\frac1{x_{ij}}\int_0^{x_{ij}}\int_0^s\delta(u)\,du\,ds.
$$
It then follows by $0\le\delta(u)\le0.86$, and the numerical fact that $0.219\le\int_0^\infty\delta(u)\,du\le0.22$.

For the final constraint, we have
\begin{align*}
  0.07c_{ij}-r_i
   & \le\int_0^\infty f_i(s)
  \left(0.07\delta(s)-0.3\int_s^\infty\delta(u)\,du\right)ds
  \tag{monotonicity of $\delta$} \\
   & \le\int_0^\infty
  \max\left\{0.07\delta(s)-0.3\int_s^\infty\delta(u)\,du,\,0\right\}ds
  \tag{$0\le f_i\le1$}           \\
   & <0.004.
  \tag{numerical integration}
\end{align*}

\paragraph{Numerical Verification.}
We verify that the left side of \eqref{eqn:load-polynomial-check} minus its right side is negative after substituting \eqref{eqn:load-normalized-increment}.
Cover the feasible domain by the substitutions
\begin{gather*}
  a=
  \begin{cases}
    (1-x_{ij}/2)\xi_1 & 0\le a\le1-x_{ij}/2 \\
    1-(x_{ij}/2)\xi_1 & 1-x_{ij}/2\le a\le1 \\
    1+\tfrac34\xi_1   & 1\le a\le7/4
  \end{cases}, \\[2pt]
  x_{ij}=0.15\xi_2,\quad
  c_{ij}=\kappa_{ij}\xi_3,\quad
  \kappa_{ij}=0.22-(0.43x_{ij}+0.001)\xi_4,\quad
  r_i=0.07c_{ij}-0.004+0.01478\xi_5,
\end{gather*}
for $0\le\xi_1,\ldots,\xi_5\le1$.
The substitutions cover the feasible domain.
Each of the three cases for $a$ gives a polynomial on $[0,1]^5$.

For each polynomial, start with the box $[0,1]^5$.
Rescale the current box to $y\in[0,1]^5$ and expand the polynomial in monomials.
Since $0\le y^m\le1$, we have
$$
  b_0+\sum_{m\ne0}b_m y^m
  \le b_0+\sum_{m\ne0}\max\{b_m,0\}.
$$
If this bound is not below $0$, bisect the box and repeat on both halves, cycling through coordinates $1,\ldots,5$ along each branch.
With exact rational arithmetic, this procedure terminates with $8056$ boxes across the three cases, proving \eqref{eqn:load-polynomial-check}.

\end{document}